\documentclass[
amsmath,
prd,nofootinbib,floatfix,11pt,
]{revtex4}

\usepackage{epsf}
\usepackage{setspace}
\usepackage{bm}
\usepackage{accents}

\usepackage[dvips]{color}
\definecolor{Red}{cmyk}{0,1,1,0}
\definecolor{BrickRed}{cmyk}{0,0.89,0.94,0.28}
\definecolor{Blue}{cmyk}{1,1,0,0}

\DeclareMathSymbol{\widetildesym}{\mathord}{largesymbols}{"65}
\newcommand\lowerwidetildesym{%
  \text{\smash{\raisebox{-1.3ex}{%
    $\widetildesym$}}}}
\newcommand\fixwidetilde[1]{%
  \mathchoice
    {\accentset{\displaystyle\lowerwidetildesym}{#1}}
    {\accentset{\textstyle\lowerwidetildesym}{#1}}
    {\accentset{\scriptstyle\lowerwidetildesym}{#1}}
    {\accentset{\scriptscriptstyle\lowerwidetildesym}{#1}}
}

\newcommand\beq{\begin{eqnarray}}
\newcommand\eeq{\end{eqnarray}}
\def\lsim{\mathrel{\rlap{\lower4pt\hbox{$\sim$}}
    \raise1pt\hbox{$<$}}}                
\def\gsim{\mathrel{\rlap{\lower4pt\hbox{$\sim$}}
    \raise1pt\hbox{$>$}}}

\newcommand\IIxyz{{\bf I}(x,y,z) }
\newcommand\AAx{{\bf A}(x) }
\newcommand\AAy{{\bf A}(y) }
\newcommand\AAz{{\bf A}(z) }
\newcommand\sigmabar{\overline\sigma}
\newcommand\xiprime{\widetilde\xi}
\newcommand\xiprimesq{{\widetilde \xi}^2}
\allowdisplaybreaks
\newcommand\metric{\eta}
\newcommand\lnbar{\overline{\ln}}
\newcommand\Fbar{\overline{F}}
\newcommand\Vbar{\overline{V}}
\newcommand\MSbar{$\overline{\rm{MS}}$ }

\newcommand\DRbarprime{$\overline{\rm{DR}}'$ }
\newcommand\eps{\epsilon}
\newcommand\epsp{{\epsilon'}}
\newcommand\epspp{{\epsilon''}}

\begin{document}

\renewcommand{\theequation}{\arabic{section}.\arabic{equation}}
\renewcommand{\thefigure}{\arabic{section}.\arabic{figure}}
\renewcommand{\thetable}{\arabic{section}.\arabic{table}}

\title{\Large \baselineskip=20pt 
Two-loop effective potential for generalized gauge fixing}

\author{Stephen P.~Martin$^{1}$ and Hiren H.~Patel$^2$}
\affiliation{\baselineskip=15pt
\mbox{\it $^1$Department of Physics, 
Northern Illinois University, DeKalb IL 60115}\\
\mbox{\it $^2$Amherst Center for Fundamental Interactions, Department of Physics,}\\ 
\mbox{University of Massachusetts, Amherst, MA 01003}
}

\begin{abstract}\normalsize \baselineskip=16pt 
We obtain the two-loop effective potential for general renormalizable theories, using 
a generalized gauge-fixing scheme that includes as 
special cases the background-field $R_\xi$ gauges, the Fermi gauges, and 
the familiar Landau gauge, and using dimensional regularization
in the bare and \MSbar renormalization schemes. As examples, the results are then 
specialized to the Abelian Higgs model and to the Standard Model. 
In the case of the Standard Model, we study how the vacuum 
expectation value and the minimum vacuum energy depend numerically on 
the gauge-fixing parameters. 
The results at fixed two-loop order exhibit non-convergent behavior 
for sufficiently large gauge-fixing parameters; this can presumably
be addressed by a resummation of higher-order contributions.
\end{abstract}

\maketitle

\vspace{-0.4in}

\baselineskip=14pt

\tableofcontents

\newpage

\baselineskip=15.4pt

\section{Introduction\label{sec:intro}}
\setcounter{equation}{0}
\setcounter{figure}{0}
\setcounter{table}{0}
\setcounter{footnote}{1}

The effective potential \cite{Coleman:1973jx,Jackiw:1974cv,Sher:1988mj} 
is a useful tool for the quantitative understanding of spontaneous 
symmetry breaking, with the most obvious application being to 
electroweak symmetry breaking in the Standard Model and its extensions.

In gauge theories, the effective potential is 
simplest and easiest to compute in Landau gauge. The 2-loop order effective potential 
was originally obtained for the Standard Model in \cite{Ford:1992pn}, and 
extended to general theories in \cite{Martin:2001vx}. The leading 
3-loop contributions for the Standard Model were obtained in ref.~\cite{Martin:2013gka} 
in the approximation that the QCD and top-quark 
Yukawa couplings are treated as much larger than the other dimensionless couplings. 
These results were then extended to full 3-loop order for a general 
theory in ref.~\cite{Martin:2017lqn}, where they were written in terms of
the basis of 3-loop vacuum integral functions with arbitrary masses, 
as given in \cite{Martin:2016bgz}.
(For an alternative treatment of the necessary basis integral 
functions, see \cite{Bauberger:2017nct}.)
When the tree-level Goldstone boson squared mass is small or negative, as 
indeed occurs in the Standard Model,
infrared (IR) divergences or spurious imaginary parts arise in the 
effective potential, but it has been shown that a resummation of Goldstone boson 
propagator contributions 
cures this issue \cite{Martin:2014bca,Elias-Miro:2014pca}; 
for further development and related perspectives, see
\cite{Pilaftsis:2015cka,Pilaftsis:2015bbs,Kumar:2016ltb,Braathen:2016cqe,Pilaftsis:2017enx,Braathen:2017izn,Espinosa:2017aew}.
The 4-loop contributions to the Standard Model effective
potential at leading order in QCD are also known \cite{Martin:2015eia}. 
One application of these results is to precision calculations of physical masses
and other observables 
in the Standard Model using the tadpole-free scheme, which means that perturbation
theory is organized around a vacuum expectation value (VEV) defined as the minimum
of the effective potential. This contrasts with 
the choice of expanding around the minimum of the tree-level potential,
which is often done but then requires inclusion of tadpole diagrams 
and has formally slower convergence properties.
Full two-loop electroweak corrections to the Higgs boson, $W$, $Z$, and top-quark masses
in this tadpole-free scheme have been give in 
refs.~\cite{Martin:2014cxa,Martin:2015lxa,Martin:2015rea,Martin:2016xsp}; these 
rely on the two-loop Standard Model effective potential result.
Softly-broken supersymmetric theories require
a different renormalization scheme based on dimensional reduction
rather than dimensional regularization, and the 2-loop effective potential
for the minimal supersymmetric extension of the Standard Model 
has been obtained accordingly in 
refs.~\cite{Hempfling:1993qq,Zhang:1998bm,Espinosa:1999zm,Espinosa:2000df},
\cite{Martin:2001vx}, \cite{Martin:2002iu}. 
All of these multi-loop results have been obtained in Landau 
gauge and no other, up to now. 
We think it is reasonable to assert that Landau gauge is the preferred choice 
whenever the effective potential plays a central role in high precision calculations.

However, it is also sometimes considered beneficial to make use of gauge 
invariance as a check of both calculations and conceptual understanding. 
This can be done by 
considering the effective potential obtained with other gauge-fixing 
schemes. It has long been understood \cite{Jackiw:1974cv,Dolan:1974gu} 
that the effective potential, and the vacuum expectation values of 
scalar fields defined by its minimum, does depend on the gauge-fixing choice. 
This is not a problem, because physical observables following from the 
effective potential, including its values at local minima, pole masses 
of particles, and properly defined transition rates, are independent of 
the choice of gauge fixing. Important results and a variety of 
perspectives on the issues related to the gauge dependence of the 
effective potential and the gauge independence of physical observables 
can be found in
\cite{Jackiw:1974cv,
Dolan:1974gu,
Kang:1974yj,
Fischler:1974ue,
Frere:1974ia,
Nielsen:1975fs,
Fukuda:1975di,
Aitchison:1983ns,
Metaxas:1995ab,
Loinaz:1997td,
DelCima:1999gg,
Gambino:1999ai,
Alexander:2008hd,
Patel:2011th,
Lewandowski:2013uha,
DiLuzio:2014bua,
Nielsen:2014spa,
Andreassen:2014eha,
Andreassen:2014gha,
Espinosa:2015qea,
Plascencia:2015pga,
Lalak:2016zlv,
Espinosa:2016uaw,
Espinosa:2016nld,
Endo:2017gal,
Irges:2017ztc}.
The Nielsen identities \cite{Nielsen:1975fs,Fukuda:1975di} parameterize 
the fact that the gauge-fixing dependence of the effective potential can 
always be absorbed into a redefinition of the scalar fields. However, 
these identities hold to all orders in perturbation theory, and 
practical results that are truncated at finite order often require a 
careful treatment in order to demonstrate gauge-fixing 
independence of physical quantities. In some cases, there are subtleties
involved in verifying that a particular version of a
calculated quantity of interest is really a physical observable. 
Recently, it has been argued that resummations of diagrams to all orders
in perturbation theory are necessary to make manifest the 
gauge-fixing independence \cite{Andreassen:2014eha} and to cure 
\cite{Espinosa:2016uaw} 
related infrared (IR) divergence problems 
\cite{Aitchison:1983ns,Loinaz:1997td}
that occur in Fermi gauges.

One of the uses of the effective potential is to study the 
stability of the Standard Model vacuum with respect to the Higgs field
\cite{Lindner:1988ww,
Arnold:1991cv,
Ford:1992mv,
Casas:1994qy,
Espinosa:1995se,
Casas:1996aq},
\cite{Loinaz:1997td},
\cite{Isidori:2001bm,
Espinosa:2007qp,
ArkaniHamed:2008ym,
Bezrukov:2009db,
Ellis:2009tp,
EliasMiro:2011aa,
Bezrukov:2012sa,
Degrassi:2012ry,
Alekhin:2012py,
Buttazzo:2013uya},
\cite{DiLuzio:2014bua,
Espinosa:2015qea,
Espinosa:2016nld},
\cite{Andreassen:2017rzq,
Chigusa:2017dux}.
The observed value of the Higgs boson mass near 125 GeV implies that the 
electroweak vacuum is metastable, if one assumes that the Standard Model 
holds without extension up to very high energy scales. As noted particularly
in \cite{Loinaz:1997td,DiLuzio:2014bua}, it is non-trivial to identify
an instability scale that is gauge-independent. Care is needed to identify physical
observables correlated with the vacuum instability problem, and to 
ensure that practical calculations of them in perturbation theory 
maintain the gauge invariance that in principle should govern an all-orders 
calculation, as dictated by the Nielsen identities.

In this paper, we provide a calculation of the 2-loop effective potential
in a general linear gauge-fixing scheme, but leave aside such issues as
resummation. We will provide results for a 
general gauge theory, and then
specialize to the Abelian Higgs model and the Standard Model as examples.
 
To establish notations and conventions,
let us write the bosonic 
degrees of freedom in the Lagrangian as a list of real 
gauge vector bosons $A^a_\mu(x)$ and a list of real scalar fields 
$\Phi_j(x)$. The latter transform under the gauge group with generators 
$t^{a}_{jk}$, which are Hermitian, antisymmetric, and therefore 
purely imaginary matrices. The indices $j,k,\ldots$ label the real scalars, 
and $a,b,\ldots$ are adjoint representation indices for the real vector 
fields $A^a_\mu$, with coupling constants $g_a$ and totally 
antisymmetric structure constants $f^{abc}$, determined by $[t^a, t^b] = 
i f^{abc} t^c$. 
Before gauge fixing, the Lagrangian is:
\beq
{\cal L}_{\rm YM} &=& -\frac{1}{4} F^{\mu\nu a} F_{\mu\nu}^a
- \frac{1}{2} D^\mu \Phi_j D_\mu \Phi_j - V(\Phi_j)
+ {\cal L}_{\rm fermions}
,
\label{eq:LYM}
\eeq
where $V(\Phi_j)$
is the tree-level scalar potential, 
and\footnote{The metric signature is ($-$,$+$,$+$,$+$).
Throughout this paper, by convention, repeated indices in each term
are implicitly summed over, unless they appear on both sides of an equation. 
Thus, $a$ is summed over in the last term of eq.~(\ref{eq:covariantDPhi}), 
but not in eq.~(\ref{eq:Fmunu}).}
\beq
F_{\mu\nu}^a &=& \partial_\mu A_\nu^a - \partial_\nu A_\mu^a
+ g_a f^{abc} A_\mu^b A_\nu^c,
\label{eq:Fmunu}
\\
D_\mu \Phi_j &=& \partial_\mu \Phi_j - i g_a A^a_\mu t^{a}_{jk} \Phi_k .
\label{eq:covariantDPhi}
\eeq
Now we write each real scalar field as the sum of a constant background field 
$\phi_j$ and a dynamical field $R_j$,
\beq
\Phi_j(x) &=& \phi_j + R_j(x).
\eeq 
In this background, the fermion Lagrangian for a general renormalizable theory
can be written as
\beq
{\cal L}_{\rm fermions} &=&
i \psi^{\dagger I} \sigmabar^\mu D_\mu \psi_I 
- \frac{1}{2} (M^{II'} \psi_I \psi_{I'} + 
Y^{jIJ} R_j \psi_I \psi_J + {\rm c.c.})
.
\label{eq:Lfermions}
\eeq
Here $\psi_I$ are two-component left-handed fermion fields,
labeled by capital letters from the middle of the alphabet, $I,J,K,\ldots$.
The covariant derivative acting on fermions is
\beq
D_\mu \psi_I &=& \partial_\mu \psi_I - i g_a A^a_\mu T^{aJ}_I \psi_J ,
\label{eq:fermioncovariantderivative}
\eeq
with gauge group generator Hermitian matrices $T^{a J}_I$, which also satisfy
$[T^a, T^b] = i f^{abc} T^c$.
In eq.~(\ref{eq:Lfermions}), 
$Y^{jIJ}$ are Yukawa couplings, and $M^{II'}$ 
are $\phi_j$-dependent fermion masses.
It is assumed that (by performing an appropriate 
unitary rotation on the fermion indices) 
the fields $\psi_I$ have been arranged to be 
eigenstates of the background field-dependent squared masses 
\beq
M_I^2 = M_{I'}^2 = |M^{II'}|^2,
\eeq
such that the mass matrix $M^{II'}$ connects 
pairs of fermion fields with opposite conserved charges. 
Thus, it is understood that primed indices $I', J', K'\ldots$ label
the mass partners of fermions with the opposite
charges labeled $I,J,K,\ldots$ when they form
a Dirac pair, while $I'=I$ for each 
fermion with a Majorana mass and no conserved
charge left unbroken by the background fields $\phi_j$. 
Because two-component fermion fields are intrinsically complex, 
the heights of the fermion indices are significant,
and raising and lowering them is taken to indicate 
complex conjugation, so that:
\beq
M_{II'} = (M^{II'})^*;\qquad\quad Y_{jIJ} = (Y^{jIJ})^*;\qquad\quad
T^{aI}_J = (T^{aJ}_I)^*. 
\eeq

The effective potential is then a function of the constant 
background fields $\phi_j$, and can be evaluated in a loop expansion:
\beq
V_{\rm eff}(\phi_j) &=&
V^{(0)} (\phi_j)
+ \frac{1}{16 \pi^2} V^{(1)}(\phi_j)
+ \frac{1}{(16 \pi^2)^2} V^{(2)}(\phi_j) +
\ldots ,
\label{eq:Veffexp}
\eeq
where $V^{(0)} (\phi_j) = V(\phi_j)$ is the tree-level part, 
and the contribution $V^{(n)}$ is obtained for $n\geq 1$
from the sum of 1-particle irreducible $n$-loop Feynman diagrams with no external legs.
Carrying out the evaluation of the loop corrections
requires gauge fixing and regularization of divergences.

A useful consistency check is obtained from renormalization group invariance
of the \MSbar form of the effective potential. 
Writing the loop expansion of the beta function for each 
\MSbar parameter $X$ (including the background fields $\phi_j$, and the gauge-fixing 
parameters discussed below) as
\beq
Q \frac{dX}{dQ} \equiv \beta_X &=& \frac{1}{16\pi^2}  \beta^{(1)}_{X} 
+ \frac{1}{(16\pi^2)^2} \beta^{(2)}_{X} + \ldots,
\label{eq:betagenform}
\eeq
then the requirement
\beq
Q\frac{dV_{\rm eff}}{dQ} = \Bigl (Q\frac{\partial}{\partial Q} + 
\sum_X \beta_X \frac{\partial}{\partial X} \Bigr ) V_{\rm eff} &=& 0
\eeq
yields
\beq
Q\frac{\partial}{\partial Q} V^{(\ell)} + 
\sum_{n=0}^{\ell-1} \Bigl (
\sum_X \beta_X^{(\ell-n)} \frac{\partial}{\partial X} V^{(n)}
\Bigr ) &=& 0
\label{eq:RGcheck}
\eeq
at each loop order $\ell = 1,2,\ldots$.

\section{Generalized gauge fixing\label{sec:gaugefixing}}
\setcounter{equation}{0}
\setcounter{figure}{0}
\setcounter{table}{0}
\setcounter{footnote}{1}

To treat the gauge fixing, consider an off-shell 
BRST \cite{BRST}
formalism for the 
gauge invariance, with Grassmann-odd ghost and anti-ghost fields
$\eta^a$ and $\overline \eta^a$, and bosonic
Nakanishi-Lautrup \cite{NakanishiLautrup} auxiliary fields 
$b^a$. The BRST transformations of the fields are essentially gauge transformations
parameterized by the ghost fields $\eta^a$:
\beq
\delta_{\rm BRST} A_\mu^a &=& \partial_\mu \eta^a - g_a f^{abc} \eta^b A^c_\mu,
\\
\delta_{\rm BRST} R_j &=& i g_a \eta^a t^{a}_{jk} (\phi_k + R_k),
\\
\delta_{\rm BRST} \psi_I &=& i g_a \eta^a T^{aJ}_I \psi_I,
\\
\delta_{\rm BRST} \eta^a &=& -\frac{1}{2} g_a f^{abc} \eta^b \eta^c,
\\
\delta_{\rm BRST} \overline \eta^a &=& b^a ,
\\
\delta_{\rm BRST} b^a &=& 0.
\eeq
From these one can check the nilpotency of the BRST transformations:
\beq
\delta_{\rm BRST}(\delta_{\rm BRST} X) = 0
\eeq
for any field $X$. (Note that $\delta_{\rm BRST}$ is Grassmann-odd; 
it obtains a minus sign when moved past a fermion or ghost field.) 
The Lagrangian in eq.~(\ref{eq:LYM}) is invariant under this BRST transformation. 
Together, these facts mean that 
we can obtain a BRST-invariant gauge-fixed Lagrangian by:
\beq
{\cal L} &=& {\cal L}_{\rm YM} + {\cal L}_{\rm g.f.} + {\cal L}_{\rm ghost},
\eeq
where the gauge-fixing plus ghost part is obtained as a BRST variation:
\beq
{\cal L}_{\rm g.f.} + {\cal L}_{\rm ghost} &=& 
\delta_{\rm BRST} 
\Bigl (\overline \eta^a \Bigl [\frac{1}{2} \xi_a b^a - \partial^\mu A^{a}_{\mu}
- i g_a \fixwidetilde{\phi}_j^a t^a_{jk} R_k\Bigr] \Bigr ). 
\eeq
Here $\xi_a$ and $\fixwidetilde{\phi}_j^a$ are gauge-fixing parameters; in general
the latter may or may not be related to the background scalar fields $\phi_j$ that
the effective potential depends on. It follows that
\beq
{\cal L}_{\rm g.f.} &=& \frac{1}{2} \xi_a b^a b^a - b^a \Bigl (\partial^\mu A^{a}_{\mu}
+ i g_a \fixwidetilde{\phi}^a_j t^a_{jk} R_k  \Bigr )
\label{eq:gengaugefixingwithb}
\eeq
and
\beq
{\cal L}_{\rm ghost} &=& 
-\partial_\mu \overline \eta^a \partial^\mu \eta^a + 
g_a f^{abc} \partial^\mu \overline \eta^a \eta^b A_\mu^c
+[g_a t^a_{kj} \fixwidetilde{\phi}^a_j] 
[g_b t^b_{kl} (\phi_l + R_l)] \overline \eta^a \eta^b
.
\eeq
By integrating out the auxiliary fields $b^a$, one can re-write eq.~(\ref{eq:gengaugefixingwithb}) as:
\beq
{\cal L}_{\rm g.f.} &=& -\frac{1}{2 \xi_a} \left (
\partial^\mu A^{a}_{\mu} + i g_a \fixwidetilde{\phi}^a_j 
t^a_{jk} R_k \right )^2
.
\label{eq:gengaugefixingwithoutb}
\eeq

There are various special cases of the above general gauge-fixing condition that are of interest:
\begin{itemize}
\item \underline{Landau gauge:}~ 
$\fixwidetilde{\phi}^a_j = 0$ and $\xi_a \rightarrow 0$. This condition is 
renormalization group invariant, and avoids kinetic mixing between scalar and 
vector fields. The resulting simplicity is why this gauge condition is by far 
the most popular one for practical applications involving the effective potential.
\item \underline{Fermi gauges:}~ $\fixwidetilde{\phi}^a_j = 0$. 
This condition is renormalization group invariant.
However, the parameters $\xi_a$ do run with the renormalization scale 
(except when they vanish). A further complication is that 
when $\xi_a \not=0$, the scalar and vector fields have
propagator mixing with each other, which 
arises due to cross-terms 
$A^a_\mu \partial^\mu R_j$ 
in the scalar kinetic term in eq.~(\ref{eq:LYM}).
In the Landau gauge 
limit $\xi_a \rightarrow 0$, the effects of this cross-term disappear from
the scalar and vector propagators.
\item \underline{``Standard" $R_\xi$ gauges:}~  
$\fixwidetilde{\phi}^a_j =  \xi_a \phi_j^{{\rm cl}}$, 
where the $\phi_j^{{\rm cl}}$ are the classical VEVs that minimize the
tree-level scalar potential. 
This gauge-fixing condition is not renormalization group invariant. 
In applications other than the effective potential, one can also set the background
fields $\phi_j$ to be equal to $\phi_j^{{\rm cl}}$, which results in cancellation
of the scalar-vector propagator kinetic mixing.
However, when calculating the effective potential $V_{\rm eff}(\phi_j)$,
the whole point is to allow variation of the 
background scalar fields $\phi_j$ that appear in the scalar kinetic terms, 
the scalar potential, and in the fermion Lagrangian, so they
cannot be fixed equal to the tree-level VEVs $\phi^{{\rm cl}}_j$ that appear
in the gauge-fixing term. 
Therefore the  
$A^a_\mu \partial^\mu R_j$ cross-terms in the
scalar kinetic term in eq.~(\ref{eq:LYM}) do not cancel against those in
eq.~(\ref{eq:gengaugefixingwithoutb}),   
so that there is kinetic mixing between the
scalar and vector fields. 
\item \underline{Background-field $R_\xi$ gauges:}~ 
$\fixwidetilde{\phi}^a_j = \xi_a \phi_j$. 
This avoids kinetic mixing
between scalar and vector fields, by canceling the cross-terms 
$A^a_\mu \partial^\mu R_j$ 
in the
scalar kinetic term in eq.~(\ref{eq:LYM}) against those in
the gauge-fixing term eq.~(\ref{eq:gengaugefixingwithoutb}), after integration by parts. 
However, this condition 
is not renormalization group invariant, as noted immediately below.
\item \underline{Generalized background-field $R_{\xi,\xiprime}$ gauges:}~ 
$\fixwidetilde{\phi}^a_j = \xiprime_a \phi_j$
where $\xiprime_a$ is a gauge-fixing parameter that is taken to be independent of 
$\xi_a$. As a result, there is propagator kinetic mixing between the scalars and vectors,
proportional to $\xi_a - \xiprime_a$.
Also, it turns out that $\xi_a$ and $\xiprime_a$ have different counterterms, and
run differently with the renormalization scale (except in the Landau gauge case 
$\xiprime_a = \xi_a = 0$). To understand this, note that 
invariance of the Lagrangian under
the BRST symmetry does not require any special relationship 
between $\xi_a$ and $\xiprime_a$.
Therefore, they are free to be renormalized differently, and explicit calculation
(given below for the Abelian Higgs model and the Standard Model) shows that 
indeed they are. In contrast, while $\xiprime_a$ 
appears in both ${\cal L}_{\rm g.f.}$ and ${\cal L}_{\rm ghost}$, those 
instances of $\xiprime_a$ are required to be the same by the BRST invariance.
\end{itemize}

In this paper, we choose to specialize slightly to a particular
version of the last, 
generalized background-field $R_{\xi,\xiprime}$ gauge-fixing condition.
However, the 37 two-loop effective potential functions 
that we will use to write the results [listed below in eq.~(\ref{eq:listfs}), and
evaluated in eqs.~(\ref{eq:fSSSbare})-(\ref{eq:ffFGbare}) and 
(\ref{eq:deffSSS})-(\ref{eq:defffFG})]
are actually generally applicable, because they correspond to 
the complete set of two-loop vacuum Feynman diagram topologies, and 
so in principle are sufficient to evaluate the two-loop effective potential 
even in the case of arbitrary $\tilde \phi^a_j$, or if the 
parameter $\xi_a$ is generalized to a matrix $\xi_{ab}$.

To see why the qualifier ``particular version" appears in the preceding paragraph,
note that when the rank of the gauge group is larger than 1, the gauge fixing actually 
depends on a choice of basis for the gauge generators,
because the form of eq.~(\ref{eq:gengaugefixingwithoutb}) is not 
invariant\footnote{We will discuss this further in the concrete example of 
the Standard Model, in section \ref{subsec:StandardModel}.}  
under an arbitrary orthogonal rotation of
the real vector labels $a$. To choose a nice basis, 
consider the real rectangular matrix:
\beq
F^a{}_j &\equiv& i g_a t^a_{jk} \phi_k .
\label{eq:defFaj}
\eeq
The Singular Value Decomposition Theorem of linear algebra says that a 
real rectangular
matrix can be put into a diagonal form by an invertible change of basis, so that
for some (perhaps background field-dependent) 
orthogonal matrices $({\cal O}_V)^{ab}$ and $({\cal O}_S)_{kj}$,
\beq
({\cal O}_V)^{ab} F^b{}_k ({\cal O}_S)_{kj} = M_a \delta^a_j .
\label{eq:rotateFaj}
\eeq
Assume that we have already rotated to the diagonal basis, which will be distinguished
from now on by boldfaced indices ${\bf a}, {\bf b}, \bf{c}, \ldots$ for the vectors,
and ${\bf j}, {\bf k}, {\bf l},\ldots$ for the scalars, so that:
\beq
F^{\bf a}{}_{\bf j} \>=\> M_{\bf a} \delta^{\bf a}_{\bf j},
\label{eq:diagbasis}
\eeq
where the $M_{\bf a}$ are the singular values, with magnitudes equal to 
the gauge boson masses. In general, this basis will mix 
vector bosons belonging to different simple or $U(1)$ factors of the gauge 
Lie algebra; in particular, this occurs in the Standard Model, where the 
mass eigenstate $Z$ boson and photon are mixtures of the $SU(2)_L$ and 
$U(1)_Y$ gauge eigenstate vector fields.

In this basis, eq.~(\ref{eq:diagbasis}) provides a natural 
correspondence between the massive vector bosons and a subset of the dynamical
scalar bosons.
The members of this subset of the scalar bosons will be 
called Goldstone scalars because of this association with 
massive vector bosons and therefore with broken generators. 
However, the 
contributions to the Goldstone scalar tree-level squared masses from the scalar 
potential $V$ do not vanish, because we are not expanding around the 
minimum of the tree-level potential.

It is convenient to split the lists of 
real vector fields and real scalar
fields into those
which have non-zero $M_{\bf a}$, denoted by $Z^A_\mu$ and $G_A$, respectively, and 
the remaining ones, which will be denoted by $A^{a}_\mu$ and $R_{j}$.
Thus,
indices $A,B,C,\ldots$ are used to span the sub-spaces corresponding to 
massive vectors and their corresponding Goldstone scalars, while from now on
non-boldfaced indices $a,b,c,\ldots$ span only the complementary subspace 
for massless vectors,
and non-boldfaced $j,k,l,\ldots$ now span only the complementary 
subspace of non-Goldstone scalars.
Thus the lists of vectors and scalars split up as:
\beq
\{ A^{\bf a}_\mu \} = \{Z^A_\mu,\, A^{a}_\mu\},\qquad\quad
\{ R_{\bf j} \} = \{G_A,\, R_{j}\}.
\eeq
The ghosts and anti-ghosts also split into these sectors 
in the same way as the vectors:
\beq
\{ \eta^{\bf a} \} = \{\eta^A,\, \eta^a\},
\qquad\quad
\{ \overline\eta^{\bf a} \} = \{\overline\eta^A,\, \overline\eta^a\},
\eeq
where the same orthogonal rotation on the adjoint representation
indices has been used as for the vector fields.
One can also write:
\beq
\{M_{\bf a}\} &=& \{M_A,\> 0\},
\label{eq:Mbreakdown}
\\
\{\xi_{\bf a}\} &=& \{\xi_A,\>\xi_a\},
\\
\{\xiprime_{\bf a}\} &=& \{\xiprime_A,\> 0\}.
\label{eq:xiprimebreakdown}
\eeq
The vanishing of $\xiprime_a$ in eq.~(\ref{eq:xiprimebreakdown}) follows from 
eq.~(\ref{eq:Mbreakdown}), 
because the $\xiprime_{\bf a}$ always appear multiplied 
by the corresponding $M_{\bf a}$. In the 
following, the gauge interaction terms in the Lagrangian will be written 
in terms of couplings:
\beq
g^{\bf abc},\qquad
g^{\bf a}_{\bf jk},\qquad
g^{{\bf a} J}_I,
\eeq
which are obtained respectively from the couplings
$g_a f^{abc}$, 
$i g_{a} t^{a}_{jk}$, 
and 
$g_a T^{aJ}_I$ appearing in eq.~(\ref{eq:fermioncovariantderivative}),
by performing the same basis change via
orthogonal rotations on vector and scalar
indices as in eq.~(\ref{eq:rotateFaj}). Note that
we rely on the index height to distinguish these
vector-vector-vector, vector-scalar-scalar, and vector-fermion-fermion
interaction couplings, 
because they all use the letter $g$, 
and because scalar and vector indices can both be $A,B,\ldots$.

The gauge-fixing and ghost terms in the Lagrangian then become:
\beq
{\cal L}_{\rm g.f.} &=&
-\frac{1}{2\xi_A} \bigl (\partial^\mu Z_\mu^A - \xiprime_A M_A G_A \bigr )^2
- \frac{1}{2\xi_{a}}\bigl (\partial^\mu A_\mu^{a} \bigr )^2 ,
\\
{\cal L}_{\rm ghost} &=&
-\partial^\mu \overline \eta^{{\bf a}} \partial_\mu \eta^{{\bf a}} 
- \xiprime_A M_A^2 \overline \eta^A \eta^A
+ g^{\bf abc} A^{\bf a}_\mu  \eta^{\bf b} \partial^\mu \overline \eta^{\bf c} 
- \xiprime_A g^{\bf a}_{A{\bf j}} M_A  R_{\bf j} 
\overline \eta^A \eta^{\bf a}.
\eeq
This gauge-fixing can be specialized to the Landau gauge (by taking 
$\xiprime_A =0$
and  $\xi_A, \xi_{a} \rightarrow 0$), or the Fermi gauges 
(by taking $\xiprime_A = 0$), or the background-field $R_\xi$
gauges either in the bare theory or at some particular renormalization scale 
(by taking $\xiprime_A= \xi_A$).

There are contributions to the scalar squared masses from the tree-level
potential: 
\beq
\mu^2_{\bf jk} = \frac{\partial^2 V}{\partial R_{\bf j} \partial R_{\bf k}} 
\Bigl |_{R_{\bf n} = 0},  
\label{eq:defmu2jk}
\eeq
which, in the basis we are using, can be divided into sectors as:
\beq
\begin{pmatrix} \mu^2_{jk}\phantom{x}   & \mu^2_{Bj} \\
\mu^2_{Ak}\phantom{x} & \mu^2_{AB}
\end{pmatrix}
.
\label{eq:defmu2jksectors}
\eeq
One can always specify a basis consistent with the one chosen so far, 
by doing a further rotation (if necessary)
among only the non-Goldstone scalar fields $R_{j}$, with the result that 
\beq
\mu^2_{jk} = \mu^2_{j} \delta_{jk}
\eeq 
is diagonal. However, in the most general case $\mu^2_{AB}$ is not diagonal
and $\mu^2_{Ak}$ need not vanish. In the remainder of this section 
we will discuss this general case, and in Section \ref{sec:examples} 
we will discuss the simplifications
that occur in the favorable case 
$\mu^2_{AB} = \delta_{AB} \mu^2_A$ and $\mu^2_{Ak} = 0$, 
with examples including the Abelian Higgs model and the Standard Model. 

\begin{figure}[!t]
\epsfxsize=0.75\linewidth
\epsffile{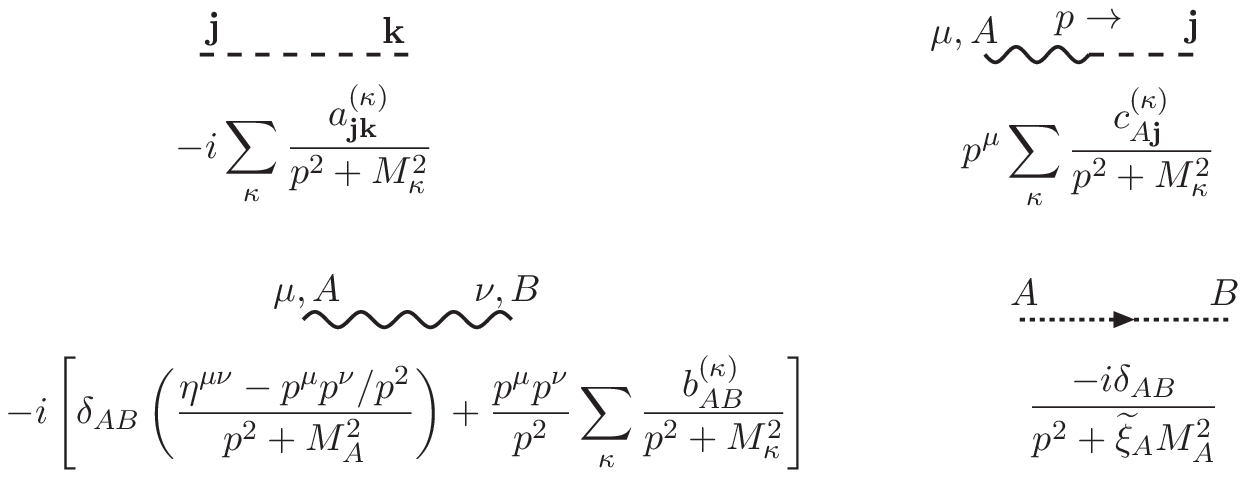}
\vspace{-0.4cm}
\begin{center}\begin{minipage}[]{0.95\linewidth}
\caption{\label{fig:propagatorsverygeneral}The scalar, massive vector 
$Z_A^\mu$, and corresponding ghost propagators, 
in the general case of arbitrary mixing between Goldstone scalars 
and other scalar and vector bosons.
The squared mass poles $M^2_\kappa$ arise as the roots of a polynomial of order $N$
in $-p^2$, where $N$ is the total number of real scalar bosons and massive vector bosons,
with $\kappa = 1,\ldots,N$.}
\end{minipage}\end{center}
\end{figure}

\begin{figure}[!t]
\epsfxsize=0.575\linewidth
\epsffile{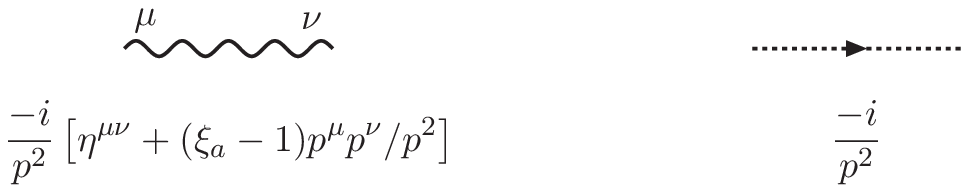}
\vspace{-0.4cm}
\begin{center}\begin{minipage}[]{0.95\linewidth}
\caption{\label{fig:propagatorsverygeneralmassless}Feynman rules for the propagators of the 
massless vectors $A_\mu^{a}$ 
(wavy lines), and the corresponding massless ghosts $\overline \eta^{a},  \eta^{a}$ 
(dotted lines with arrows), each carrying 4-momentum $p^\mu$.}
\end{minipage}\end{center}
\end{figure}

\begin{figure}[!t]
\epsfxsize=0.84\linewidth
\epsffile{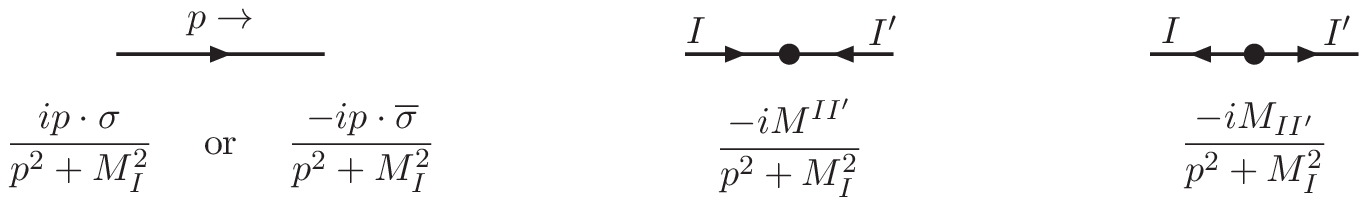}
\vspace{-0.4cm}
\begin{center}\begin{minipage}[]{0.95\linewidth}
\caption{\label{fig:fermionprops}Feynman rules for the propagators of the 
two-component fermions (using the conventions of 
\cite{Dreiner:2008tw,Martin:2012us}),
each carrying 4-momentum $p^\mu$. 
The arrows follow the helicity, and the large dots represent
fermion mass insertions.}
\end{minipage}\end{center}
\end{figure}

The part of the Lagrangian quadratic in the bosonic and ghost fields 
is, after integration by parts:
\beq
{\cal L} &=& 
\frac{1}{2} R_{j} \left [ \partial^2 - \mu_{j}^2 \right ] R_{j}
+ \frac{1}{2} G_A \left [ \partial^2 - (\xiprime_A^2/\xi_A) M_A^2\right ] G_A
- \frac{1}{2} \mu^2_{AB} G_A G_B - \mu^2_{Aj} G_A R_j  
\nonumber \\ &&
+ \frac{1}{2} A_\mu^{a} \left [\metric^{\mu\nu} \partial^2 + (1/\xi_{a} - 1)
\partial^\mu\partial^\nu \right ] A_\nu^{a}
+
\frac{1}{2} Z_\mu^{A} \left [\metric^{\mu\nu} (\partial^2 - M_A^2) + (1/\xi_{A} - 1)
\partial^\mu\partial^\nu  \right ] Z_\nu^{A}
\nonumber \\ &&
+ M_A (1 - \xiprime_A/\xi_A) Z_\mu^A \partial^\mu G_A 
+ \overline \eta_A [\partial^2 - \xiprime_A M_A^2 ] \eta_A
+ \overline \eta_{a} \partial^2 \eta_{a}
\eeq
By taking the inverse of the quadratic kinetic differential operator,
one obtains propagator Feynman rules of the form shown in 
Figures \ref{fig:propagatorsverygeneral} 
and \ref{fig:propagatorsverygeneralmassless}.
The propagators for scalars and the massive vector bosons both
involve the same unphysical 
squared mass poles $M^2_\kappa$, labeled by $\kappa = 1,\ldots,N$,
with $N$ the total number of real scalars plus massive vector bosons. The 
$M^2_\kappa$ are the roots of a polynomial  in $-p^2$ of order $N$, involving
the quantities $M^2_A$, $\xi_A$, $\xiprime_A$, $\mu^2_j$, $\mu^2_{Aj}$, and $\mu^2_{AB}$.
The $M^2_\kappa$ 
may well be complex, and are not always obtainable in closed algebraic 
form, but can be solved for numerically on a case-by-case basis. 
The propagator Feynman rules also
involve residue coefficients $a_{\bf jk}^{(\kappa)}$, $b_{AB}^{(\kappa)}$, and
$c_{A\bf j}^{(\kappa)}$, which similarly require numerical 
evaluation in the most 
general case.
The massive vector boson propagators also have poles at the physical
squared masses $M_A^2$. 
The massless vectors and their corresponding ghosts are 
unmixed, and their propagator Feynman rules are shown in Figure \ref{fig:propagatorsverygeneralmassless}.
The two-component fermion propagators follow from eq.~(\ref{eq:Lfermions}) 
in the usual way \cite{Dreiner:2008tw,Martin:2012us}, 
and are shown in Figure \ref{fig:fermionprops}.

\begin{figure}[!tb]
\epsfxsize=0.94\linewidth
\epsffile{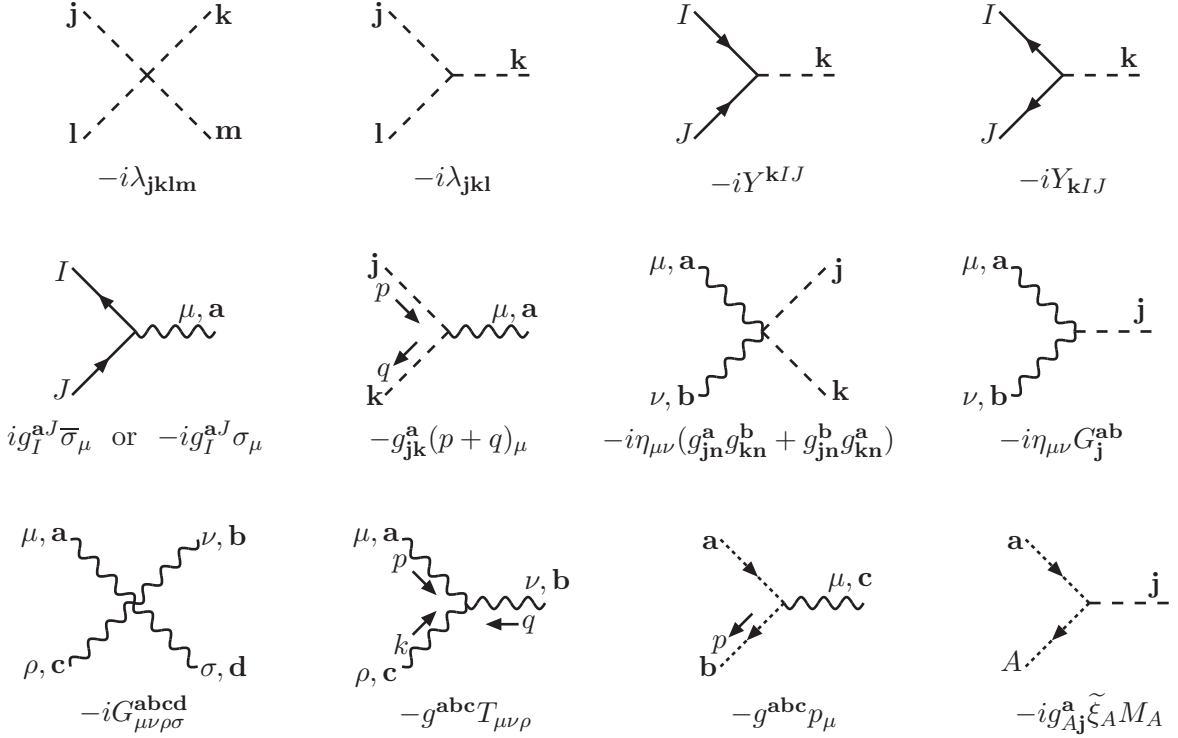}
\vspace{-0.25cm}
\begin{center}\begin{minipage}[]{0.95\linewidth}
\caption{\label{fig:geninteractions}
\baselineskip=15pt
Feynman rules for interactions. Dashed 
lines represent scalars, solid lines with arrows represent fermions,
wavy lines represent vectors, and dotted lines with arrows represent ghosts.
Bold-faced letters from the beginning of the alphabet 
($\bf a$, $\bf b$, $\bf c$, $\ldots$) run over all real vectors in the theory
and their corresponding ghosts and anti-ghosts.
Bold-faced letters from the middle of the alphabet 
($\bf j$, $\bf k$, $\bf l$, $\ldots$) run over all of the 
real scalars in the theory.
Capital letters from the middle of the alphabet $(I,J,\ldots)$ represent
two-component fermions. 
The vector-vector-scalar coupling $G^{\bf ab}_{\bf j}$ is given by
eqs.~(\ref{eq:defGabj})-(\ref{eq:defGABj}).
The (vector)$^4$ coupling is defined by 
$
G^{\bf abcd}_{\mu\nu\rho\sigma} = 
g^{\bf abe} g^{\bf cde} [\metric_{\mu\rho} \metric_{\nu\sigma} - 
\metric_{\mu\sigma} \metric_{\nu\rho}]
+ 
g^{\bf ace} g^{\bf bde} [\metric_{\mu\nu} \metric_{\rho\sigma} - 
\metric_{\mu\sigma} \metric_{\nu\rho}]
+ 
g^{\bf ade} g^{\bf bce} [\metric_{\mu\nu} \metric_{\rho\sigma} - 
\metric_{\mu\rho} \metric_{\nu\sigma}]. 
$
The (vector)$^3$ coupling tensor is defined by
$T_{\mu\nu\rho} = \metric_{\mu\nu} (p-q)_\rho + \metric_{\nu\rho} (q-k)_\mu
+ \metric_{\mu\rho} (k-p)_\nu$.
In the last, ghost-antighost-scalar, interaction, the index $A$ corresponds to a
vector with non-zero physical mass.}
\end{minipage}\end{center}
\end{figure}

The interaction part of the Lagrangian  
can now be written in the form:
\beq
{\cal L}_{\rm int} 
&=& -\frac{1}{6} \lambda^{\bf jkl} R_{\bf j} R_{\bf k} R_{\bf l} 
- \frac{1}{24} \lambda^{\bf jklm} R_{\bf j} R_{\bf k} R_{\bf l} R_{\bf m} 
- \frac{1}{2} (Y^{{\bf j}IJ} R_{\bf j} \psi_I \psi_J + {\rm c.c.})
\nonumber \\ &&
+ g_I^{{\bf a}J} A^{\bf a}_\mu \psi^{\dagger I} \overline \sigma^\mu \psi_J
- g^{\bf a}_{\bf jk} A_\mu^{\bf a} R_{\bf j} 
\partial^\mu\hspace{-0.2pt} R_{\bf k}
- \frac{1}{2} g^{\bf a}_{\bf jn} g^{\bf b}_{\bf kn} 
A_\mu^{\bf a} A^{\mu \bf b} R_{\bf j} R_{\bf k}
\nonumber \\ &&
- g^{\bf a}_{A\bf j} M_A R_{\bf j} Z_\mu^A A^{\mu \bf a} 
- g^{\bf a}_{A\bf j} \xiprime_A M_A R_{\bf j} \overline \eta^A \eta^{\bf a}
\nonumber \\ &&
-g^{\bf abc} A^{\mu \bf a} A^{\nu \bf b} \partial_\mu A^{\bf c}_\nu
- \frac{1}{4} g^{\bf abe} g^{\bf cde} A^{\mu \bf a} A^{\nu \bf b} 
A_\mu^{\bf c} A_\nu^{\bf d}
+ g^{\bf abc} A_\mu^{\bf a} \eta^{\bf b} \partial^\mu \overline \eta^{\bf c} ,
\label{eq:Lintgeneral}
\eeq
where the $\phi$-dependent (scalar)$^3$ and (scalar)$^4$ couplings
are defined from the tree-level scalar potential by
\beq
\lambda^{\bf jkl} &=& \frac{\partial^3 V}{\partial R_{\bf j} \partial R_{\bf k}
\partial R_{\bf l}} 
\Bigl |_{R_{\bf n} = 0}, 
\\
\lambda^{\bf jklm} &=& \frac{\partial^4 V}{\partial R_{\bf j} \partial R_{\bf k}
\partial R_{\bf l} \partial R_{\bf m} } .  
\eeq
The interaction vertex Feynman rules can be obtained 
in the usual way, and are shown in Figure \ref{fig:geninteractions}.
Here we have defined a vector-vector-scalar coupling $G^{\bf ab}_{\bf j}$ 
in terms of the scalar-scalar-vector coupling, according to:
\beq
G^{ab}_{\bf j} &=& 0,
\label{eq:defGabj}
\\
G^{Aa}_{\bf j} &=& G^{aA}_{\bf j} \>=\> g^a_{A\bf j} M_A,
\\
G^{AB}_{\bf j} &=& g^B_{A\bf j} M_A + g^A_{B\bf j} M_B.
\label{eq:defGABj}
\eeq 
\baselineskip=15.4pt

\section{Effective potential at two-loop order\label{sec:effpot}}
\setcounter{equation}{0}
\setcounter{figure}{0}
\setcounter{table}{0}
\setcounter{footnote}{1}

\subsection{General form\label{subsec:generalform}}

In this section we present the results for the effective potential, with 
the general gauge fixing described above.

The 1-loop effective potential contribution is:
\beq
V^{(1)} &=& 
\sum_\kappa f(\kappa)  
-2 \sum_I f(I) 
+ \sum_A [3 f_V(A) - 2 f(A_\eta)].
\label{eq:V1loopform}
\eeq
where $f(x)$ and $f_V(x)$ are renormalization
scheme-dependent loop integral functions, which will be 
given below in the bare and \MSbar renormalization schemes. 
Here and below, we use a notation in which an index is used as a synonym for the
squared mass whenever it appears as the
argument of a loop integral function. For example, 
in eq.~(\ref{eq:V1loopform}), 
$\kappa$ stands for $M_\kappa^2$, and $I$ stands for $M_I^2$, and 
$A$ for $M_A^2$, and we also use 
\beq 
A_\eta &=& \xiprime_A M_A^2,
\\
a_\eta &=& 0
\eeq
for the ghost squared masses.

\begin{figure}[p]
\epsfxsize=0.98\linewidth
\epsffile{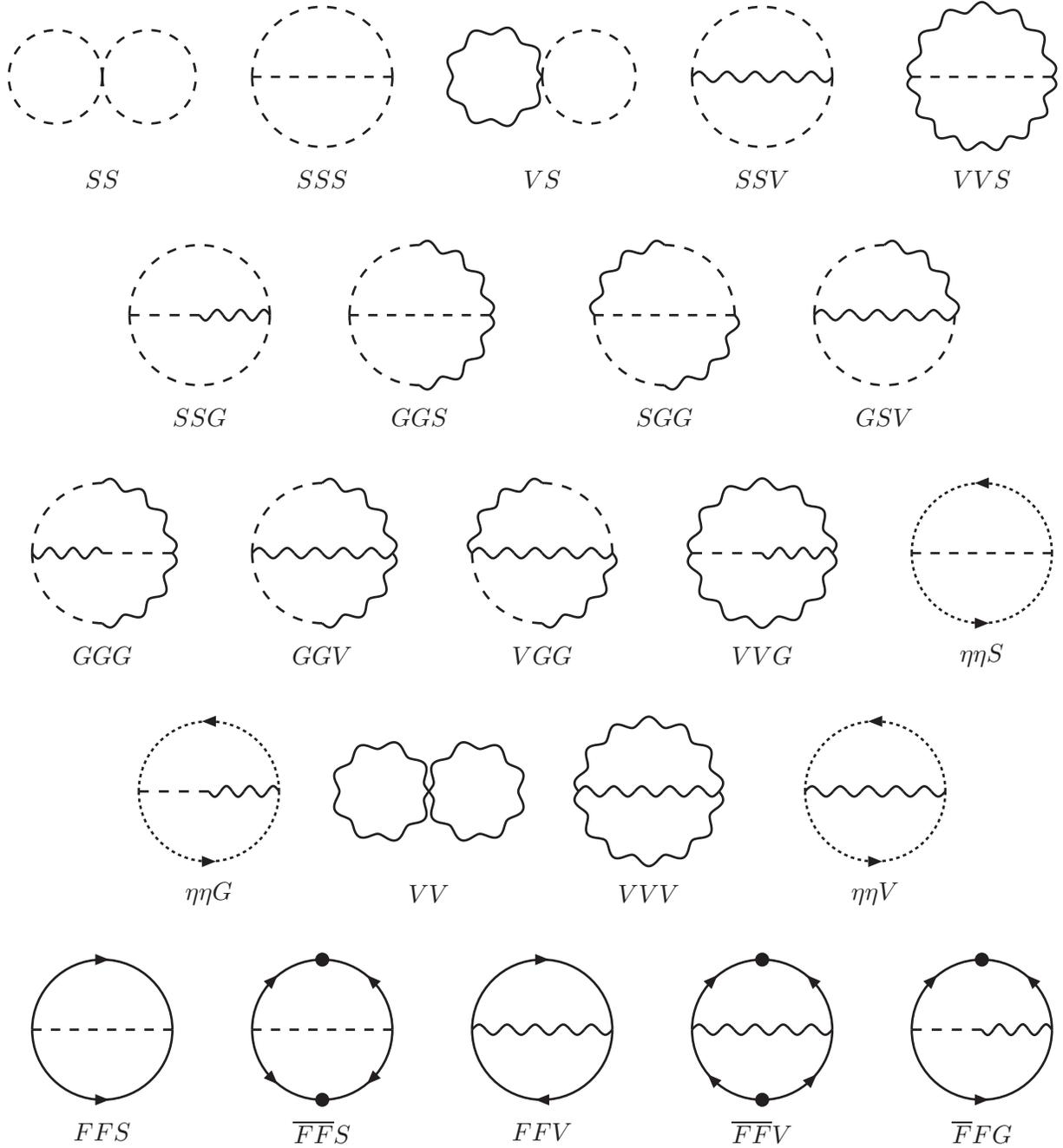}

\caption{\label{fig:feynmandiagrams}The non-vanishing 2-loop Feynman 
diagrams for the effective potential, for gauge-fixing choices that have 
propagator mixing between massive vectors and Goldstone scalars. Scalar 
bosons, fermions, vector bosons, and ghosts are represented by dashed, 
solid, wavy, and dotted lines, respectively. The arrows on fermion lines 
indicate the helicity, and large dots represent fermion mass insertions. 
For the $\Fbar\Fbar S$ and $\Fbar FG$ diagrams, there are also diagrams 
with all fermion arrows reversed.}
\end{figure}

For the 2-loop effective potential, there are 23 
non-vanishing Feynman diagrams,
shown in Figure \ref{fig:feynmandiagrams}.
It follows that the two-loop contributions to the effective potential 
are given, in terms of the couplings and propagator parameters defined above, by:
\beq
V^{(2)}_{SS} &=&
  \frac{1}{8} \lambda^{\bf jklm} a_{\bf jk}^{(\kappa)} a_{\bf lm}^{(\sigma)} 
  f_{SS}(\kappa, \sigma)
\label{eq:V2formSS}
,
\\
V^{(2)}_{SSS} &=&
  \frac{1}{12}  \lambda^{\bf jkl} \lambda^{\bf mnp}
  a_{\bf jm}^{(\kappa)} a_{\bf kn}^{(\sigma)} a_{\bf lp}^{(\rho)}
  f_{SSS}(\kappa, \sigma, \rho)
,
\\
V^{(2)}_{VS} &=&
  \frac{1}{2} g^{\bf a}_{\bf jk} g^{\bf a}_{\bf jl}
   a^{(\kappa)}_{\bf kl} f_{VS}({\bf a}, \kappa) 
+   \frac{1}{2} g^{A}_{\bf jk} g^{B}_{\bf jl}
   a^{(\kappa)}_{\bf kl} b^{(\sigma)}_{AB} f_{\Vbar S}(\sigma, \kappa) 
   , 
\\
V^{(2)}_{SSV} &=&
  \frac{1}{4} g^{\bf a}_{\bf jk} g^{\bf a}_{\bf lm}
  a_{\bf jl}^{(\kappa)} a_{\bf km}^{(\sigma)} 
  f_{SSV}(\kappa, \sigma, {\bf a}) +
 \frac{1}{4} g^{A}_{\bf jk} g^{B}_{\bf lm}
  a_{\bf jl}^{(\kappa)} a_{\bf km}^{(\sigma)} b_{AB}^{(\rho)}
  f_{SS\Vbar}(\kappa, \sigma, \rho) 
,  
\\
V^{(2)}_{VVS} &=& 
  \frac{1}{4} G^{\bf ab}_{\bf j} G^{\bf ab}_{\bf k} a_{\bf jk}^{(\kappa)} 
  f_{VVS}({\bf a},{\bf b}, \kappa) 
+ \frac{1}{2} G^{{\bf a}A}_{\bf j} G^{{\bf a}B}_{\bf k} a_{\bf jk}^{(\kappa)} b_{AB}^{(\sigma)} 
  f_{\Vbar VS}(\sigma, {\bf a},\kappa)
\nonumber \\ &&
+ \frac{1}{4} G^{AB}_{\bf j} G^{CD}_{\bf k} a_{\bf jk}^{(\kappa)} 
  b_{AC}^{(\sigma)} b_{BD}^{(\rho)} 
  f_{\Vbar\Vbar S}(\sigma, \rho,\kappa)
,
\\
V^{(2)}_{SSG} &=& \frac{1}{2} \lambda^{\bf jkl} g^A_{\bf mn}
  a_{\bf jm}^{(\kappa)} a_{\bf kn}^{(\sigma)} c_{A\bf l}^{(\rho)} 
  f_{SSG}(\kappa, \sigma, \rho)
,
\\
V^{(2)}_{GGS} &=& \frac{1}{2} \lambda^{\bf jkl} G^{AB}_{\bf m} 
  a_{\bf jm}^{(\kappa)} c_{A\bf k}^{(\sigma)} c_{B\bf l}^{(\rho)}
  f_{GGS}(\sigma, \rho, \kappa)
,
\\
V^{(2)}_{SGG} &=& \frac{1}{2}
  g^A_{\bf jk} g^B_{\bf lm} a_{\bf jl}^{(\kappa)} c_{A\bf l}^{(\sigma)} c_{B\bf k}^{(\rho)}  
  f_{SGG}(\kappa, \sigma, \rho)
,
\\
V^{(2)}_{GSV} &=& 
  g^{\bf a}_{\bf jk} G^{A\bf a}_{\bf l} a_{\bf jl}^{(\kappa)} c_{A\bf k}^{(\sigma)} 
  f_{GSV}(\sigma, \kappa, {\bf a}) 
+ g^{B}_{\bf jk} G^{AC}_{\bf l} a_{\bf jl}^{(\kappa)} c_{A\bf k}^{(\sigma)} b_{BC}^{(\rho)} 
  f_{GS\Vbar}(\sigma, \kappa, \rho) 
,
\\
V^{(2)}_{GGG} &=& \frac{1}{2} g^{C}_{\bf jk} G^{AB}_{\bf l} 
  c_{A\bf j}^{(\kappa)} c_{B\bf k}^{(\sigma)} c_{C\bf l}^{(\rho)}
  f_{GGG}(\kappa, \sigma, \rho)
,
\\
V^{(2)}_{GGV} &=& \frac{1}{2}g^{{\bf a}AB} g^{\bf a}_{\bf jk} 
  c_{A\bf j}^{(\kappa)} c_{B\bf k}^{(\sigma)}
  f_{GGV}(\kappa, \sigma, {\bf a}) 
,
\label{eq:V2GGV}
\\
V^{(2)}_{VGG} &=& \frac{1}{2} G^{A\bf a}_{\bf j} G^{B\bf a}_{\bf k} 
  c_{A\bf k}^{(\kappa)} c_{B\bf j}^{(\sigma)}
  f_{VGG} ({\bf a}, \kappa, \sigma) 
+ \frac{1}{2} G^{AC}_{\bf j} G^{BD}_{\bf k} 
  c_{A\bf k}^{(\kappa)} c_{B\bf j}^{(\sigma)} b_{CD}^{(\rho)}
  f_{\Vbar GG} (\rho, \kappa, \sigma)  
,
\\
V^{(2)}_{VVG} &=& \frac{1}{2} g^{{\bf ab}A} G^{\bf ab}_{\bf j} c_{A\bf j}^{(\kappa)} 
  f_{VVG}({\bf a},{\bf b},\kappa) 
+ g^{AB{\bf a}} G^{C\bf a}_{\bf j} c_{A\bf j}^{(\kappa)} b_{BC}^{(\sigma)} 
  f_{\Vbar VG}(\sigma,{\bf a},\kappa)    
,
\label{eq:V2VVG}
\\
V^{(2)}_{\eta\eta S} &=&
  \frac{1}{2} g^A_{B\bf j} g^B_{A\bf k} \xiprime_A \xiprime_B M_A M_B
  a_{\bf jk}^{(\kappa)} f_{\eta\eta S}(A_\eta, B_\eta, \kappa) 
,
\\
V^{(2)}_{\eta\eta G} &=& g^{{\bf a}AB} g^{\bf a}_{A\bf j} \xiprime_A M_A 
  c_{B\bf j}^{(\kappa)} 
  f_{\eta\eta G}({\bf a}_\eta, A_\eta, \kappa)
,
\\
V^{(2)}_{VV} &=& \frac{1}{4} \bigl ( g^{\bf abc} \bigr )^2 f_{VV}({\bf a}, {\bf b})
+  \frac{1}{2} g^{{\bf ab}A} g^{{\bf ab}B} b_{AB}^{(\kappa)} f_{\Vbar V}(\kappa, {\bf a})
\nonumber \\ &&
+  \frac{1}{4} g^{{\bf a}AB} g^{{\bf a}CD} b_{AC}^{(\kappa)} b_{BD}^{(\sigma)}
   f_{\Vbar\Vbar}(\kappa, \sigma) 
,
\\
V^{(2)}_{VVV} &=&
  \frac{1}{12} \bigl (g^{\bf abc} \bigr )^2 f_{VVV}({\bf a}, {\bf b}, {\bf c})
+  \frac{1}{4} g^{{\bf ab}A} g^{{\bf ab}B} b_{AB}^{(\kappa)}  
  f_{\Vbar VV}(\kappa, {\bf a}, {\bf b})
\nonumber \\ &&
+  \frac{1}{4} g^{{\bf a}AB} g^{{\bf a}CD} b_{AC}^{(\kappa)} b_{BD}^{(\sigma)}  
  f_{\Vbar\Vbar V}(\kappa, \sigma, {\bf a}) 
,
\label{eq:V2VVV}
\\
V^{(2)}_{\eta\eta V} &=&
  \frac{1}{4} \bigl (g^{\bf abc}\bigr )^2 f_{\eta\eta V}({\bf a}_\eta, {\bf b}_\eta, {\bf c})
+ \frac{1}{4} g^{{\bf ab}A} g^{{\bf ab}B} b_{AB}^{(\kappa)} 
   f_{\eta\eta\Vbar}({\bf a}_\eta, {\bf b}_\eta, \kappa)
,
\\
V^{(2)}_{FFS} &=&
\frac{1}{2} Y^{{\bf j}IJ} Y_{{\bf k}IJ} a_{\bf jk}^{(\kappa)} f_{FFS}(I,J,\kappa)
,
\\
V^{(2)}_{\Fbar\Fbar S} &=&
\frac{1}{4} (Y^{{\bf j}IJ} Y^{{\bf k}I'J'} M_{II'} M_{JJ'}  + {\rm c.c.}) 
  a_{\bf jk}^{(\kappa)}
  f_{\Fbar\Fbar S}(I,J,\kappa)
,
\\
V^{(2)}_{FFV} &=&
\frac{1}{2} g^{{\bf a}J}_I g^{{\bf a}I}_J f_{FFV}(I,J,{\bf a}) 
+ 
\frac{1}{2} g^{AJ}_I g^{BI}_J b_{AB}^{(\kappa)} f_{FF\Vbar}(I,J,\kappa)
,
\\
V^{(2)}_{\Fbar\Fbar V} &=&
\frac{1}{2} g^{{\bf a}J}_I g^{{\bf a}J'}_{I'} M^{II'} M_{JJ'} 
f_{\Fbar\Fbar V}(I,J,{\bf a})
+
\frac{1}{2} g^{AJ}_I g^{BJ'}_{I'} M^{II'} M_{JJ'} b_{AB}^{(\kappa)} 
f_{\Fbar\Fbar\Vbar}(I,J,\kappa)  
,
\\
V^{(2)}_{\Fbar FG} &=& i\left (
g^{AJ}_I Y_{jI'J} M^{II'} - {\rm c.c.} \right ) c_{A\bf j}^{(\kappa)} 
f_{\Fbar FG}(I,J,\kappa) .
\label{eq:V2formfFG}
\eeq
In these equations, all indices (including $\kappa,\sigma,\rho$)
are summed over in each term.

It remains to find the following\footnote{One might
naively expect functions
$f_{GG\Vbar}(x,y,z)$, 
$f_{\Vbar\Vbar G}(x,y,z)$, and
$f_{\Vbar\Vbar\Vbar}(x,y,z)$
to appear in eqs.~(\ref{eq:V2GGV}), 
(\ref{eq:V2VVG}), and
(\ref{eq:V2VVV}), respectively. 
However, those three contributions turn out to vanish identically.}
37 two-loop integral functions: 
\beq
&&
f_{SS}(x,y),\>\>\>\>
f_{SSS}(x,y,z),\>\>\>\>
f_{VS}(x,y),\>\>\>\>
f_{\Vbar S}(x,y),\>\>\>\>
f_{SSV}(x,y,z),\>\>\>\>
\nonumber \\ &&
f_{SS\Vbar}(x,y,z),\>\>\>\>
f_{VVS}(x,y,z),\>\>\>\>
f_{\Vbar VS}(x,y,z),\>\>\>\>
f_{\Vbar\Vbar S}(x,y,z),\>\>\>\>
f_{SSG}(x,y,z),\>\>\>\>
\nonumber \\ &&
f_{GGS}(x,y,z),\>\>\>\>
f_{SGG}(x,y,z),\>\>\>\>
f_{GSV}(x,y,z),\>\>\>\>
f_{GS\Vbar}(x,y,z),\>\>\>\>
f_{GGG}(x,y,z),\>\>\>\>
\nonumber \\ &&
f_{GGV}(x,y,z),\>\>\>\>
f_{VGG}(x,y,z),\>\>\>\>
f_{\Vbar GG}(x,y,z),\>\>\>\>
f_{VVG}(x,y,z),\>\>\>\>
\nonumber \\ &&
f_{\Vbar VG}(x,y,z),\>\>\>\>
f_{\eta\eta S}(x,y,z),\>\>\>\>
f_{\eta\eta G}(x,y,z),\>\>\>\>
f_{VV}(x,y),\>\>\>\>
\nonumber \\ &&
f_{\Vbar V}(x,y),\>\>\>\>
f_{\Vbar\Vbar}(x,y),\>\>\>\>
f_{VVV}(x,y,z),\>\>\>\>
f_{\Vbar VV}(x,y,z),\>\>\>\>
f_{\Vbar\Vbar V}(x,y,z),\>\>\>\>
\nonumber \\ &&
f_{\eta\eta V}(x,y,z),\>\>\>\>
f_{\eta\eta \Vbar}(x,y,z),\>\>\>\>
f_{FFS}(x,y,z),\>\>\>\>
f_{\Fbar\Fbar S}(x,y,z),\>\>\>\>
\nonumber \\ &&
f_{FFV}(x,y,z),\>\>\>\>
f_{FF\Vbar}(x,y,z),\>\>\>\>
f_{\Fbar\Fbar V}(x,y,z),\>\>\>\>
f_{\Fbar\Fbar \Vbar}(x,y,z),\>\>\>\>
f_{\Fbar FG}(x,y,z).
\label{eq:listfs}
\eeq
In the next subsection \ref{subsec:barefunctions}, 
we present the results for the loop integration functions in
the case that all parameters of the theory are taken to be bare parameters
using dimensional regularization 
\cite{Bollini:1972ui,Ashmore:1972uj,Cicuta:1972jf,tHooft:1972fi,tHooft:1973mm}. In subsection
\ref{subsec:MSbarfunctions}, we present the result in the more 
practically relevant case that all
parameters are renormalized in the \MSbar \cite{Bardeen:1978yd,Braaten:1981dv} scheme. 
In both cases, we write the
results in terms of 1-loop and 2-loop
basis vacuum integrals following the conventions of 
refs.~\cite{Martin:2001vx,Martin:2017lqn}; these are reviewed
for convenience in Appendix A below.

\subsection{Results for two-loop effective potential 
functions in terms of bare parameters\label{subsec:barefunctions}}

In this section, we report the results for the 2-loop effective 
potential in terms of bare parameters. This means that all of the masses 
and couplings appearing in eqs.~(\ref{eq:V1loopform}) and 
(\ref{eq:V2formSS})-(\ref{eq:V2formfFG}) are the bare ones, and the 
corresponding loop integral functions will be distinguished by using 
$\bf f$ in place of $f$ in the names of the functions. 
Then then 1-loop 
integrals appearing in eq.~(\ref{eq:V1loopform}) are:
\beq
{\bf f}(x) &=& x \AAx /d,
\\
{\bf f}_V(x) &=& (d-1) x \AAx/3d.
\eeq
(The notations for the basis integrals $\AAx$ and $\IIxyz$ are reviewed in 
Appendix A.)
For the two-loop integrals appearing in 
eqs.~(\ref{eq:V2formSS})-(\ref{eq:V2formfFG}), we obtain:  
\beq
{\bf f}_{SSS}(x,y,z) &=& -\IIxyz ,
\label{eq:fSSSbare}
\\
{\bf f}_{SS}(x,y) &=&  \AAx \AAy ,
\\
{\bf f}_{VS}(x,y) &=& (d-1) \AAx \AAy ,
\\
{\bf f}_{\overline{V}S}(x,y) &=& \AAx \AAy ,
\\
{\bf f}_{SSV}(x,y,z) &=& \frac{1}{z} \Bigl [
-\lambda(x,y,z) \IIxyz  
+ (x-y)^2 {\bf I}(0,x,y)
+ z \AAx  \AAy 
\nonumber \\ &&
+ (y-x-z) \AAx  \AAz 
+ (x-y-z) \AAy  \AAz 
\Bigr ]
,
\\
{\bf f}_{SS\overline{V}}(x,y,z) &=& \frac{1}{z} \Bigl \{
(x-y)^2 [ \IIxyz  - {\bf I}(0,x,y) ]
+ (x-y-z) \AAx  \AAz 
\nonumber \\ && 
+ (y-x-z) \AAy  \AAz 
\Bigr \}
,
\\
{\bf f}_{\overline{V}\overline{V}S}(x,y,z) &=& 
  \frac{1}{4 x y} \Bigl [ -(x+y-z)^2 \IIxyz 
  + (x-z)^2 {\bf I}(0,x,z) 
  + (y-z)^2 {\bf I}(0,y,z)
\nonumber \\ &&
  -z^2 {\bf I}(0,0,z) + (z-x-y) \AAx  \AAy 
  + y \AAx  \AAz  + x \AAy  \AAz 
\Bigr ] 
,
\\
{\bf f}_{\overline{V}VS}(x,y,z) &=& 
  -{\bf f}_{\overline{V}\overline{V}S}(x,y,z)
  -\IIxyz  
,
\\
{\bf f}_{VVS}(x,y,z) &=&  
  {\bf f}_{\overline{V}\overline{V}S}(x,y,z) +
  (2-d) \IIxyz 
,
\\
{\bf f}_{SSG}(x,y,z) &=& (x-y) \IIxyz  
  + [\AAx  - \AAy ] \AAz 
,
\\
{\bf f}_{GGS}(x,y,z) &=& \frac{1}{2} \Bigl [ (x+y-z) \IIxyz  
  + \AAx  \AAy 
  - \AAx  \AAz 
  - \AAy  \AAz 
  \Bigr ] 
,
\\
{\bf f}_{SGG}(x,y,z) &=& -(x-y)(x-z) \IIxyz  
  + (y-x) \AAx  \AAy  
  + (z-x) \AAx  \AAz 
\nonumber \\ &&
  + x \AAy \AAz 
,
\\
{\bf f}_{GSV}(x,y,z) &=& -\frac{1}{2} {\bf f}_{SSV}(x,y,z),
\\
{\bf f}_{GS\overline{V}}(x,y,z) &=& \frac{1}{2 z} \Bigl [
(y-x) (x - y + z) \IIxyz  
+ (x-y)^2 {\bf I}(0,x,y)
\nonumber \\ &&
+ (x-y+2z) \AAy  \AAz 
+ (y-x) \AAx  \AAz 
\Bigr ]
,
\\
{\bf f}_{GGG}(x,y,z) &=& \frac{1}{2} \bigl \{ (x-y) [
  (z-x-y) \IIxyz  - \AAx  \AAy  ] 
\nonumber \\ &&	
  + (x+y) [\AAy  - \AAx ] \AAz  \bigr \}
,
\\
{\bf f}_{GGV}(x,y,z) &=&  \frac{1}{2} \bigl [
  \lambda(x,y,z) \IIxyz  
  + (x+y-z) \AAx  \AAy 
\nonumber \\ &&	
  + (x+z-y) \AAx  \AAz 
  + (y+z-x) \AAy  \AAz 
\bigr ]  
,
\\
{\bf f}_{VGG}(x,y,z) &=& \frac{1}{4x} \bigl [
\lambda(x,y,z) \IIxyz   - (y-z)^2 {\bf I}(0,y,z)
  + (x+y-z) \AAx  \AAy 
\nonumber \\ &&	
  + (x-y+z) \AAx  \AAz 
  - x \AAy  \AAz 
\bigr ]
,
\\
{\bf f}_{\Vbar GG}(x,y,z) &=& \frac{1}{4x} \bigl [
  (x+y-z)(x+z-y) \IIxyz  + (y-z)^2 {\bf I}(0,y,z)
  - x \AAy  \AAz 
\nonumber \\ &&	
  + (x+y-z) \AAx \AAz
  + (x+z-y) \AAx \AAy
\bigr ]
,
\\
{\bf f}_{VVG}(x,y,z) &=& \frac{1}{4xy} \Bigl \{
  (x-y)[\lambda(x,y,z) + 4(d-1) x y] \IIxyz 
  - x (x - z)^2 {\bf I}(0,x,z) 
\nonumber \\ &&
  + y (y - z)^2 {\bf I}(0,y,z)
  + (x - y) (x + y - z)\AAx  \AAy 
\nonumber \\ &&
  + y [(4 d - 6) x + y - z] \AAx  \AAz 
  - x [(4 d - 6) y + x - z] \AAy  \AAz   \Bigr \}
,
\\
{\bf f}_{\Vbar VG}(x,y,z) &=& -\frac{1}{4} {\bf f}_{SSV}(z,y,x) ,
\\
{\bf f}_{\eta\eta S}(x,y,z) &=& \IIxyz ,
\\
{\bf f}_{\eta\eta G}(x,y,z) &=& -{\bf f}_{GGS}(x,z,y) ,
\\
{\bf f}_{VV}(x,y) &=& \frac{(d-1)^3}{d} \AAx  \AAy ,
\\
{\bf f}_{\Vbar V}(x,y) &=& \frac{(d-1)^2}{d} \AAx  \AAy ,
\\
{\bf f}_{\Vbar\Vbar}(x,y) &=& \frac{(d-1)}{d} \AAx  \AAy ,
\\
{\bf f}_{VVV}(x,y,z) &=& \frac{1}{4xyz}  \Bigl \{
  -\lambda(x,y,z) [\lambda(x,y,z) + 4(d-1) (x y + x z + y z)] \IIxyz 
\nonumber \\ &&
  + (x-y)^2 [x^2 + y^2 + (4d-6) x y] {\bf I}(0,x,y) - z^4 {\bf I}(0,0,z)
\nonumber \\ &&
  + (x-z)^2 [x^2 + z^2 + (4d-6) x z] {\bf I}(0,x,z) - y^4 {\bf I}(0,0,y)
\nonumber \\ &&
  + (y-z)^2 [y^2 + z^2 + (4d-6) y z] {\bf I}(0,y,z) - x^4 {\bf I}(0,0,x)
\nonumber \\ &&  
  + [(7 -4 d) (x^2 + y^2 - x z - y z) + 
  (2 - 4d + 4/d) x y + z^2 ] z \AAx  \AAy 
\nonumber \\ &&  
  + [(7 -4 d) (x^2 + z^2 - x y - y z) + 
  (2 - 4d + 4/d) x z + y^2 ] y \AAx  \AAz 
\nonumber \\ &&  
  + [(7 -4 d) (y^2 + z^2 - x y - x z) + 
  (2 - 4d + 4/d) y z + x^2 ] x \AAy  \AAz   
\Bigr \} ,
\\
{\bf f}_{\Vbar VV}(x,y,z) &=& \frac{1}{4xyz}  \Bigl \{
  (y-z)^2 [\lambda(x,y,z) + 4 (d-1) y z ] \IIxyz  
\nonumber \\ &&
  -(y-z)^2 [y^2 + z^2 + (4d-6) y z] {\bf I}(0,y,z)
  -(x-z)^2 z^2 {\bf I}(0,x,z)
\nonumber \\ &&
  - (x-y)^2 y^2 {\bf I}(0,x,y)
  + y^4 {\bf I}(0,0,y)
  + z^4 {\bf I}(0,0,z)
  - x (y-z)^2 \AAy  \AAz 
\nonumber \\ &&
  + [(7 - 4 d) y (z-y) + (x-z) z + (6 - 4 d - 4/d) x y ] 
    z \AAx  \AAy 
\nonumber \\ &&
  + [(7  - 4 d) z (y-z) + (x-y) y + (6 - 4 d - 4/d) x z] 
    y \AAx  \AAz 
\Bigr \}
,
\\
{\bf f}_{\Vbar\Vbar V}(x,y,z) &=& \frac{1}{4xy} \Bigl \{
  -z \lambda(x,y,z) \IIxyz 
  + (x-z)^2 z {\bf I}(0,x,z)
  + (y-z)^2 z {\bf I}(0,y,z)
\nonumber \\ &&
  - z^3 {\bf I}(0,0,z)
  + [z^2 - x z - y z + 4 x y (1/d-1)] \AAx  \AAy 
\nonumber \\ &&
  + z [x \AAy  + y \AAx ] \AAz  \Bigr \}
,
\\
{\bf f}_{\eta\eta V}(x,y,z) &=& \frac{1}{2z} \bigl [
\lambda(x,y,z) \IIxyz  - (x-y)^2 {\bf I}(0,x,y)
\nonumber \\ &&
-z \AAx  \AAy  
+ (x - y + z) \AAx  \AAz  
+ (y - x + z) \AAy  \AAz 
\bigr ]
,
\\
{\bf f}_{\eta\eta \Vbar}(x,y,z) &=& \frac{1}{2z} \bigl [
(y-x-z)(x-y-z) \IIxyz  + (x-y)^2 {\bf I}(0,x,y)
-z \AAx  \AAy  
\nonumber \\ &&
+ (y - x + z) \AAx  \AAz  
+ (x - y + z) \AAy  \AAz 
\bigr ]
,
\\
{\bf f}_{FFS}(x,y,z) &=&  (x+y-z) \IIxyz  
  + \AAx  \AAy 
  - \AAx  \AAz 
  - \AAy  \AAz 
,
\\
{\bf f}_{\Fbar\Fbar S}(x,y,z) &=& 2 \IIxyz 
,
\\
{\bf f}_{FFV}(x,y,z) &=&  \frac{1}{z} \Bigl \{
[\lambda(x,y,z) + (d-1) z (x+y-z)]  \IIxyz  - (x-y)^2 {\bf I}(0,x,y)
\nonumber \\ &&
+ [x-y + (2-d) z] \AAx  \AAz  
+ [y-x + (2-d) z] \AAy  \AAz 
\nonumber \\ &&
+ (d-2) z \AAx  \AAy 
\Bigr \}
,
\\
{\bf f}_{FF\Vbar}(x,y,z) &=&  \frac{1}{z} \bigl \{
  (x z + y z - x^2 + 2 x y - y^2) \IIxyz  + (x-y)^2 {\bf I}(0,x,y)
\nonumber \\ &&
  + (x-y) [\AAy  - \AAx ] \AAz 
  \bigr \}
,
\\
{\bf f}_{\Fbar\Fbar V}(x,y,z) &=&  2 (d-1) \IIxyz  
,
\\
{\bf f}_{\Fbar\Fbar \Vbar}(x,y,z) &=&  2 \IIxyz  
,
\\
{\bf f}_{\Fbar FG}(x,y,z) &=& 
(y+z-x) \IIxyz  
  - \AAx  \AAy 
  - \AAx  \AAz 
  + \AAy  \AAz .
\label{eq:ffFGbare}
\eeq

\subsection{Results in terms of \MSbar parameters\label{subsec:MSbarfunctions}}

In this subsection, we provide the results for the effective potential loop integral
functions, this time as they appear in
in the \MSbar scheme 
with renormalization scale $Q$, and $\lnbar(x) \equiv \ln(x/Q^2)$,
and renormalized basis integrals $A(x)$ and $I(x,y,z)$ given in Appendix A. 

The one-loop functions for the \MSbar scheme 
can be obtained from the ones for the bare scheme by
including counterterms for the ultraviolet 1-loop sub-divergences, and then
taking the limit as $\epsilon \rightarrow 0$. 
One has 
\beq
f(x) &=& \lim_{\eps \rightarrow 0} \Bigl [ {\bf f}(x) + \frac{x^2}{4\eps} \Bigr ],
\\
f_V(x) &=& \lim_{\eps \rightarrow 0} \Bigl [ {\bf f}_V(x) + \frac{x^2}{4\eps} \Bigr ],
\eeq
with the results:
\beq
f(x) &=& \frac{x^2}{4} \bigl [ \lnbar(x) - 3/2],
\label{eq:deff}
\\
f_V(x) &=& \frac{x^2}{4} \bigl [ \lnbar(x) - 5/6],
\label{eq:deffV}
\eeq
which should be used in eq.~(\ref{eq:V1loopform}) for the \MSbar scheme. 

Similarly, the two-loop functions appearing in 
eqs.~(\ref{eq:V2formSS})-(\ref{eq:V2formfFG})
in the \MSbar scheme can be obtained
by taking the limit $\eps \rightarrow 0$ after including 
counterterms 
for the 1-loop and 2-loop sub-divergences. 
The 2-loop counterterms are determined by modified minimal subtraction
and the requirement that the resulting
functions are finite as $\epsilon \rightarrow 0$. The inclusions of counterterms
are as follows: 
\beq
f_{SSS}(x,y,z) &=& \lim_{\eps \rightarrow 0} \Bigl \{
{\bf f}_{SSS}(x,y,z) 
+ \frac{1}{2\eps^2} (x + y + z)
- \frac{1}{2\eps} (x + y + z)
\nonumber \\ &&
+ \frac{1}{\eps} \left [ \AAx + \AAy + \AAz \right ]
\Bigr \}
,
\\
f_{SS}(x,y) &=& \lim_{\eps \rightarrow 0} \Bigl \{
{\bf f}_{SS} + \frac{1}{\eps^2} x y + \frac{1}{\eps} \left [ y \AAx + x \AAy \right]
\Bigr \},
\\
f_{VS}(x,y) &=& 
\lim_{\eps \rightarrow 0} \Bigl \{
{\bf f}_{VS} + 
\frac{3}{\eps^2} x y + \frac{1}{\eps} \left [(d-1) y \AAx + 3 x \AAy \right]
\Bigr \},
\\
f_{\overline{V}S}(x,y) &=& 
\lim_{\eps \rightarrow 0} \Bigl \{
{\bf f}_{\overline{V}S} + \frac{1}{\eps^2} x y 
+ \frac{1}{\eps} \left [ y \AAx + x \AAy \right ]
\Bigr \}
,
\\
f_{SSV}(x,y,z) &=& 
\lim_{\eps \rightarrow 0} \Bigl \{
{\bf f}_{SSV}(x,y,z) +
\frac{1}{2 \eps^2} \left (
-3 x^2 - 3 y^2 - 3 x z - 3 y z + z^2
\right )
\nonumber \\ &&
+
\frac{1}{6 \eps} \left (
3 x^2 + 3 y^2 - 7 z^2 + 18 x y + 15 x z + 15 y z 
\right )
\nonumber \\ &&
+ \frac{1}{\eps} \left [ - 3 x \AAx - 3 y \AAy + (1-d) (x+y-z/3) \AAz \right ]
\Bigr \}
,
\\
f_{SS\overline{V}}(x,y,z) &=& 
\lim_{\eps \rightarrow 0} \Bigl \{
{\bf f}_{SS\overline{V}}(x,y,z)
+ \frac{1}{2\eps^2} [\lambda(x,y,z) - z^2]
+ \frac{1}{2\eps} (x-y)^2
\nonumber \\ && 
+ \frac{1}{\eps} \bigl [
(x-y-z) \AAx + (y-x-z) \AAy - (x+y) \AAz
\bigr ] \Bigr \}
,
\\
f_{VVS}(x,y,z) &=&  
\lim_{\eps \rightarrow 0} \Bigl \{
{\bf f}_{VVS}(x,y,z)
+ \frac{1}{8\eps^2} (9x + 9 y + 12 z)
+ \frac{1}{8\eps} (2z - 15x - 15 y)
\nonumber \\ &&
+ \frac{3}{4\eps} \left [(d-1) \AAx + (d-1) \AAy + 4 \AAz \right ]
\Bigr \}
,
\\
f_{\overline{V}VS}(x,y,z) &=& 
\lim_{\eps \rightarrow 0} \Bigl \{
{\bf f}_{\overline{V}VS}(x,y,z) + 
\frac{3}{8 \eps^2} (x+y) - \frac{1}{8 \eps} (x + 5 y + 6z)
\nonumber \\ &&
+ \frac{1}{4\eps} \left [3 \AAx + (d-1) \AAy \right ] \Bigr \}
,
\\
f_{\overline{V}\overline{V}S}(x,y,z) &=& 
\lim_{\eps \rightarrow 0} \Bigl \{
{\bf f}_{\overline{V}\overline{V}S}(x,y,z)
+ \frac{1}{8 \eps^2} (x+y+4z)
+ \frac{1}{8\eps} (-3x-3y+2z)
\nonumber \\ &&
+ \frac{1}{4\eps} \left [
\AAx + \AAy + 4 \AAz
\right ] 
\Bigr \}
,
\\
f_{SSG}(x,y,z) &=& 
\lim_{\eps \rightarrow 0} \Bigl \{
{\bf f}_{SSG}(x,y,z) + \frac{1}{2\eps^2} (x-y)(z-x-y) + \frac{1}{2\eps} (x-y) (x+y+z)
\nonumber \\ &&
+ \frac{1}{\eps} \left [
(y-x+z) \AAx + (y-x-z) \AAy 
\right ] \Bigr \}
,
\\
f_{GGS}(x,y,z) &=& 
\lim_{\eps \rightarrow 0} \Bigl \{
{\bf f}_{GGS}(x,y,z) + 
\frac{1}{4\eps^2} (z^2 -x^2 - y^2 - 2 x z - 2 y z)
+ \frac{1}{4\eps} (x+y-z)(x+y+z)
\nonumber \\ &&
+ \frac{1}{2\eps} \left [ -x \AAx - y \AAy + (z - 2 x - 2 y) \AAz \right ] 
\Bigr \}
,
\\
f_{SGG}(x,y,z) &=& 
\lim_{\eps \rightarrow 0} \Bigl \{
{\bf f}_{SGG}(x,y,z)
+ \frac{1}{2\eps^2} (y+z-x)(x y + x z + y z - x^2)
\nonumber \\ &&
+ \frac{1}{2\eps}(x-y)(z-x)(x+y+z)
\nonumber \\ &&
+ \frac{1}{\eps}
\left [
(\lambda(x,y,z) + 3 y z) \AAx + y z \AAy + y z \AAz 
\right ]
\Bigr \}
,
\\
f_{GSV}(x,y,z) &=& 
\lim_{\eps \rightarrow 0} \Bigl \{
{\bf f}_{GSV}(x,y,z) +
\frac{1}{4 \eps^2} (3 x^2 + 3 y^2 + 3 x z + 3 y z - z^2 ) 
\nonumber \\ && 
+ \frac{1}{4 \eps} (7 z^2/3 - x^2 - 6 x y - y^2 - 5 x z - 5 y z)
\nonumber \\ &&
+ \frac{1}{2 \eps}
\left [
3 x \AAx + 3 y \AAy + (d-1) (x + y - z/3) \AAz
\right ]
\Bigr \}
,
\\
f_{GS\overline{V}}(x,y,z) &=& 
\lim_{\eps \rightarrow 0} \Bigl \{
{\bf f}_{GS\overline{V}}(x,y,z) + 
\frac{1}{4 \eps^2} (2 x y + x z + 3 y z -2 y^2) 
+ \frac{1}{4 \eps} (y-x)(2x+z)
\nonumber \\ &&
+ \frac{1}{\eps}
\left [
(x-y+z) \AAy + (x+y) \AAz/2
\right ]
\Bigr \}
,
\\
f_{GGG}(x,y,z) &=& 
\lim_{\eps \rightarrow 0} \Bigl \{
{\bf f}_{GGG}(x,y,z) + 
\frac{1}{4\eps^2} (x-y)(x^2 + y^2 - z^2 - 2 x z - 2 y z)
\nonumber \\ &&
+ \frac{1}{4\eps} (x-y)(z-x-y)(x+y+z)
\nonumber \\ &&
+ \frac{1}{2 \eps} \left [
x (x-y-2z) \AAx + y (x-y+2z) \AAy + (y-x) z \AAz 
\right ]
\Bigr \}
,
\\
f_{GGV}(x,y,z) &=& 
\lim_{\eps \rightarrow 0} \Bigl \{
{\bf f}_{GGV}(x,y,z)
\nonumber \\ &&
+
\frac{1}{4 \eps^2} 
(-x^3 - y^3 - z^3 + 3 x^2 y + 3 x y^2 + 3 x^2 z + 3 x z^2 + 3 y^2 z + 3 y z^2)
\nonumber \\ &&
+ \frac{1}{4 \eps}
(x^3 + y^3 + 7 z^3/3 -x^2 y - x y^2 - x^2 z - y^2 z  - 5 x z^2 - 5 y z^2 - 6 x y z)  
\nonumber \\ &&
+ \frac{1}{2\eps} \bigl [
x (3y+3z-x) \AAx
+ y (3x+3z-y) \AAy 
\nonumber \\ &&
+ (d-1) z (x+y -z/3) \AAz 
\bigr ]
\Bigr \}
,
\\
f_{VGG}(x,y,z) &=& \lim_{\eps \rightarrow 0} \Bigl \{
{\bf f}_{VGG}(x,y,z)
+ \frac{1}{8\eps^2} (3 y^2 + 3 z^2 - x^2 + 3 x y + 3 x z)
\nonumber \\ &&
+ \frac{1}{8 \eps} (7 x^2/3 -y^2 - z^2  - 5 x y - 5 x z - 6 y z)
\nonumber \\ &&
+ \frac{1}{4 \eps} \left [
(d-1) (y + z - x/3) \AAx + 3 y \AAy + 3 z \AAz 
\right ]
\Bigr \}
,
\\
f_{\Vbar GG}(x,y,z) &=& 
\lim_{\eps \rightarrow 0} \Bigl \{
{\bf f}_{\Vbar GG}(x,y,z) + 
\frac{1}{8\eps^2} (x y + x z - x^2 - y^2 - z^2)
\nonumber \\ &&
+ \frac{1}{8\eps} (x^2 - y^2 - z^2 + x y + x z + 2 y z )
\nonumber \\ &&
+ \frac{1}{4\eps} \left [
(y+z-x) \AAx - y \AAy - z \AAz 
\right ]
\Bigr \}
,
\\
f_{VVG}(x,y,z) &=& \lim_{\eps \rightarrow 0} \Bigl \{
{\bf f}_{VVG}(x,y,z)
+ \frac{1}{\eps^2} (x-y)(9z/8 - x - y)
\nonumber \\ &&
+ \frac{1}{24 \eps} (x-y)(38 x + 38 y + 9 z)
\nonumber \\ &&
+ \frac{1}{12 \eps} (d-1) \left [
(9y +9z - 8x) \AAx - (9x + 9z - 8y) \AAy
\right ]
\Bigr \}
,
\\
f_{\Vbar VG}(x,y,z) &=& 
\lim_{\eps \rightarrow 0} \Bigl \{
{\bf f}_{\Vbar VG}(x,y,z)
+ \frac{1}{8 \eps^2} (3 y^2 + 3 z^2 - x^2 + 3 x y + 3 x z)  
\nonumber \\ &&
+ \frac{1}{8 \eps} (x^2 - 5 y^2 - z^2 - x y - x z - 6 y z)
\nonumber \\ &&
+ \frac{1}{4 \eps} \left [(3 y + 3 z - x) \AAx + (d-1) y \AAy + 3 z \AAz \right ]
\Bigr \}
,
\\
f_{\eta\eta S}(x,y,z) &=& 
\lim_{\eps \rightarrow 0} \Bigl \{
{\bf f}_{\eta\eta S}(x,y,z)
- \frac{1}{2\eps^2} (x + y + z)
+ \frac{1}{2\eps} (x + y + z)
\nonumber \\ &&
- \frac{1}{\eps} \left [ \AAx + \AAy + \AAz \right ]
\Bigr \}
,
\\
f_{\eta\eta G}(x,y,z) &=& 
\lim_{\eps \rightarrow 0} \Bigl \{
{\bf f}_{\eta\eta G}(x,y,z) 
+ \frac{1}{4\eps^2} (x^2 - y^2 + z^2 + 2 x y + 2 y z)
+ \frac{1}{4\eps} (y-x-z)(x+y+z)
\nonumber \\ &&
+ \frac{1}{2\eps} \left [ x \AAx + (2 x - y+ 2 z) \AAy + z \AAz \right ] 
\Bigr \}
,
\\
f_{VV}(x,y) &=& 
\lim_{\eps \rightarrow 0} \Bigl \{
{\bf f}_{VV}(x,y)
+ \frac{27}{4 \eps^2} x y 
+ \frac{9}{8 \eps} x y
+ \frac{9}{4 \eps} (d-1) \left [y \AAx + x \AAy \right ] 
\Bigr \}
,
\\
f_{\Vbar V}(x,y) &=& 
\lim_{\eps \rightarrow 0} \Bigl \{
{\bf f}_{\Vbar V}(x,y)
+ \frac{9}{4 \eps^2} x y 
+ \frac{3}{8 \eps} x y
+ \frac{3}{4 \eps} \left [ 3 y \AAx + (d-1) x \AAy \right ] 
\Bigr \}
,
\\
f_{\Vbar\Vbar}(x,y) &=& \lim_{\eps \rightarrow 0} \Bigl \{
{\bf f}_{\Vbar\Vbar}(x,y)
+ \frac{3}{4 \eps^2} x y 
+ \frac{1}{8 \eps} x y
+ \frac{3}{4 \eps} \left [ y \AAx + x \AAy \right ] 
\Bigr \}
,
\\
f_{VVV}(x,y,z) &=& 
\lim_{\eps \rightarrow 0} \Bigl \{
{\bf f}_{VVV}(x,y,z) 
- \frac{1}{4 \eps^2} \left [ 25 (x^2 + y^2 + z^2) + 36 (x y + x z + y z) \right ]
\nonumber \\ &&
+ \frac{1}{24 \eps} \left [128 (x^2 + y^2 + z^2) + 387 (x y + x z + y z) \right ]
\nonumber \\ &&
+ \frac{1}{6 \eps} (1-d) \bigl [
(25 x + 18 y + 18 z) \AAx 
+ (25 y + 18 x + 18 z) \AAy
\nonumber \\ &&
+ (25 z + 18 x + 18 y) \AAz
\bigr ]
\Bigr \}
,
\\
f_{\Vbar VV}(x,y,z) &=& \lim_{\eps \rightarrow 0} \Bigl \{
{\bf f}_{\Vbar VV}(x,y,z) 
+ \frac{1}{8 \eps^2} ( 2 x^2 + 6 y^2 + 6 z^2 - 21 x y - 21 x z - 18 y z) 
\nonumber \\ &&
+ \frac{1}{4 \eps} (4 y^2 + 4 z^2 -x^2 - x y - x z - 3 y z)
+ \frac{1}{4 \eps} \bigl [
2 (x-6y-6z) \AAx 
\nonumber \\ &&
+ (d-1)(2y - 3 x - 3 z) \AAy
+ (d-1)(2z - 3 x - 3 y) \AAz
\bigr ]
\Bigr \}
,
\\
f_{\Vbar\Vbar V}(x,y,z) &=& 
\lim_{\eps \rightarrow 0} \Bigl \{
{\bf f}_{\Vbar\Vbar V}(x,y,z) 
- \frac{3}{8 \eps^2} (2 x y + x z + y z)
+ \frac{1}{8 \eps} (6 z^2 + x z + y z - x y)
\nonumber \\ &&
- \frac{3}{4 \eps} \left [(y + z) \AAx + (x+z) \AAy \right ]
\Bigr \}
,
\\
f_{\eta\eta V}(x,y,z) &=& 
\lim_{\eps \rightarrow 0} \Bigl \{
{\bf f}_{\eta\eta V}(x,y,z)
+ \frac{1}{4 \eps^2} (3 x^2 + 3 y^2 - z^2 + 3 x z + 3 y z)
\nonumber \\ &&
+ \frac{1}{12 \eps} (7z^2 - 3 x^2 - 3 y^2 - 18 x y - 15 x z - 15 y z)
\nonumber \\ &&
+ \frac{1}{2 \eps} \left [
3 x \AAx + 3 y \AAy + (d-1) (x+y-z/3) \AAz
\right ]
\Bigr \},
\\
f_{\eta\eta \Vbar}(x,y,z) &=& 
\lim_{\eps \rightarrow 0} \Bigl \{
{\bf f}_{\eta\eta \Vbar}(x,y,z)
+ \frac{1}{4 \eps^2} (x z + y z - x^2 - y^2 - z^2)
\nonumber \\ &&
+ \frac{1}{4 \eps} (z^2 - x^2 - y^2 + 2 x y + x z + y z)
\nonumber \\ &&
+ \frac{1}{2 \eps} \left [
- x \AAx - y \AAy + (x+y-z) \AAz
\right ]
\Bigr \}
,
\\
f_{FFS}(x,y,z) &=&  
\lim_{\eps \rightarrow 0} \Bigl \{
{\bf f}_{FFS}(x,y,z)
+ \frac{1}{2 \eps^2} (z^2 - x^2 - y^2 - 2 x z - 2 y z)
+ \frac{1}{2 \eps} (x+y-z)(x+y+z)
\nonumber \\ &&
+ \frac{1}{\eps} \left [
-x \AAx - y \AAy + (z-2x-2y) \AAz
\right ]
\Bigr \}
,
\\
f_{\Fbar\Fbar S}(x,y,z) &=& 
\lim_{\eps \rightarrow 0} \Bigl \{
{\bf f}_{\Fbar\Fbar S}(x,y,z) 
- \frac{1}{\eps^2} (x + y + z)
+ \frac{1}{\eps} (x + y + z)
\nonumber \\ &&
- \frac{2}{\eps} \left [ \AAx + \AAy + \AAz \right ]
\Bigr \}
,
\\
f_{FFV}(x,y,z) &=&  
\lim_{\eps \rightarrow 0} \Bigl \{
{\bf f}_{FFV}(x,y,z) 
+ \frac{1}{2 \eps^2} z (2z - 3 x - 3 y)
- \frac{1}{2 \eps} z (x + y + 8z/3)
\nonumber \\ &&
+ \frac{1}{\eps} (1-d) (x+y-2z/3) \AAz 
\Bigr \}
,
\\
f_{FF\Vbar}(x,y,z) &=&  
\lim_{\eps \rightarrow 0} \Bigl \{
{\bf f}_{FF\Vbar}(x,y,z) 
- \frac{1}{\eps^2} (x^2 + y^2 + x z/2 + y z/2)
+ \frac{1}{2 \eps} (4 x y + x z + y z)
\nonumber \\ &&
- \frac{1}{\eps} \left [ 2 x \AAx + 2 y \AAy + (x+y) \AAz \right ]
\Bigr \}
,
\\
f_{\Fbar\Fbar V}(x,y,z) &=&  
 \lim_{\eps \rightarrow 0} \Bigl \{
{\bf f}_{\Fbar\Fbar V}(x,y,z) 
- \frac{3}{\eps^2} (x + y + z)
+ \frac{1}{\eps} (x + y + 5 z)
\nonumber \\ &&
- \frac{6}{\eps} \left [ \AAx + \AAy + (d-1) \AAz/3 \right ]
\Bigr \}
,
\\
f_{\Fbar\Fbar\Vbar}(x,y,z) &=&  
 \lim_{\eps \rightarrow 0} \Bigl \{
{\bf f}_{\Fbar\Fbar\Vbar}(x,y,z) 
- \frac{1}{\eps^2} (x + y + z)
+ \frac{1}{\eps} (x + y + z)
\nonumber \\ &&
- \frac{2}{\eps} \left [ \AAx + \AAy + \AAz \right ]
\Bigr \}
,
\\
f_{\Fbar FG}(x,y,z) &=& 
 \lim_{\eps \rightarrow 0} \Bigl \{
{\bf f}_{\Fbar FG}(x,y,z) 
+ \frac{1}{2 \eps^2} (x^2 - y^2 - z^2 - 2 x y - 2 x z)
+ \frac{1}{2 \eps} (y+z-x) (x+y+z)
\nonumber \\ &&
+ \frac{1}{\eps} \left [ (x-2y-2z) \AAx -y \AAy -z \AAz \right ]
\Bigr \}
.
\eeq
Using eqs.~(\ref{eq:defineAA}) and (\ref{eq:defineII}), we thus obtain the \MSbar
two-loop functions:
\beq
f_{SSS}(x,y,z) &=& -I(x,y,z),
\label{eq:deffSSS}
\\
f_{SS}(x,y) &=&  A(x) A(y),
\label{eq:deffSS}
\\
f_{VS}(x,y) &=& 3 A(x) A(y) + 2 x A(y),
\label{eq:deffVS}
\\
f_{\overline{V}S}(x,y) &=& A(x) A(y),
\label{eq:deffvS}
\\
f_{SSV}(x,y,z) &=& \frac{1}{z} \Bigl [
-\lambda(x,y,z) I(x,y,z) 
+ (x-y)^2 I(0,x,y)
+ (y-x-z) A(x) A(z)
\nonumber \\ &&
+ (x-y-z) A(y) A(z)
\Bigr ] 
+ A(x) A(y)
+ 2 (3 x + 3 y - z) A(z)/3
,
\label{eq:deffSSV}
\\
f_{SS\overline{V}}(x,y,z) &=& \frac{1}{z} \Bigl \{
(x-y)^2 [ I(x,y,z) - I(0,x,y) ]
+ (x-y-z) A(x) A(z)
\nonumber \\ && 
+ (y-x-z) A(y) A(z)
\Bigr \}
,
\label{eq:deffSSv}
\\
f_{\overline{V}\overline{V}S}(x,y,z) &=& 
  \frac{1}{4 x y} \Bigl [ -(x+y-z)^2 I(x,y,z)
  + (x-z)^2 I(0,x,z) 
  + (y-z)^2 I(0,y,z)
\nonumber \\ &&
  -z^2 I(0,0,z) + (z-x-y) A(x) A(y)
  + y A(x) A(z) + x A(y) A(z)
\Bigr ] 
,
\label{eq:deffvvS}
\\
f_{\overline{V}VS}(x,y,z) &=& 
  -f_{\overline{V}\overline{V}S}(x,y,z) -I(x,y,z) - A(y)/2 
,
\label{eq:deffvVS}
\\
f_{VVS}(x,y,z) &=&  
  f_{\overline{V}\overline{V}S}(x,y,z)
  -2 I(x,y,z) + A(x)/2 + A(y)/2 + 2 A(z) -x-y-z,
\label{eq:deffVVS}
\\
f_{SSG}(x,y,z) &=& (x-y) I(x,y,z) + [A(x) - A(y)] A(z)
,
\label{eq:deffSSG}
\\
f_{GGS}(x,y,z) &=& \frac{1}{2} \Bigl [ (x+y-z) I(x,y,z) 
  + A(x) A(y)
  - A(x) A(z)
  - A(y) A(z)
  \Bigr ] 
,
\label{eq:deffGGS}
\\
f_{SGG}(x,y,z) &=& -(x-y)(x-z) I(x,y,z) 
  + (y-x) A(x) A(y)
  + (z-x) A(x) A(z) 
\nonumber \\ &&
  + x A(y) A(z)
,
\label{eq:deffSGG}
\\
f_{GSV}(x,y,z) &=& -\frac{1}{2} f_{SSV}(x,y,z),
\label{eq:deffGSV}
\\
f_{GS\overline{V}}(x,y,z) &=& \frac{1}{2 z} \Bigl [
(y-x) (x - y + z) I(x,y,z) 
+ (x-y)^2 I(0,x,y)
\nonumber \\ &&
+ (x-y+2z) A(y) A(z)
+ (y-x) A(x) A(z)
\Bigr ]
,
\label{eq:deffGSv}
\\
f_{GGG}(x,y,z) &=& \frac{1}{2} \bigl \{ (x-y) [
  (z-x-y) I(x,y,z) - A(x) A(y) ] 
\nonumber \\ &&	
  + (x+y) [A(y) - A(x)] A(z) \bigr \}
,
\label{eq:deffGGG}
\\
f_{GGV}(x,y,z) &=&  \frac{1}{2} \bigl [
  \lambda(x,y,z) I(x,y,z) 
  + (x+y-z) A(x) A(y)
  + (x+z-y) A(x) A(z)
\nonumber \\ &&	
  + (y+z-x) A(y) A(z)
\bigr ] 
+ z (z/3-x-y) A(z)
,
\label{eq:deffGGV}
\\
f_{VGG}(x,y,z) &=& \frac{1}{4x} \bigl [
\lambda(x,y,z) I(x,y,z)  - (y-z)^2 I(0,y,z)
  + (x+y-z) A(x) A(y)
 \nonumber \\ &&	
  + (x-y+z) A(x) A(z)
  - x A(y) A(z)
\bigr ] 
\nonumber \\ &&	
+ (x/6 - y/2 - z/2) A(x)
,
\label{eq:deffVGG}
\\
f_{\Vbar GG}(x,y,z) &=& \frac{1}{4x} \bigl [
  (x + y - z) (x - y + z) I(x,y,z) + (y-z)^2 I(0,y,z)
  + (x-y+z) A(x) A(y)
\nonumber \\ &&	
  + (x+y-z) A(x) A(z)
  - x A(y) A(z)
\bigr ]
,
\label{eq:deffvGG}
\\
f_{VVG}(x,y,z) &=& \frac{1}{4xy} \Bigl \{
  (x-y)[\lambda(x,y,z) + 12 x y] I(x,y,z)
  - x (x - z)^2 I(0,x,z) 
\nonumber \\ &&
  + y (y - z)^2 I(0,y,z)
  + (x - y) (x + y - z) A(x) A(y)
\nonumber \\ &&
  + y [10 x + y - z] A(x) A(z)
  - x [10 y + x - z] A(y) A(z)  \Bigr \}
\nonumber \\ &&
+ (y/2 + z/2 - 2 x/3) A(x)
- (x/2 + z/2 - 2 y/3) A(y)
\nonumber \\ &&
+ (x-y)(x+y+z),
\label{eq:deffVVG}
\\
f_{\Vbar VG}(x,y,z) &=& -\frac{1}{4} f_{SSV}(z,y,x) 
+ (y/2 + z/2 - x/6) A(x) - y A(y)/2,
\label{eq:deffvVG}
\\
f_{\eta\eta S}(x,y,z) &=& I(x,y,z),
\label{eq:deffggS}
\\
f_{\eta\eta G}(x,y,z) &=& -f_{GGS}(x,z,y) ,
\\
f_{VV}(x,y) &=& 
  \frac{27}{4} A(x) A(y) 
  + \frac{45}{8} [ y A(x) + x A(y)] 
  + \frac{63}{16} x y,
\label{eq:deffVV}
\\
f_{\Vbar V}(x,y) &=& 
  \frac{9}{4} A(x) A(y)
  + \frac{15}{8} y A(x) 
  + \frac{3}{8} x A(y) 
  + x y/16,
\label{eq:deffvV}
\\
f_{\Vbar\Vbar}(x,y) &=& 
  \frac{3}{4} A(x) A(y) 
  + \frac{1}{8}[y A(x) + x A(y)]
  - x y/16 ,
\label{eq:deffvv}  
\\
f_{VVV}(x,y,z) &=& \frac{1}{4xyz}  \Bigl \{
  -\lambda(x,y,z) \bigl [\lambda(x,y,z) + 12 (x y + x z + y z) \bigr ] I(x,y,z)
\nonumber \\ &&
  + (x-y)^2 (x^2 + y^2 + 10 x y) I(0,x,y) - z^4 I(0,0,z)
\nonumber \\ &&
  + (x-z)^2 (x^2 + z^2 + 10 x z) I(0,x,z) - y^4 I(0,0,y)
\nonumber \\ &&
  + (y-z)^2 (y^2 + z^2 + 10 y z) I(0,y,z) - x^4 I(0,0,x)
\nonumber \\ &&  
  + (z^2 -9 x^2 -9 y^2 +9 x z +9 y z -13 x y) z A(x) A(y)
\nonumber \\ &&  
  + (y^2 -9 x^2 -9 z^2 +9 x y +9 y z -13 x z) y A(x) A(z)
\nonumber \\ &&  
  + (x^2 -9 y^2 -9 z^2 +9 x y +9 x z -13 y z) x A(y) A(z)  
\Bigr \} 
\nonumber \\ &&
- \bigl [(40 x + 3 y + 3 z) A(x) 
+ (40 y + 3 x + 3 z) A(y) 
\nonumber \\ &&
+ (40 z + 3 x + 3 y) A(z) \bigr ]/24
+ \lambda(x,y,z) + 161 (x y + x z + y z)/16
,
\label{eq:deffVVV}
\\
f_{\Vbar VV}(x,y,z) &=& \frac{1}{4xyz}  \Bigl \{
  (y-z)^2 [\lambda(x,y,z) + 12 y z ] I(x,y,z) 
\nonumber \\ &&
  -(y-z)^2 [y^2 + z^2 + 10 y z] I(0,y,z)
  -(x-z)^2 z^2 I(0,x,z)
\nonumber \\ &&
  - (x-y)^2 y^2 I(0,x,y)
  + y^4 I(0,0,y)
  + z^4 I(0,0,z)
  - x (y-z)^2 A(y) A(z)
\nonumber \\ &&
  + [x z - z^2 - 11 x y - 9 y z + 9 y^2 ] z A(x) A(y)
\nonumber \\ &&
  + [x y - y^2 - 11 x z - 9 y z + 9 z^2 ] y A(x) A(z)
\Bigr \}
\nonumber \\ &&
+[-15(y+z) A(x) + (8y-4z-3x) A(y) + (8z-4y-3x) A(z)]/8
\nonumber \\ &&
+ (y-z)^2 - x (y+z)/16 ,
\label{eq:deffvVV}
\\
f_{\Vbar\Vbar V}(x,y,z) &=& \frac{1}{4xy} \Bigl \{
  -z \lambda(x,y,z) I(x,y,z)
  + (x-z)^2 z I(0,x,z)
  + (y-z)^2 z I(0,y,z)
\nonumber \\ &&
  - z^3 I(0,0,z)
  + [z^2 - x z - y z - 3 x y] A(x) A(y)
\nonumber \\ &&
  + z [x A(y) + y A(x)] A(z) \Bigr \}
  - [y A(x) + x A(y)]/8 + x y/16,
\label{eq:deffvvV}
\\
f_{\eta\eta V}(x,y,z) &=& 
  \frac{1}{2z} \bigl [
    \lambda(x,y,z) I(x,y,z) - (x-y)^2 I(0,x,y) -z A(x) A(y) 
\nonumber \\ &&
    + (x - y + z) A(x) A(z) 
    + (y - x + z) A(y) A(z)
\bigr ]
\nonumber \\ &&
  + (z/3 - x - y) A(z)
,
\label{eq:deffggV}
\\
f_{\eta\eta \Vbar}(x,y,z) &=& 
 \frac{1}{2z} \bigl [
   (y-x-z)(x-y-z) I(x,y,z) + (x-y)^2 I(0,x,y) -z A(x) A(y) 
\nonumber \\ &&
   + (y - x + z) A(x) A(z) 
   + (x - y + z) A(y) A(z)
\bigr ]
,
\label{eq:deffggv}
\\
f_{FFS}(x,y,z) &=&  (x+y-z) I(x,y,z) 
  + A(x) A(y)
  - A(x) A(z)
  - A(y) A(z)
,
\label{eq:deffFFS}
\\
f_{\Fbar\Fbar S}(x,y,z) &=& 2 I(x,y,z),
\label{eq:deffffS}
\\
f_{FFV}(x,y,z) &=&  \frac{1}{z} \Bigl \{
[\lambda(x,y,z) + 3 z (x+y-z)] I(x,y,z) - (x-y)^2 I(0,x,y)
\nonumber \\ &&
+ [x-y -2 z] A(x) A(z) 
+ [y-x -2 z] A(y) A(z)
+ 2 z A(x) A(y)
\Bigr \}
\nonumber \\ &&
-2 x A(x) - 2 y A(y) 
+ (2z/3 - 2 x - 2 y) A(z)
+ (x + y - z) (x + y + z)
,
\label{eq:deffFFV}
\\
f_{FF\Vbar}(x,y,z) &=&  \frac{1}{z} \bigl \{
  (x z + y z - x^2 + 2 x y - y^2) I(x,y,z) + (x-y)^2 I(0,x,y)
\nonumber \\ &&
  + (x-y) [A(y) - A(x)] A(z)
  \bigr \}
,
\label{eq:deffFFv}
\\
f_{\Fbar\Fbar V}(x,y,z) &=&  6 I(x,y,z) - 4 A(x) - 4 A(y) + 2 x + 2 y + 2 z,
\label{eq:deffffV}
\\
f_{\Fbar\Fbar \Vbar}(x,y,z) &=&  2 I(x,y,z) ,
\label{eq:deffffv}
\\
f_{\Fbar FG}(x,y,z) &=& 
(y+z-x) I(x,y,z) 
  - A(x) A(y)
  - A(x) A(z)
  + A(y) A(z).
\label{eq:defffFG}  
\eeq
The results for $f_{SS}$, $f_{SSS}$, $f_{VS}$, $f_{SSV}$, $f_{VVS}$, 
$f_{VV}$, $f_{VVV}$, $f_{\eta\eta V}$, $f_{FFS}$, $f_{\Fbar\Fbar S}$, 
$f_{FFV}$, and $f_{\Fbar\Fbar V}$ agree with those found in 
refs.~\cite{Ford:1992pn,Martin:2001vx}; the other functions do not 
contribute in Landau gauge. In ref.~\cite{Martin:2001vx}, some of these
functions were combined, so that a function $f_{\rm gauge}$ 
included all of the effects of $f_{VVV}$, $f_{VV}$, and $f_{\eta\eta V}$. 
In the present paper we choose to keep them separate so that the functions are in 
correspondence with the Feynman diagrams, to keep their origins clear.

Despite the factors of $1/x$, $1/y$, or $1/z$ appearing in the above expressions,
the two-loop integral functions are 
all finite and well-defined  in the limits of massless vector 
bosons.\footnote{However, this is not true at three-loop and higher orders
for similar loop integral functions involving massless gauge bosons. 
The three-loop contribution to the
Standard Model effective potential has a (benign) IR logarithmic divergence 
due to doubled photon propagators
\cite{Martin:2017lqn}.}
To make this plain, one can take the appropriate limits $x\rightarrow 0$, 
etc. The limiting cases that are not immediately obvious are:
\beq
f_{SSV}(x,y,0) &=& 
  (x + y)^2  - 2 x A(x) - 2 y A(y) + 3 A(x) A(y) + 3 (x + y) I(0, x, y)
,
\label{eq:fSSVxy0}
\\
f_{SS\Vbar}(x,y,0) &=& 
  -(x + y)^2  + 2 x A(x) + 2 y A(y) - 2 A(x) A(y) - (x + y) I(0, x, y)
,
\\
f_{VVS}(0,y,z) &=& \frac{1}{4y} \bigl [
  (3z - 9y) I(0,y,z) - 3z I(0,0,z) + 3 A(y) A(z) 
  + 8 y A(z) 
\nonumber \\ &&
  - y (3y+2z)  
  \bigr ]
,  
\\
f_{VVS}(0,0,z) &=& - 3 I(0,0,z) + 7 A(z)/2 - 5 z/4
,
\\
f_{\Vbar VS}(0,y,z) &=& \frac{1}{4y} \bigl [
  -3 (y+z) I(0,y,z) + 3z I(0,0,z) - 3 A(y) A(z) - y (y + 2z)
\bigr ]
,  
\\
f_{\Vbar VS}(x,0,z) &=& \frac{1}{4x} \bigl [
  -3 (x+z) I(0,x,z) + 3 z I(0,0,z) - 3 A(x) A(z) + 2 x A(x) 
\nonumber \\ &&
  - x (x + 2 z)
\bigr ] 
,  
\\
f_{\Vbar VS}(0,0,z) &=& -3 A(z)/2 + z/4
,
\\
f_{\Vbar\Vbar S}(0,y,z) &=&
 \frac{1}{4y} \bigl [
  (3z-y) I(0,y,z) - 3z I(0,0,z) + 3 A(y) A(z) -2 y A(y) + y (y + 2z)
\bigr ]
, \phantom{xxx} 
\\
f_{\Vbar\Vbar S}(0,0,z) &=&
  -I(0,0,z) + 3 A(z)/2 - z/4 
, 
\\
f_{GSV}(x,y,0) &=& \bigl [
-3 (x+y) I(0,x,y) - 3 A(x) A(y) + 2 x A(x) + 2 y A(y) - (x+y)^2
\bigr ]/2 
,
\\
f_{GS\Vbar}(x,y,0) &=& 
  y I(0,x,y) + A(x) A(y) - x A(x) - y A(y) + (x+y)^2/2
,
\\
f_{VGG}(0,y,z) &=& \bigl [
-3 (y+z) I(0,y,z) - 3 A(y) A(z) + 2 y A(y) + 2 z A(z) - (y+z)^2
\bigr ]/4
,
\\
f_{\Vbar GG}(0,y,z) &=& \bigl [
(y+z) I(0,y,z) + A(y) A(z) - 2 y A(y) - 2 z A(z) + (y+z)^2
\bigr ]/4
,
\\
f_{VVG}(0,y,z) &=& \frac{1}{4y} \bigl [
  (z^2 + y z - 8 y^2) I(0,y,z) - z^2 I(0,0,z) + (z-8y) A(y) A(z) 
  \bigr ] 
\nonumber \\ &&
  + (y-3z) A(y)/6 - z A(z)/2 + (y+z)(z-3y)/4
,
\\
f_{VVG}(0,0,z) &=& 0
,
\\
f_{\Vbar VG}(0,y,z) &=& \bigl [
  -3 (y+z) I(0,y,z) - 3 A(y) A(z) + 2 z A(z) - (y+z)^2
  \bigr ]/4
,
\\
f_{VVV}(0,y,z) &=& \frac{1}{4 y z} \bigl [
  (y+z)(7y-z)(7z-y) I(0,y,z) + 7 y^3 I(0,0,y) + 7 z^3 I(0,0,z) 
\nonumber \\ &&
  + 7 (y z - y^2 - z^2) A(y) A(z)  
  \bigr ] - \frac{5}{24} \bigl [(32 y + 3 z) A(y) 
  + (32 z + 3 y) A(z) \bigr ] 
\nonumber \\ &&
  + 4 y^2 + 4 z^2 + 217 y z/16
,  
\\
f_{VVV}(0,0,z) &=& 25 z I(0,0,z)/2 - 61 z A(z)/6 + 23 z^2/4
,
\\
f_{\Vbar VV}(0,y,z) &=& \frac{1}{4 y z} \bigl [
  -3 (y+z)^3 I(0,y,z) + 3 y^3 I(0,0,y) + 3 z^3 I(0,0,z)
\nonumber \\ &&
  -3 (y^2 + 6 y z + z^2) A(y) A(z)
  \bigr ] + 6 y A(y) + 6 z A(z) 
\nonumber \\ &&
  - 2 y^2 - 2 z^2 - 15 y z/2
,
\\
f_{\Vbar VV}(x,0,z) &=& \frac{1}{4 x} \bigl [
  (x z - 2 x^2 + 7 z^2) I(0,x,z) - 7 z^2 I(0,0,z) + (7z-11x) A(x) A(z)
  \bigr ]
\nonumber \\ &&
  -11z A(x)/8 + (z-3x/8) A(z) + z (8z-5x)/16 
,
\\
f_{\Vbar VV}(x,0,0) &=& -x I(0,0,x)/2 
,
\\
f_{\Vbar VV}(0,0,z) &=& -3 z I(0,0,z)/2  + 9 z A(z)/2 -5 z^2/4
,
\\
f
_{\Vbar\Vbar V}(0,y,z) &=& \frac{1}{4 y} \bigl [
  3 (y+z) z I(0,y,z) - 3 z^2 I(0,0,z) + 3 z A(y) A(z)
  \bigr ] 
\nonumber \\ &&
  - z A(y)/2
  + z (y+2z)/4
,
\\
f_{\Vbar\Vbar V}(0,0,z) &=& 3 z A(z)/2 - z^2/4 
,
\\
f_{\eta\eta V}(x,y,0) &=& \bigl [
  -3 (x+y) I(0,x,y) - 3 A(x) A(y) + 2 x A(x) + 2 y A(y) - (x+y)^2 \bigr ]/2
,
\\
f_{\eta\eta\Vbar}(x,y,0) &=& \bigl [
  (x+y) I(0,x,y) + A(x) A(y) - 2 x A(x) - 2 y A(y) + (x+y)^2 \bigr ]/2
,
\\
f_{FFV}(x,y,0) &=& 0
,
\\
f_{FF\Vbar}(x,y,0) &=& 
  2 (x + y) I(0, x, y) + 2 A(x) A(y) - 2 x A(x) - 2 y A(y)  + (x + y)^2 
.
\label{eq:fFFvxy0}
\eeq

For convenience, the results listed in eqs.~(\ref{eq:deffSSS})-(\ref{eq:fFFvxy0}) are
also given in electronic form in an ancillary file 
distributed with this paper, called {\tt functions}.
In order to carry out the renormalization group invariance check 
of eq.(\ref{eq:RGcheck}) in specific models, it is 
useful to have the derivatives of the above integral functions with respect to
the renormalization scale $Q$. These are provided in Appendix B.

\section{Examples\label{sec:examples}}
\setcounter{equation}{0}
\setcounter{figure}{0}
\setcounter{table}{0}
\setcounter{footnote}{1}

\subsection{Simplifications for models without Goldstone boson 
mixing\label{subsec:nomixing}}

In favorable cases, the Goldstone sector scalar squared masses are 
separate from the non-Goldstone scalars, and diagonal, so 
that the contributions in eqs.~(\ref{eq:defmu2jk}) and (\ref{eq:defmu2jksectors}) satisfy: 
\beq 
\mu^2_{Aj} &=& 0
,
\label{eq:noGoldstonescalarmixing} 
\\ 
\mu^2_{AB} &=& \delta_{AB} \mu^2_{A} .
\label{eq:noGoldstonemixing} 
\eeq 
This implies a significant simplification, because now 
the propagators for each index $A$ do not mix, and the unphysical
squared masses $M^2_\kappa$ occurring in the scalar and 
massive vector propagators are obtained as the solutions 
of only quadratic equations. 
This happy circumstance occurs for theories with only one background field $\phi_j$ 
in a single irreducible representation
of the gauge group, as in the 
Abelian Higgs model and the Standard Model. However, 
eq.~(\ref{eq:noGoldstonescalarmixing}) fails to hold 
in theories such as the Minimal Supersymmetric 
Standard Model or more general two Higgs doublet models; those theories do have
mixing between the Goldstone and physical Higgs scalar bosons,
and so must be treated with the more general formalism given in section
\ref{sec:effpot} above.

\begin{figure}[!t]
\epsfxsize=0.85\linewidth
\epsffile{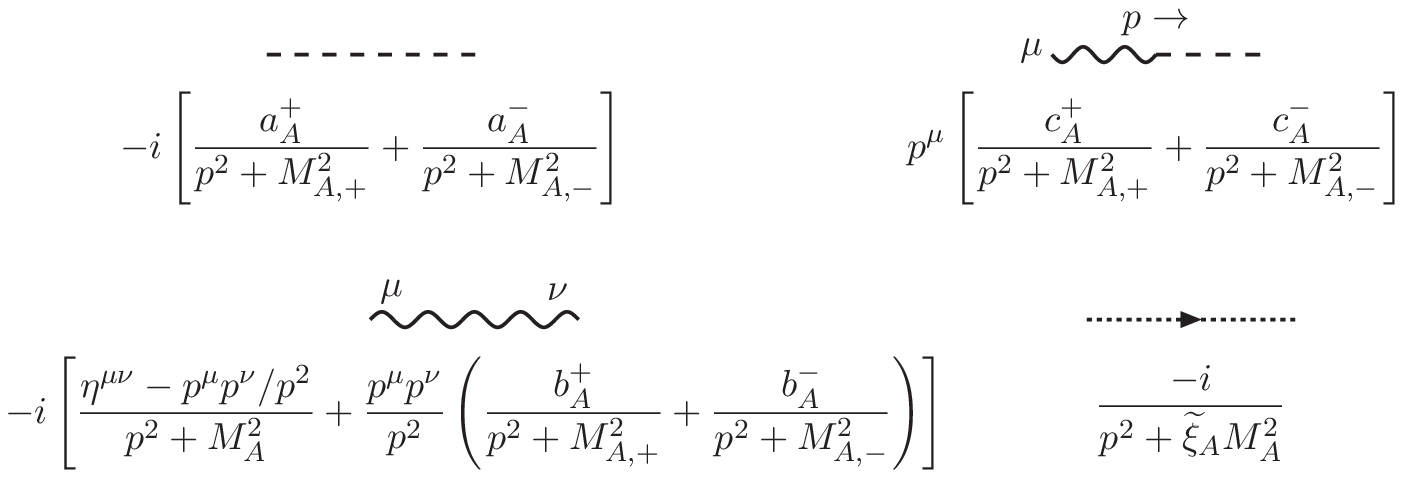}
\vspace{-0.4cm}
\begin{center}\begin{minipage}[]{0.95\linewidth}
\caption{\label{fig:propagatorsgeneralmassive}Feynman rules 
for the propagators of the 
Goldstone scalars $G_A$ (dashed lines), 
massive vectors $Z_\mu^A$ (wavy lines), 
and the corresponding ghosts and antighosts
$\eta^A,  \overline\eta^A$ 
(dotted lines with arrows), each carrying 4-momentum $p^\mu$,
in the special case that 
eqs.~(\ref{eq:noGoldstonescalarmixing}) and 
(\ref{eq:noGoldstonemixing}) hold.
The squared masses $M_{A,\pm}^2$ and the coefficients 
$a_A^\pm$, $b_A^\pm$, and $c_A^\pm$
are defined in eqs.~(\ref{eq:defM2pmgen})-(\ref{eq:defcpmcogen}).}
\end{minipage}\end{center}
\end{figure}

\begin{figure}[!t]
\epsfxsize=0.89\linewidth
\epsffile{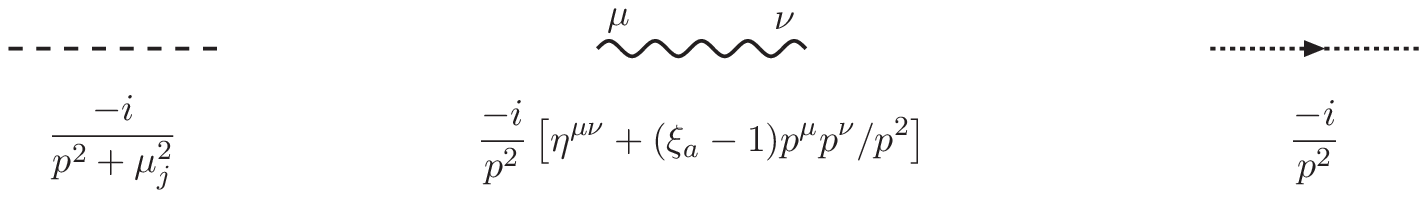}
\vspace{-0.4cm}
\begin{center}\begin{minipage}[]{0.95\linewidth}
\caption{\label{fig:propagatorsgeneralmassless}Feynman rules for 
the propagators of the 
non-Goldstone scalars $R_{j}$ (dashed lines), massless vectors $A_\mu^{a}$ 
(wavy lines), and the corresponding massless ghosts 
$\overline \eta^{a},  \eta^{a}$ 
(dotted lines with arrows), each carrying 4-momentum $p^\mu$,
in the special case that 
eqs.~(\ref{eq:noGoldstonescalarmixing}) and 
(\ref{eq:noGoldstonemixing}) hold.}
\end{minipage}\end{center}
\end{figure}

In the following, we present the results for the
case that eqs.~(\ref{eq:noGoldstonescalarmixing}) 
and (\ref{eq:noGoldstonemixing}) hold; then 
the propagator Feynman rules for the bosons 
simplify to the forms shown in Figures \ref{fig:propagatorsgeneralmassive}
and \ref{fig:propagatorsgeneralmassless}.%
The unphysical squared mass poles $M_\kappa^2$ for the 
massive vectors and Goldstone scalars are now at
\beq
M_{A,\pm}^2 &\equiv& \xiprime_A M_A^2 + \frac{1}{2} \biggl (\mu_A^2  \pm
\sqrt{\mu_A^2 \left [\mu_A^2 + 4 (\xiprime_A-\xi_A) M_A^2 \right ]} \biggr ) 
,
\label{eq:defM2pmgen}
\eeq
for each index $A$, with residue coefficients
\beq
a_A^\pm &=& \frac{M_{A,\pm}^2 - \xi_A M_A^2}{M_{A,\pm}^2 - M_{A,\mp}^2},
\label{eq:defapmcogen}
\\
b_A^\pm &=& \frac{
(2 \xi_A - \xiprime_A) \xiprime_A M_A^2
-\xi_A M_{A,\mp}^2 
}{
M_{A,\pm}^2 - M_{A,\mp}^2},
\label{eq:defbpmcogen}
\\
c_A^\pm &=& \frac{(\xi_A - \xiprime_A) M_A}{M_{A,\pm}^2 - M_{A,\mp}^2}.
\label{eq:defcpmcogen}
\eeq
Note that the superscript labels $\pm$ here correspond to the labels $\kappa$ appearing
in Figure \ref{fig:propagatorsverygeneral}, and 
\beq
a_A^+ + a_A^- &=& 1,
\\
b_A^+ + b_A^- &=& \xi_A,
\\
c_A^+ + c_A^- &=& 0.
\eeq
The massive vectors $Z_\mu^A$ and their associated Goldstone scalars $G_A$ have
propagator mixing proportional to $\xi_A - \xiprime_A$, 
and they have three distinct 
poles in $-p^2$, at 
$M_A^2$, $M_{A,+}^2$, and $M_{A,-}^2$. 
The latter two squared mass poles are real
(but not necessarily positive!) if and only if 
\beq
4 (\xi_A - \xiprime_A) M_A^2 \mu_A^2 &\leq& (\mu_A^2)^2.
\eeq
Note that care is needed to cancel $\mu_A^2$ in this inequality,
because it can have either sign. At one-loop order, complex squared
mass poles do not lead to an imaginary part of the effective potential
\cite{Espinosa:2016uaw}, but the two-loop order basis integral $I(x,y,z)$ 
has a less obvious branch cut structure when one or more of
its arguments are complex. In this paper,
we will simply avoid choices of the gauge-fixing parameters that  make the squared
mass arguments complex. 

As simple special cases, we have:
\beq
\mbox{Background-field $R_\xi$ gauge:} &&
M^2_{A,+} = \xi_A M_A^2 + \mu_A^2,\>\>\>\> \>\>\>\>
M^2_{A,-} = M^2_{A,\eta} = \xi_A M_A^2,
\nonumber \\ &&
a_A^+ = 1,\>\>\>\>\> 
a_A^- = 0,\>\>\>\>\> 
b_A^+ = 0,\>\>\>\>\> 
b_A^- = \xi_A,\>\>\>\>\> 
c_A^\pm = 0,
\eeq
and the further specialization 
\beq
\mbox{Landau gauge:}&&M^2_{A,+} = \mu_A^2,\>\>\>\> \>\>\>\> \>\> 
M^2_{A,-} = M^2_{A,\eta} = 0,
\\
&& 
a_A^+ = 1,\>\>\>\>\>
a_A^- = 0,\>\>\>\>\> b_A^\pm = 0, \>\>\>\>\> c_A^\pm = 0.
\eeq

As before, we use the index of
a field as a synonym for the squared mass whenever it appears as the argument
of a loop integral function, so that in the following,
\beq
A &=& M^2_{A},
\label{eq:Anotation}
\\
A_\pm &=& M^2_{A,\pm}, 
\label{eq:Apmnotation}
\\
A_\eta &=& \xiprime_A M^2_A.
\label{eq:Aetanotation}
\eeq
The 1-loop contribution to the effective potential can now be re-expressed as:
\beq
V^{(1)} &=& \sum_j f(j) - 2 \sum_I f(I) + \sum_A \left [
3 f_V(A) + f(A_+) + f(A_-) - 2 f(A_\eta)
\right ].
\eeq

In order to facilitate compact expressions below, 
we also extend the squared-mass
notations to the massless vector fields, so that
when appearing as the argument of a two-loop integral function, 
$\bf a$ and ${\bf a}_\pm$ and ${\bf a}_\eta$ are to be interpreted 
according to eqs.~(\ref{eq:Anotation})-(\ref{eq:Aetanotation}) when
${\bf a} = A$, and are taken to vanish when ${\bf a} = a$.
We also define residue coefficients
\beq
b_a^+ &=& 0, \label{eq:defbap}
\\
b_a^- &=& \xi_a, \label{eq:defbam}
\eeq
so that the notation $b_{\bf a}^\pm$ is to be interpreted by 
either eq.~(\ref{eq:defbpmcogen}) or eqs.~(\ref{eq:defbap})-(\ref{eq:defbam}),
depending on whether the corresponding vector field is massive or not.
Similarly, for scalar fields, the notation for coefficients $a_{\bf j}^\pm$
is to be interpreted using eq.~(\ref{eq:defapmcogen}) when ${\bf j} = A$ is a
Goldstone scalar, or
\beq
a_j^+ &=& 1,
\\
a_j^- &=& 0,
\label{eq:defajm}
\eeq
when ${\bf j} = j$ is a non-Goldstone scalar. Furthermore, when 
${\bf j}_\pm$ appears as an argument in a loop integral
function, it is to be interpreted either
according to eq.~(\ref{eq:Apmnotation})
for a Goldstone scalar or  
\beq
j_+ &=& \mu_j^2,
\eeq
for a non-Goldstone scalar. [Note that 
$j_-$ is not relevant as the argument of a loop integral
function, because of eq.~(\ref{eq:defajm}).] 
The two-loop contributions to the effective potential, 
given in the most general case in 
eqs.~(\ref{eq:V2formSS})-(\ref{eq:V2formfFG}),
now become:
\beq
V^{(2)}_{SS} &=&
  \frac{1}{8} \lambda^{\bf jjkk} a_{\bf j}^\eps a_{\bf k}^{\eps'} 
  f_{SS}({\bf j}_\eps, {\bf k}_{\eps'})
\label{eq:V2formSSsimp}
,
\\
V^{(2)}_{SSS} &=&
  \frac{1}{12} \bigl (\lambda^{\bf jkl}\bigr )^2 
  a_{\bf j}^\eps a_{\bf k}^{\eps'} a_{\bf l}^{\eps''}
  f_{SSS}({\bf j}_\eps, {\bf k}_{\eps'}, {\bf l}_{\eps''})
,
\\
V^{(2)}_{VS} &=&
  \frac{1}{2} \left (g^{\bf a}_{\bf jk}\right )^2 
  a_{\bf j}^\eps \left [  
  f_{VS}({\bf a}, {\bf j}_\eps) 
  + b_{\bf a}^{\eps'} f_{\Vbar S}({\bf a}_{\eps'}, {\bf j}_\eps) \right ]
,
\\
V^{(2)}_{SSV} &=&
  \frac{1}{4} \left ( g^{\bf a}_{\bf jk} \right )^2
  a_{\bf j}^\eps a_{\bf k}^{\eps'} \left [
  f_{SSV}({\bf j}_\eps, {\bf k}_{\eps'}, {\bf a}) +
  b_{\bf a}^{\eps''} f_{SS\Vbar}({\bf j}_\eps, {\bf k}_{\eps'}, {\bf a}_{\eps''})
  \right ]
,
\\
V^{(2)}_{VVS} &=& 
  \frac{1}{4} \bigl ( G^{\bf ab}_{\bf j} \bigr )^2 a_{\bf j}^\eps \left [
  f_{VVS}({\bf a},{\bf b}, {\bf j}_\eps) 
  + 2 b_{\bf a}^{\eps'} f_{\Vbar VS}({\bf a}_{\eps'},{\bf b}, {\bf j}_\eps) 
  + b_{\bf a}^{\eps'} b_{\bf b}^{\eps''} 
  f_{\Vbar\Vbar S}({\bf a}_{\eps'},{\bf b}_{\eps''}, {\bf j}_\eps) 
  \right ] 
,
\\
V^{(2)}_{SSG} &=& \frac{1}{2} \lambda^{A\bf jk} g^A_{\bf jk}
  a_{\bf j}^\eps a_{\bf k}^{\eps'} c_A^{\eps''} 
  f_{SSG}({\bf j}_\eps, {\bf k}_{\eps'}, A_{\eps''})
,
\\
V^{(2)}_{GGS} &=& \frac{1}{2} \lambda^{AB\bf j} G^{AB}_{\bf j} c_A^\eps c_B^{\eps'}
  a_{\bf j}^{\eps''} f_{GGS}(A_\eps, B_{\eps'}, {\bf j}_{\eps''})
,
\\
V^{(2)}_{SGG} &=& \frac{1}{2}
  g^A_{B\bf j} g^B_{A\bf j} a_{\bf j}^{\eps} c_A^{\eps'} c_B^{\eps''}  
  f_{SGG}({\bf j}_{\eps}, A_{\eps'}, B_{\eps''})
,
\\
V^{(2)}_{GSV} &=& 
  g^{\bf a}_{{\bf j}A} G^{A\bf a}_{\bf j} c_A^\eps a_{\bf j}^{\eps'} \left [
  f_{GSV}(A_\eps, {\bf j}_{\eps'}, {\bf a}) +
  b_{\bf a}^{\eps''} f_{GS\Vbar}(A_\eps, {\bf j}_{\eps'}, {\bf a}_{\eps''})
  \right ]
,
\\
V^{(2)}_{GGG} &=& \frac{1}{2} g^{C}_{AB} G^{AB}_C c_A^\eps c_B^{\eps'} c_C^{\eps''}
  f_{GGG}(A_\eps, B_{\eps'}, C_{\eps''})
,
\\
V^{(2)}_{GGV} &=& \frac{1}{2}g^{{\bf a}AB} g^{\bf a}_{AB}  c_A^\eps c_B^{\eps'}
  f_{GGV}(A_\eps, B_{\eps'}, {\bf a})   
,
\\
V^{(2)}_{VGG} &=& \frac{1}{2} G^{A\bf a}_B G^{B\bf a}_A c_A^{\eps} c_B^{\eps'}
  \left [
  f_{VGG} ({\bf a}, A_\eps, B_{\eps'}) + 
  b_{\bf a}^{\eps''} f_{\Vbar GG} ({\bf a}_{\eps''}, A_\eps, B_{\eps'})
  \right ] 
,
\\
V^{(2)}_{VVG} &=& \frac{1}{2} g^{{\bf ab}A} G^{\bf ab}_A c_A^\eps \Bigl [
  f_{VVG}({\bf a},{\bf b},A_\eps) 
  + 2 b_{\bf a}^{\eps'} f_{\Vbar VG}({\bf a}_{\eps'},{\bf b},A_\eps)
    \Bigr ]
,
\\
V^{(2)}_{\eta\eta S} &=&
  \frac{1}{2} g^A_{B\bf j} g^B_{A\bf j} \xiprime_A \xiprime_B M_A M_B
  a_{\bf j}^\eps f_{\eta\eta S}(A_\eta, B_\eta, {\bf j}_\eps) 
,
\\
V^{(2)}_{\eta\eta G} &=& g^{{\bf a}AB} g^{\bf a}_{AB} \xiprime_A M_A c_B^\eps 
  f_{\eta\eta G}({\bf a}_\eta, A_\eta, B_\eps)
,
\\
V^{(2)}_{VV} &=& \frac{1}{4} \bigl ( g^{\bf abc} \bigr )^2 \left [
  f_{VV}({\bf a}, {\bf b})
  + 2 b_{\bf a}^\eps f_{\Vbar V}({\bf a}_\eps, {\bf b})
  + b_{\bf a}^\eps b_{\bf b}^{\eps'}  f_{\Vbar\Vbar}({\bf a}_\eps, {\bf b}_{\eps'})
  \right ] 
,
\\
V^{(2)}_{VVV} &=&
  \frac{1}{12} \bigl (g^{\bf abc} \bigr )^2 \Bigl [
  f_{VVV}({\bf a}, {\bf b}, {\bf c})
  + 3 b_{\bf a}^\eps f_{\Vbar VV}({\bf a}_\eps, {\bf b}, {\bf c}) 
  + 3 b_{\bf a}^\eps b_{\bf b}^{\eps'} 
  f_{\Vbar\Vbar V}({\bf a}_\eps, {\bf b}_{\eps'}, {\bf c})
  \Bigr ]
,
\\
V^{(2)}_{\eta\eta V} &=&
  \frac{1}{4} \bigl (g^{\bf abc}\bigr )^2 \left[
  f_{\eta\eta V}({\bf a}_\eta, {\bf b}_\eta, {\bf c})
  + b_{\bf c}^\eps f_{\eta\eta\Vbar}({\bf a}_\eta, {\bf b}_\eta, {\bf c}_\eps) 
  \right ]
,
\\
V^{(2)}_{FFS} &=&
\frac{1}{2} Y^{{\bf j}IJ} Y_{{\bf j}IJ} a_{\bf j}^\eps f_{FFS}(I,J,{\bf j}_\eps)
,
\\
V^{(2)}_{\Fbar\Fbar S} &=&
\frac{1}{4} (Y^{{\bf j}IJ} Y^{{\bf j}I'J'} M_{II'} M_{JJ'} + {\rm c.c.}) 
  a_{\bf j}^\eps f_{\Fbar\Fbar S}(I,J,{\bf j}_\eps)
,
\\
V^{(2)}_{FFV} &=&
\frac{1}{2} g^{{\bf a}J}_I g^{{\bf a}I}_J 
\left [f_{FFV}(I,J,{\bf a}) + 
b_{\bf a}^\eps f_{FF\Vbar}(I,J,{\bf a}_\eps) \right ]
,
\\
V^{(2)}_{\Fbar\Fbar V} &=&
\frac{1}{2} g^{{\bf a}J}_I g^{{\bf a}J'}_{I'} M^{II'} M_{JJ'} 
\left [f_{\Fbar\Fbar V}(I,J,{\bf a}) 
+ b_{\bf a}^\eps f_{\Fbar\Fbar\Vbar}(I,J,{\bf a}_\eps) \right ]
,
\\
V^{(2)}_{\Fbar FG} &=& i\left (
g^{AJ}_I Y_{AI'J} M^{II'} - {\rm c.c.} \right ) c_A^\eps f_{\Fbar FG}(I,J,A_\eps) .
\label{eq:V2formfFGsimp}
\eeq
In these equations, all indices are summed over in each term that 
they appear in, including $\eps, \epsp, \epspp$ each summed over $\pm$.

\subsection{Abelian Higgs model\label{subsec:AHmodel}}

Consider as an example the Abelian Higgs model. The Lagrangian is:
\beq
{\cal L} = -\frac{1}{4} F^{\mu\nu} F_{\mu\nu} 
-D^\mu \Phi^* D_\mu \Phi  - \Lambda -m^2 |\Phi|^2 - \lambda |\Phi|^4
+ {\cal L}_{\rm g.f.} + {\cal L}_{\rm ghosts},
\eeq
where $\Phi$ is a complex scalar field charged under a $U(1)$ gauge symmetry
with vector field $Z_\mu$ and field strength 
\beq
F_{\mu\nu} &=& \partial_\mu Z_\nu - \partial_\nu Z_\mu,
\eeq
with covariant derivatives
\beq
D_\mu \Phi &=& (\partial_\mu - i e Z_\mu) \Phi,
\\
D^\mu \Phi^* &=& (\partial^\mu + i e Z^\mu) \Phi^*,
\eeq
and $\lambda$ is a positive scalar self-interaction coupling,
$m^2$ is a squared mass, and
$\Lambda$ is a field-independent vacuum energy 
(needed for renormalization scale invariance of the effective potential).
Now write:
\beq
\Phi(x) &=& \frac{1}{\sqrt{2}} [\phi + H(x) + i G(x)] ,
\eeq
where $\phi$ is the position-independent background scalar field,
and $H,G$ are real scalar fields.
Then:
\beq
{\cal L} &=& -\frac{1}{4} F_{\mu\nu} F^{\mu\nu} 
- \frac{1}{2} (\partial_\mu H + e Z_\mu G)^2
- \frac{1}{2} (\partial_\mu G - e Z_\mu (\phi+H))^2
\nonumber \\ &&
- \Lambda 
- \frac{1}{2} m^2 [G^2 + (\phi + H)^2] 
- \frac{\lambda}{4} [G^2 + (\phi + H)^2]^2
\nonumber \\ &&
+ {\cal L}_{\rm g.f.} + {\cal L}_{\rm ghost}
,
\eeq
where, in terms of the Nakanishi-Lautrup Lagrange multiplier field $b$ and
the ghost and anti-ghost fields $\eta$ and $\bar\eta$,
\beq
{\cal L}_{\rm g.f.} &=&  \frac{1}{2} \xi b^2  - 
b (\partial_\mu Z^\mu - e \xiprime \phi G) ,
\\
{\cal L}_{\rm ghost} &=&  -\partial_\mu \bar \eta \partial^\mu \eta 
- \xiprime e^2 \phi (\phi + H) \bar \eta \eta .
\eeq
This Lagrangian is invariant under the Grassmann-odd BRST symmetry:
\beq
\delta_{\rm BRST} Z_\mu &=& \partial_\mu \eta,
\\
\delta_{\rm BRST} G &=& e (\phi + H) \eta,
\\
\delta_{\rm BRST} H &=& -e G \eta,
\\
\delta_{\rm BRST} \bar \eta &=& b,
\\
\delta_{\rm BRST} \eta &=& 0,
\\
\delta_{\rm BRST} b &=& 0.
\eeq
Because the BRST symmetry does not require any 
particular relation for $\xi$ and $\xiprime$, 
there is no reason that they should not be renormalized 
differently, with independent counterterms. 

The parameters of the theory are: $\phi$, $e$, $\lambda$, $m^2$, $\Lambda$, $\xi$,
and $\xiprime$. 
This model can be obtained from the general case by:
\beq
g_a &\rightarrow& e,
\\
t^a_{GH} = -t^a_{HG} &=& -i,
\\
g^a_{GH} = -g^a_{HG} &=& e,
\\
\widetilde \phi^a_H &=& \xiprime \phi,
\\
\widetilde \phi^a_G &=& 0,
\\
F^Z{}_G &=& e \phi,
\\
F^Z{}_H &=& 0.
\eeq
Because there is only one Goldstone boson and it does not mix with the
non-Goldstone scalar $H$, the formalism of the previous subsection 
\ref{subsec:nomixing} applies.
The squared mass eigenvalues for use as arguments of loop integral functions
are:
\beq
H &\equiv& m^2 + 3\lambda \phi^2,
\label{eq:defHAH}
\\
Z &\equiv& e^2 \phi^2,
\\
Z_\pm &\equiv& 
\xiprime Z +
\frac{1}{2} \left [G  \pm
\sqrt{G [G + 4 (\xiprime-\xi) Z]} \right ] 
,
\label{eq:defM2pm}
\\
\eta &\equiv& \xiprime Z
,
\label{eq:defetaAH}
\eeq
where
\beq
G &\equiv& \mu^2_G = m^2 + \lambda \phi^2.
\eeq
We also have bosonic propagator residue coefficients:
\beq
a_\pm &=& \frac{Z_{\pm} - \xi Z}{Z_{\pm} - Z_{\mp}},
\label{eq:defapmcogenAH}
\\
b_\pm &=& \frac{(2 \xi - \xiprime) \xiprime Z -\xi Z_{\mp} }{
Z_{\pm} - Z_{\mp}},
\label{eq:defbpmcogenAH}
\\
c_\pm &=& \frac{(\xi - \xiprime) e \phi}{Z_{\pm} - Z_{\mp}}.
\label{eq:defcpmcogenAH}
\eeq

The effective potential in terms of bare parameters can be written as
\beq
V_{\rm eff} &=& V^{(0)}_B + \frac{1}{16\pi^2} V^{(1)}_B + 
\frac{1}{(16\pi^2)^2} V^{(2)}_B + 
\ldots 
,
\label{eq:Veffbare}
\eeq
where the subscript $B$ stands for bare.
The tree-level and 1-loop contributions are:
\beq
V_B^{(0)} &=& \mu^{2\eps} 
\Bigl (
\Lambda_B + \frac{1}{2} m^2_B \phi^2_B + \frac{1}{4} \lambda_B \phi^4_B
\Bigr )
,
\\
V_B^{(1)} &=& 
\Bigl [ 
H_B {\bf A}(H_B) 
+ Z_{+,B} {\bf A}(Z_{+,B}) 
+ Z_{-,B} {\bf A}(Z_{-,B}) 
-2 \eta_B {\bf A}(\eta_B)
\nonumber \\ &&
+ (3- 2 \eps) Z_B {\bf A}(Z_{B}) 
\Bigr ]/(4 - 2 \eps),
\phantom{xxx}
\eeq
where $H_B$, $Z_B$, $Z_{\pm,B}$, and $\eta_B$ are 
obtained from eqs.~(\ref{eq:defHAH})-(\ref{eq:defetaAH}) 
by substituting bare parameters everywhere, and $\eps = (4-d)/2$ in $d$ spacetime
dimensions, and $\mu$ is the regularization scale (see Appendix A).
The 2-loop contributions to the effective potential in the bare scheme can also
be obtained from 
eqs.~(\ref{eq:V2formSSsimp})-(\ref{eq:V2formfFGsimp}), yielding:
\beq
V^{(2),B}_{SS} &=& \frac{3 \lambda}{4} \Bigl [
{\bf f}_{SS}(H, H)
+ a_+^2 {\bf f}_{SS}(Z_+,Z_+)
+ a_-^2 {\bf f}_{SS}(Z_-,Z_-)
+ 2 a_+ a_- {\bf f}_{SS}(Z_+, Z_-)
\Bigr ]
\nonumber \\ &&
+ \frac{\lambda}{2}  
\Bigl [a_+ {\bf f}_{SS}(H,Z_+) + a_- {\bf f}_{SS}(H,Z_-) \Bigr ] 
,
\label{eq:V2AHSS}
\\
V^{(2),B}_{SSS} &=& \lambda^2 \phi^2 \,
\bigl [ 3 {\bf f}_{SSS}(H,H,H)
+ a_+^2 {\bf f}_{SSS}(H, Z_+, Z_+)
\nonumber \\ &&
+ a_-^2 {\bf f}_{SSS}(H, Z_-, Z_-)
+ 2 a_+ a_- {\bf f}_{SSS}(H, Z_+, Z_-)
\bigr ] 
,
\\
V^{(2),B}_{VS} &=& \frac{e^2}{2}  
\Bigl [{\bf f}_{VS}(Z,H) 
+ b_+ {\bf f}_{\overline{V}S}(Z_+,H)  
+ b_- {\bf f}_{\overline{V}S}(Z_-,H)
\nonumber \\ && 
+ a_+ {\bf f}_{VS}(Z, Z_+) 
+ a_- {\bf f}_{VS}(Z, Z_-)
+ a_+ b_+ {\bf f}_{\overline{V}S}(Z_+, Z_+)
\nonumber \\ &&
+ a_- b_- {\bf f}_{\overline{V}S}(Z_-, Z_-)
+ a_- b_+ {\bf f}_{\overline{V}S}(Z_+, Z_-)
+ a_+ b_- {\bf f}_{\overline{V}S}(Z_-, Z_+)
\Bigr ]
,
\\
V^{(2),B}_{SSV} &=& \frac{e^2}{2} \Bigl [ 
  a_+ {\bf f}_{SSV}(H, Z_+, Z)
+ a_- {\bf f}_{SSV}(H, Z_-, Z)
\nonumber \\ &&
+ a_+ b_+ {\bf f}_{SS\overline{V}}(H, Z_+, Z_+)
+ a_- b_- {\bf f}_{SS\overline{V}}(H, Z_-, Z_-)
\nonumber \\ &&
+ a_+ b_- {\bf f}_{SS\overline{V}}(H, Z_+, Z_-)
+ a_- b_+ {\bf f}_{SS\overline{V}}(H, Z_-, Z_+)
\Bigr ]
,
\\
V^{(2),B}_{VVS} &=& e^4 \phi^2 \Bigl [
{\bf f}_{VVS}(Z, Z, H) 
+ 2 b_+ {\bf f}_{\overline{V}VS}(Z_+, Z, H) 
\nonumber \\ &&
+ 2 b_- {\bf f}_{\overline{V}VS}(Z_-, Z, H) 
+ b_+^2 {\bf f}_{\overline{V}\overline{V}S}(Z_+, Z_+, H)
\nonumber \\ &&
+ b_-^2 {\bf f}_{\overline{V}\overline{V}S}(Z_-, Z_-, H)
+ 2 b_+ b_- {\bf f}_{\overline{V}\overline{V}S}(Z_+, Z_-, H)
\Bigr ]
,
\\
V^{(2),B}_{SSG} &=& -2 \lambda e \phi \Bigl [
  a_+ c_+ {\bf f}_{SSG}(H,Z_+,Z_+)
+ a_- c_- {\bf f}_{SSG}(H,Z_-,Z_-)
\nonumber \\ &&
+ a_+ c_- {\bf f}_{SSG}(H,Z_+,Z_-)
+ a_- c_+ {\bf f}_{SSG}(H,Z_-,Z_+)
\Bigr ] 
,
\\
V^{(2),B}_{GGS} &=& 2 \lambda e^2 \phi^2 \Bigl [
  c_+^2 {\bf f}_{GGS} (Z_+, Z_+, H)
+ c_-^2 {\bf f}_{GGS} (Z_-, Z_-, H)
\nonumber \\ &&
+ 2 c_+ c_- {\bf f}_{GGS} (Z_+, Z_-, H)
\Bigr ]
,
\\
V^{(2),B}_{SGG} &=& \frac{1}{2} e^2 \Bigl [
  c_+^2 {\bf f}_{SGG} (H, Z_+, Z_+)
+ c_-^2 {\bf f}_{SGG} (H, Z_-, Z_-)
\nonumber \\ &&
+ 2 c_+ c_- {\bf f}_{SGG} (H, Z_+, Z_-)
\Bigr ]
,
\\
V^{(2),B}_{GSV} &=& -2 e^3 \phi \Bigl [
  c_+ {\bf f}_{GSV} (Z_+, H, Z)
+ c_- {\bf f}_{GSV} (Z_-, H, Z)
\nonumber \\ &&
+ c_+ b_+ {\bf f}_{GS\overline{V}} (Z_+, H, Z_+)
+ c_- b_- {\bf f}_{GS\overline{V}} (Z_-, H, Z_-)
\nonumber \\ &&
+ c_+ b_- {\bf f}_{GS\overline{V}} (Z_+, H, Z_-)
+ c_- b_+ {\bf f}_{GS\overline{V}} (Z_-, H, Z_+)
\Bigr ]
,
\\
V^{(2),B}_{\eta\eta S} &=& 
\frac{1}{2} \xiprimesq e^4 \phi^2 {\bf f}_{\eta\eta S}(\eta, \eta, H)
,
\label{eq:V2AHggS}
\eeq
where the model-independent integral functions were
given above in subsection \ref{subsec:barefunctions}.
There is no need to distinguish between bare and 
renormalized parameters 
in the 2-loop part,
because the difference is of higher order in the loop expansion. 

Now we can derive the \MSbar version of $V_{\rm eff}$, using an alternative 
but equivalent method
to that described above in the general case. To do so, consider the
relationships between bare and \MSbar parameters:
\beq
m_B^2 &=& m^2 + \frac{1}{16\pi^2} \frac{c^{m^2}_{1,1}}{\eps} + 
+ \frac{1}{(16\pi^2)^2} \Bigl[
\frac{c^{m^2}_{2,2}}{\eps^2} + \frac{c^{m^2}_{2,1}}{\eps} \Bigr ]
+ \ldots ,
\\
\lambda_B &=& \mu^{2\eps} \Bigl (
\lambda + \frac{1}{16\pi^2} \frac{c^{\lambda}_{1,1}}{\eps} 
+ \frac{1}{(16\pi^2)^2} \Bigl[
\frac{c^\lambda_{2,2}}{\eps^2} + \frac{c^\lambda_{2,1}}{\eps} \Bigr ]
+ \ldots \Bigr ) ,
\\
\Lambda_B &=& \mu^{-2 \eps} \Bigl (\Lambda + 
\frac{1}{16\pi^2} \frac{c^{\Lambda}_{1,1}}{\eps} 
+ \frac{1}{(16\pi^2)^2} \Bigl[
\frac{c^\Lambda_{2,2}}{\eps^2} + \frac{c^\Lambda_{2,1}}{\eps} \Bigr ]
+ \ldots \Bigr ),
\\
e_B &=& \mu^{\eps} \Bigl (e + 
\frac{1}{16\pi^2} \frac{c^{e}_{1,1}}{\eps} 
+ \ldots \Bigr ),
\\
\xi_B &=& \xi + 
\frac{1}{16\pi^2} \frac{c^{\xi}_{1,1}}{\eps} 
+ \ldots ,
\\
\xiprime_B &=& \xiprime + 
\frac{1}{16\pi^2} \frac{c^{\xiprime}_{1,1}}{\eps} 
+ \ldots ,
\\
\phi_B^2 &=& \mu^{-2\eps} \phi^2 \Bigl (
1 + \frac{1}{16\pi^2} \frac{c^\phi_{1,1}}{\eps}
+ \frac{1}{(16\pi^2)^2} \Bigl[
\frac{c^\phi_{2,2}}{\eps^2} + \frac{c^\phi_{2,1}}{\eps} \Bigr ]
+ \ldots \Bigr ) ,
\eeq
with counterterm coefficients:
\beq
c^{m^2}_{1,1} &=& (4 \lambda - 3 e^2) m^2,
\\
c^{m^2}_{2,1} &=& (-10 \lambda^2 + 16 \lambda e^2 + 43 e^4/6) m^2,
\\
c^{m^2}_{2,2} &=& (28 \lambda^2 - 24 \lambda e^2 + 10 e^4) m^2,
\\
c^{\lambda}_{1,1} &=& 10 \lambda^2 - 6 \lambda e^2 + 3 e^4,
\\
c^{\lambda}_{2,1} &=&
-60 \lambda^3 + 28 \lambda^2 e^2 + 79 \lambda e^4/3 - 52 e^6/3,
\\
c^{\lambda}_{2,2} &=&
100 \lambda^3 - 90\lambda^2 e^2 + 47 \lambda e^4 - 8 e^6,
\\
c^{\Lambda}_{1,1} &=& (m^2)^2/2,
\\
c^{\Lambda}_{2,1} &=& 2 e^2  (m^2)^2,
\\
c^{\Lambda}_{2,2} &=& (2 \lambda - 3 e^2/2) (m^2)^2 ,
\\
c^{e}_{1,1} &=& e^3/6,
\eeq
which can be obtained from existing results in the 
literature \cite{MVI,MVII,MVIII}, and
\beq
c^{\xi}_{1,1} &=& -e^2 \xi/3,
\label{eq:cxi11AH}
\\
c^{\xiprime}_{1,1} &=& e^2 \xiprime (\xi - \xiprime - 10/3),
\\
c^\phi_{1,1} &=& e^2 (3 - \xi + 2 \xiprime),
\\
c^\phi_{2,1} &=& -2 \lambda^2 + e^4 \left (-5/3 + \xiprime [1 + \xi] \right ) ,
\\
c^\phi_{2,2} &=& e^4 \left (5 - 3 \xi + \xi^2/2 + \xiprime [3 - \xi + \xiprime] \right) 
.
\label{eq:cphi22AH}
\eeq
We obtained eqs.~(\ref{eq:cxi11AH})-(\ref{eq:cphi22AH}) 
by requiring no $1/\eps$ or $1/\eps^2$ 
poles survive in $V_{\rm eff}$ when written in terms of the \MSbar parameters.
This involves re-expanding $V_{\rm eff}$ from eq.~(\ref{eq:Veffbare}) 
both in $1/16\pi^2$ and in $\eps$ to get the \MSbar version of the expansion:
\beq
V_{\rm eff} &=& V^{(0)} + \frac{1}{16\pi^2} V^{(1)} 
+ \frac{1}{(16\pi^2)^2} V^{(2)} + \ldots .
\eeq
The tree-level and 1-loop contributions in the \MSbar expansion are:
\beq
V^{(0)} &=& 
\Lambda + \frac{1}{2} m^2 \phi^2 + \frac{1}{4} \lambda \phi^4
,
\\
V^{(1)} &=&   f(H) + 3 f_V(Z) + f(Z_+) + f(Z_-)
- 2 f(\eta),
\eeq
where $f(x)$ and $f_V(x)$ were given in eqs.~(\ref{eq:deff}) and (\ref{eq:deffV}) above.
The results obtained
for the 2-loop \MSbar contribution $V^{(2)}$ are just given by 
eqs.~(\ref{eq:V2AHSS})-(\ref{eq:V2AHggS}) with each function 
${\bf f}$ substituted by the corresponding function $f$
from subsection \ref{subsec:MSbarfunctions}. 
Using eqs.~(\ref{eq:deffSSS})-(\ref{eq:deffvVG}) 
and then combining the coefficients of basis functions, one obtains:
\beq
V^{(2)} &=& \sum_j C^{\cal I}_j {\cal I}_j 
+ \sum_{j,k} C^{\cal AA}_{j,k} {\cal A}_j {\cal A}_k
+ \sum_{j} C^{\cal A}_{j} {\cal A}_j  
+ C
\label{eq:V2AHform}
\eeq
where\footnote{The basis integrals $I(0,0,H)$, $I(0,H,Z)$, $I(0,H,Z_+)$, 
and $I(0,H,Z_-)$ appear in individual diagram contributions, but 
cancel completely in the total.} 
\beq
{\cal I} &=& \{ 
I(H,H,H), \>\>
I(H,Z,Z), \>\>
I(H, Z, Z_-), \>\>
I(H, Z, Z_+), \>\>
I(H, Z_-, Z_-), \>\>
\nonumber \\ && 
I(H, Z_-, Z_+),\>\> 
I(H, Z_+, Z_+), \>\>
I(H, \eta, \eta)
\},
\\
{\cal A} &=& \{
A(H), \>\>
A(Z), \>\>
A(Z_+), \>\>
A(Z_-)
\},
\eeq
and the coefficients $C^{\cal I}_j$, $C^{\cal AA}_{j,k}$, $C^{\cal A}_{j}$, and
$C$ are rational functions of the \MSbar parameters of the theory.
Although there is significant simplification in the coefficients after combining 
diagrams, some of them are still somewhat complicated, so the
explicit result for $V^{(2)}$ is relegated to an ancillary electronic file 
{\tt V2AH} distributed with this paper, in a form suitable for evaluation
by computers.

The beta functions of the parameters of the theory 
in the general form of
eq.~(\ref{eq:betagenform}),
at the orders needed to check
renormalization group invariance, are: 
\beq
\beta^{(1)}_{\Lambda} &=& (m^2)^2,
\\
\beta^{(2)}_{\Lambda} &=& 8 e^2 (m^2)^2,
\\
\beta^{(1)}_{m^2} &=& (8 \lambda - 6 e^2) m^2,
\\
\beta^{(2)}_{m^2} &=& (86 e^4/3 + 64 \lambda e^2 - 40 \lambda^2) m^2,
\\
\beta^{(1)}_{\lambda} &=& 6 e^4 - 12 e^2 \lambda + 20 \lambda^2, 
\\
\beta^{(2)}_{\lambda} &=& -208 e^6/3 +  316 e^4 \lambda/3 + 
112 e^2\lambda^2 - 240 \lambda^3,
\\
\beta^{(1)}_{e} &=& e^3/3,
\\
\beta^{(1)}_{\phi} &=& (3 - \xi + 2 \xiprime) e^2 \phi,
\\
\beta^{(2)}_{\phi} &=&  
[e^4 (-10/3 + 2 \xiprime + 2 \xi \xiprime) -4 \lambda^2] \phi,
\\
\beta^{(1)}_{\xi} &=& -2 e^2 \xi/3,
\label{eq:betaxiAH}
\\
\beta^{(1)}_{\xiprime} &=& 2 e^2 (\xi - \xiprime - 10/3) \xiprime.
\label{eq:betaxiprimeAH}
\eeq
These can be obtained from the counterterms provided above.

The background field $R_\xi$ gauge-fixing result is obtained by setting $\xiprime=\xi$,
which simplifies $V^{(2)}$ greatly, resulting in:
\beq
V^{(2)}_{\xiprime = \xi} &=& \Bigl \{
[H Z -H^2/4 - 3 Z^2] I(H,Z,Z) 
+ [(H+Z-G) \eta - \eta^2/2 - \lambda(G,H,Z)/2 ] I(H,Z,Z_+)
\nonumber \\ &&
- [(H-G)^2/4] I(H,Z_+,Z_+) 
+ [(H + \eta - Z)^2/2 - 2 \eta H ] I(H,Z,\eta)
\nonumber \\ &&
+ [(H - Z_+)^2/2] I(H,\eta,Z_+)
+ [\eta H - \eta^2/2 -H^2/4] I(H,\eta,\eta)
- [3 (H-G)^2/4] I(H,H,H) 
\nonumber \\ &&
+ [H/4 - Z/2] A(Z)^2 + [H/4 - \eta/2] A(\eta)^2 + [3 (H-G)/8] A(Z_+)^2
\nonumber \\ &&
+ [3 (H-G)/8] A(H)^2
+ [(H + 2 Z - \eta - G)/2] A(Z) A(Z_+)
+ [(Z_+ - H)/2] A(\eta) A(Z_+)
\nonumber \\ &&
+ [(Z + \eta -H)/2] A(Z) A(\eta)
+ [(3Z - H+G)/2] A(H) A(Z) 
\nonumber \\ &&
+ [(H+2Z-G)/4] A(H) A(Z_+) 
+ [(H-Z-G)/2] A(H) A(\eta)
\nonumber \\ &&
+ Z [G+H+2Z/3] A(Z) 
+ 3 Z^2 A(H)
+ Z^2 A(Z_+)
- Z^2 (2Z+H)
\Bigr \}/\phi^2,
\eeq
where now $Z_- = \eta = \xi Z$ and $Z_+ = \eta + G$. 
This gauge has the nice property that all squared mass arguments are
real and positive as long as $\xi$ is positive with $\xi Z > -G$,
in which case there are no infrared problems for small $G$.
However, as noted above, 
this gauge-fixing condition is not respected by renormalization, as can 
be seen from eqs.~(\ref{eq:betaxiAH}) and (\ref{eq:betaxiprimeAH}), 
which clearly do not preserve $\xi=\xiprime$ if imposed as an initial 
condition. Moreover, if the \MSbar gauge fixing parameters obey 
$\xi=\xiprime$ at some particular choice of renormalization scale, then 
the corresponding bare parameters will not obey this condition.

\subsection{The Standard Model\label{subsec:StandardModel}} 

In this section we obtain the Standard Model results as a special case 
of the results above. The parameters of the theory consist of 
the constant background Higgs scalar field $\phi$, a field-independent 
vacuum energy $\Lambda$, a Higgs scalar squared mass parameter $m^2$, a 
Higgs self-interaction coupling $\lambda$, gauge couplings $g_3, g, g'$, 
the top-quark Yukawa coupling $y_t$, and gauge-fixing parameters 
$\xi_\gamma$, $\xi_Z$, $\xiprime_Z$, $\xi_W$, $\xiprime_W$. The 2-loop 
effective potential does depend on the QED gauge-fixing parameter 
$\xi_\gamma$, but not on the corresponding QCD 
$SU(3)_c$ gauge-fixing parameter $\xi_{QCD}$. There is no 
parameter $\xiprime_\gamma$, because the photon is massless. The Yukawa 
couplings of all fermions other than the top quark are negligible, and neglected.

The field content with $n_G$ generations consists of: 
\beq 
\mbox{Real vectors:}&\>\>\>\>&A, Z, W_R, W_I, 
\\ 
\mbox{Real scalars:}&\>\>\>\>&H, G_0, G_I, G_R, 
\\ 
\mbox{2-component fermions:} &\>\>\>\>&t,\,\bar t,\, 
b,\, \bar b,\, \tau,\, \bar \tau,\, \nu_\tau \>+\> (n_G - 1) \times 
\left ( u,\, \bar u,\, d,\, \bar d,\, e,\, \bar e,\, \nu_e \right 
),\phantom{xxxx} 
\eeq 
and the color octet gluons, which do not pose any 
problems with respect to gauge fixing. The charged $W$ bosons and 
charged Goldstone scalars have been split into real and imaginary parts 
$W_R$, $W_I$ and $G_R$, $G_I$ respectively. We now list all of the 
(non-QCD) interactions of the Standard Model.\footnote{The conventions 
for the couplings used in the present paper differ in certain minus 
signs from those listed in section V.A [eqs.~(5.2), (5.15)-(5.18), 
(5.20), (5.22), (5.23), (5.26), (5.28), and (5.29)] of 
ref.~\cite{Martin:2017lqn}. The two conventions are related by field 
redefinitions, specifically, flipping the signs of $W_R$, $Z$, $A$, and 
$G_0$. The convention chosen here avoids minus signs in 
eqs.~(\ref{eq:FWGSM}) and (\ref{eq:FZGSM}) below. The resulting 
effective potential is of course independent of this convention choice.}

The scalar cubic interactions are
\beq
\lambda^{HHH} &=& 6 \lambda \phi,
\label{eq:SMhhhcoupling}
\\
\lambda^{H G_0 G_0} &=& \lambda^{H G_R G_R} \>=\> \lambda^{H G_I G_I} \>=\> 2 \lambda \phi,
\eeq
and the scalar quartic couplings are
\beq
\lambda^{HHHH} &=& 
\lambda^{G_0 G_0 G_0 G_0} \>=\> 
\lambda^{G_R G_R G_R G_R} \>=\> 
\lambda^{G_I G_I G_I G_I} \>=\> 6 \lambda ,
\\
\lambda^{HH G_0 G_0} &=& 
\lambda^{HH G_R G_R} = 
\lambda^{HH G_I G_I} =
\lambda^{G_0 G_0 G_R G_R} = \lambda^{G_0 G_0 G_I G_I} =
\lambda^{G_R G_R G_I G_I} = 2 \lambda ,\phantom{xxx} 
\eeq
with both of these lists supplemented by all cases  
dictated by the symmetry under interchange of any two scalars.
The Yukawa couplings (neglecting all fermion mass effects other than the top quark) 
are given by
\beq
Y^{H t\bar t} = -Y^{G_R b\bar t} = 
i Y^{G_0 t\bar t} = i Y^{G_I b \bar t} &=& y_t/\sqrt{2},
\eeq
with symmetry under interchange of the fermion (last two) indices. 
The electroweak gauge boson interactions with the fermions 
are given by couplings of the type $g^{{\bf a}J}_I$:
\beq
g^{Z f}_f &=& I_f g c_W - Y_f g^{\prime} s_W
,
\\
g^{Z \bar f}_{\bar f} &=& Q_f g^{\prime} s_W
,
\\
g^{A f}_f &=& -g^{A \bar f}_{\bar f} = Q_f e
,
\eeq
where 
\beq
e  &=& g g'/\sqrt{g^2 + g^{\prime 2}}
,
\\
s_W &=& g'/\sqrt{g^2 + g^{\prime 2}}
,
\\
c_W &=& g/\sqrt{g^2 + g^{\prime 2}} ,
\eeq
and
$Q_u = 2/3$ and $Q_d = -1/3$ and $Q_{\nu} = 0$ and $Q_e = -1$, and
$I_u = I_{\nu} = 1/2$ and $I_d = I_e = -1/2$, and $Y_f = Q_f - I_f$ for each $f$,
and
\beq
g^{W_R u}_d = g^{W_R d}_u = g^{W_R \nu}_e = g^{W_R e}_\nu = g/2,
\\
g^{W_I u}_d = -g^{W_I d}_u = g^{W_I \nu}_e = -g^{W_I e}_\nu = ig/2.
\eeq
The non-zero vector-scalar-scalar interaction couplings 
of the type $g^{\bf a}_{\bf jk}$ are
\beq
g^A_{G_I G_R } &=& e,
\\
g^Z_{G_0 H} &=& \sqrt{g^2 + g^{\prime 2}}/2,
\\
g^Z_{G_I G_R } &=& (g^2 - g^{\prime 2})/(2 \sqrt{g^2 + g^{\prime 2}}),
\\
g^{W_R}_{G_R G_0} &=& g^{W_R}_{G_I H} \>=\> 
g^{W_I}_{G_0 G_I} \>=\> g^{W_I}_{G_R H} \>=\>  g/2,
\label{eq:SMVSScouplings}
\eeq
with antisymmetry under interchange of the scalar (lowered) indices. 
The vector-vector-scalar-scalar interactions are
determined in terms of these 
[see eq.~(\ref{eq:Lintgeneral}) and fig.~\ref{fig:geninteractions}], 
and so there is no need to list them separately.
The non-zero vector-vector-scalar couplings of the type $G^{\bf ab}_{\bf j}$ 
follow from eqs.~(\ref{eq:defGabj})-(\ref{eq:defGABj}), 
and are given by:
\beq
G^{A W_R}_{G_R} \>=\> -G^{A W_I}_{G_I} &=& 
g e \phi/2,
\\
G^{Z W_I}_{G_I} \>=\>  -G^{Z W_R}_{G_R} &=& 
g' e \phi/2,
\\
G^{W_R W_R}_{H} \>=\> G^{W_I W_I}_{H} &=& g^2 \phi/2,
\\
G^{ZZ}_{H} &=& (g^2 + g^{\prime 2}) \phi/2 ,
\eeq
and others determined by symmetry under interchanging the vector (raised) indices.
Finally there are the totally anti-symmetric vector-vector-vector couplings defined by:
\beq
g^{A W_R W_I} &=& e,
\\
g^{Z W_R W_I} &=& g^2/\sqrt{g^2 + g^{\prime 2}}.
\eeq

The matrix $F^{\bf a}{}_{\bf j}$ of gauge boson masses, 
using the ordered bases $(W_R, W_I, Z, A)$ and $(G_I, G_R, G_0, H)$,
is diagonal, and positive in the convention chosen here when $\phi$ is positive, 
with non-zero entries:
\beq
F^{W_R}{}_{G_I} = F^{W_I}{}_{G_R} &=& M_W = g \phi/2,
\label{eq:FWGSM}
\\
F^{Z}{}_{G_0} &=& M_Z = \sqrt{g^2 + g^{\prime 2}} \phi/2.
\label{eq:FZGSM}
\eeq
The gauge-fixing part of the Lagrangian is:
\beq
{\cal L} &=& -\frac{1}{2\xi_\gamma} (\partial_\mu A^\mu)^2 
- \frac{1}{2\xi_Z} (\partial_\mu Z^\mu - \xiprime_Z M_Z G_0)^2
\nonumber \\ &&
- \frac{1}{2\xi_W} (\partial_\mu W_R^\mu - \xiprime_W M_W G_I)^2
- \frac{1}{2\xi_W} (\partial_\mu W_I^\mu - \xiprime_W M_W G_R)^2
.
\eeq

As an aside, we note that our choice of basis for the gauge-fixing terms 
differs from the choice made in ref.~\cite{DiLuzio:2014bua}, in which the neutral
bosons have a gauge fixing Lagrangian that is instead equivalent to the form:
\beq
-\frac{1}{2\xi_1} (\partial_\mu B^\mu - \xiprime_1 M_B G_0)^2
-\frac{1}{2\xi_2} (\partial_\mu W_0^\mu - \xiprime_2 M_W G_0)^2
\label{eq:gaugefixingDiLuzioMihaila}
\eeq
where $M_B = g' \phi/2$ and
$B^\mu = c_W A^\mu - s_W Z^\mu$ and $W_0^\mu = c_W Z^\mu + s_W A^\mu$
are the gauge-eigenstate neutral vector fields for 
$U(1)_Y$ and $SU(2)_L$ respectively.
Note that there is no redefinition of gauge-fixing parameters that can make 
this choice equivalent to ours in general, 
because the cross-terms are different; in particular,
eq.~(\ref{eq:gaugefixingDiLuzioMihaila}) implies a mixing between the photon
and the $Z$ boson (unless $\xi_1 = \xi_2$) 
and between the photon and the neutral Goldstone boson (unless 
$\xi_1 \xiprime_2 = - \xi_2 \xiprime_1$).
We prefer our choice of a mass-eigenstate basis for the gauge fixing terms 
because it avoids this tree-level gauge-dependent mixing of the photon. 
This inequivalence illustrates the general remark 
made just before eq.~(\ref{eq:defFaj}) 
above, concerning the fact that
the form of the gauge-fixing terms depends on the choice of basis. (The equivalence
could be restored if the gauge fixing parameter $\xi_{\bf a}$ were generalized to a matrix
$\xi_{\bf ab}$.) 

The squared mass poles associated with the electroweak bosons and their ghosts 
are at 0 and
\beq
H &=& m^2 + 3 \lambda \phi^2,
\\
Z &=& (g^2 + g^{\prime 2}) \phi^2/4,
\\
Z_\pm &=& \xiprime_Z Z + \frac{1}{2} \left [ G \pm \sqrt{G [G + 4 (\xiprime_Z-\xi_Z) Z]} \right ]
,
\\
\eta_Z &=& \xiprime_Z Z,
\\
W &=& g^2 \phi^2/4,
\\
W_\pm &=& \xiprime_W W + \frac{1}{2} \left [ G \pm \sqrt{G [G + 4 (\xiprime_W-\xi_W) W]} \right ]
,
\\
\eta_W &=& \xiprime_W W,
\eeq
where
\beq
G = m^2 + \lambda \phi^2,
\eeq
which coincides the Landau gauge version of the common Goldstone squared mass.
The only other non-zero squared mass is that of the top quark,
\beq
T = y_t^2 \phi^2/2.
\eeq

Because there is no mixing among the Goldstone bosons or between them and $H$, 
the results of subsection 
\ref{subsec:nomixing} apply.
Using those results, and combining coefficients of basis functions,
the tree-level and one-loop results for 
the Standard Model in the \MSbar scheme 
are
\beq
V^{(0)} &=& \Lambda + m^2 \phi^2/2 + \lambda \phi^4/4,
\\
V^{(1)} &=& 
f(H) - 12 f(T) + 6 f_V(W) + 2 f(W_+) + 2 f(W_-) - 4 f(\eta_W) 
\nonumber \\ &&
+ 3 f_V(Z) + f(Z_+) + f(Z_-) - 2 f(\eta_Z),
\eeq
and, using eqs.~(\ref{eq:deffSSS})-(\ref{eq:fFFvxy0}), the two-loop part
$V^{(2)}$ can be written in the same form as
eq.~(\ref{eq:V2AHform}), but now with:
\beq
{\cal A} &=& \{
A(H), \>
A(t), \>
A(W), \>
A(W_+), \>
A(W_-), \>
A(\eta_W), \>
A(Z), \>
A(Z_+), \>
A(Z_-), \>
A(\eta_Z)
\},\phantom{xx}
\eeq
and\footnote{The following basis integrals appear in individual diagram contributions, 
but cancel completely from the total: 
$
I(0, 0, H), \>\>
I(0, 0, T), \>\>
I(0, 0, \eta_W), \>\> 
I(0, 0, W_-), \>\>
I(0, 0, W_+), \>\>
I(0, H, W), \>\>
I(0, H, W_-), \>\>
I(0, H, W_+), \>\>
I(0, H, Z), 
\\
I(0, H, Z_-), \>\>
I(0, H, Z_+), \>\>
I(0, W, Z), \>\>
I(0, W, Z_-), \>\> 
I(0, W, Z_+), \>\>
I(0, W_-, Z), \>\>
I(0, W_-, Z_-), \>\>
I(0, W_-, Z_+), \>\>
\\
I(0, W_+, Z), \>\>
I(0, W_+, Z_-), \>\>
I(0, W_+, Z_+), \>\>
I(0, \eta_W, \eta_Z), \>\>
I(W_-, W_-, Z_-), \>\>
I(W_-, W_-, Z_+), \>\>
I(W_+, W_+, Z_-), \>\>
\\
I(W_+, W_+, Z_+)
$.}
\beq
{\cal I} &=& \{
I(H, T, T), \>\>
I(T, T, Z), \>\>
I(T, T, Z_-), \>\>
I(T, T, Z_+), \>\>
I(0, T, W), \>\>
I(0, T, W_-), \>\>
\nonumber \\ &&
I(0, T, W_+), \>\>
I(0, 0, W), \>\>
I(0, 0, Z), \>\>
I(0, W, W_-), \>\>
I(0, W, W_+), \>\>
I(0, W_-, W_+), \>\>
\nonumber \\ &&
I(0, \eta_W, W),\>\> 
I(0, \eta_W, W_-),\>\> 
I(0, \eta_W, W_+), \>\>
I(H, H, H), \>\>
I(H, W, W), \>\>
\nonumber \\ &&
I(H, W, W_-), \>\>
I(H, W, W_+), \>\>
I(H, W_-, W_-), \>\>
I(H, W_-, W_+), \>\>
I(H, W_+, W_+), \>\>
\nonumber \\ &&
I(H, \eta_W, \eta_W), \>\> 
I(H, Z, Z), \>\>
I(H, Z, Z_-), \>\>
I(H, Z, Z_+), \>\>
I(H, Z_-, Z_-), \>\>
I(H, Z_-, Z_+), \>\>
\nonumber \\ &&
I(H, Z_+, Z_+), \>\>
I(H, \eta_Z, \eta_Z), \>\>
I(W, W, Z), \>\>
I(W, W_-, Z), \>\>
I(W, W_-, Z_-), \>\>
I(W, W_-, Z_+), \>\>
\nonumber \\ &&
I(W, W_+, Z), \>\>
I(W, W_+, Z_-), \>\>
I(W, W_+, Z_+), \>\>
I(W_-, W_-, Z), \>\>
I(W_-, W_+, Z), \>\>
\nonumber \\ &&
I(W_-, W_+, Z_-), \>\>
I(W_-, W_+, Z_+), \>\>
I(W_+, W_+, Z), \>\>
I(\eta_W, \eta_Z, W), \>\> 
I(\eta_W, \eta_W, Z), \>\> 
\nonumber \\ &&
I(\eta_W, \eta_W, Z_-), \>\>
I(\eta_W, \eta_W, Z_+), \>\>
I(\eta_W, \eta_Z, W_-), \>\>
I(\eta_W, \eta_Z, W_+)
\}
.
\eeq
The coefficients in this result for $V^{(2)}$ are rather complicated,
so they are again relegated to an electronic ancillary file 
{\tt V2SM} distributed with this paper, in a form suitable for evaluation
by computers. For convenience, we also include separate files 
{\tt V2SMFermi} and {\tt V2SMbackgroundRxi} and {\tt V2SMLandau}
for the specializations to Fermi gauges (with $\xiprime_Z = \xiprime_W = 0$) and to 
background field $R_\xi$ gauges (with $\xiprime_Z = \xi_Z$ and $\xiprime_W = \xi_W$)
and Landau gauge (with $\xi_A = \xiprime_Z = \xi_Z = \xiprime_W = \xi_W = 0$),
respectively.

The check of 
renormalization group invariance of the effective potential can now be carried 
out as in eq.~(\ref{eq:RGcheck}), with the beta functions:
\beq
\beta_\Lambda^{(1)} &=& 2 (m^2)^2
,
\label{eq:betaLambda1SM}
\\
\beta_\Lambda^{(2)} &=& (12 g^2 + 4 g^{\prime 2} - 12 y_t^2)(m^2)^2
,
\label{eq:betaLambda2SM}
\\
\beta_{m^2}^{(1)} &=& m^2 ( 6 y_t^2 + 12 \lambda - 9 g^2/2 - 3 g^{\prime 2}/2)
,
\label{eq:betam21SM}
\\
\beta_{m^2}^{(2)} &=& m^2 (
40 g_3^2 y_t^2 
- 27 y_t^4/2
+ 45 y_t^2 g^2/4 
+ 85 y_t^2 g^{\prime 2} /12 
- 72 y_t^2 \lambda 
+ (5 n_G -385/16) g^4 
\nonumber \\ &&
+ 15 g^2 g^{\prime 2}/8 
+ (25 n_G/9 + 157/48) g^{\prime 4} 
+ 72 g^2 \lambda 
+ 24 g^{\prime 2} \lambda 
- 60 \lambda^2 
)
,
\label{eq:betam22SM}
\\
\beta_{\lambda}^{(1)} &=&  
-6 y_t^4 + 12 y_t^2 \lambda 
+ 9 g^4/8 + 3 g^2 g^{\prime 2}/4 + 3 g^{\prime 4}/8
- 9 g^2 \lambda - 3 g^{\prime 2}\lambda
+ 24 \lambda^2 
,
\label{eq:betak1SM}
\\
\beta_{\lambda}^{(2)} &=& 
- 32 g_3^2 y_t^4 
+ 80 g_3^2 y_t^2 \lambda 
+ 30 y_t^6
- 8 y_t^4 g^{\prime 2}/3 
- 3 y_t^4 \lambda   
- 9 y_t^2 g^4 /4 
+ 21 y_t^2 g^2 g^{\prime 2} /2 
\nonumber \\ &&
- 19 y_t^2 g^{\prime 4} /4 
+ 45 y_t^2 g^2 \lambda /2 
+ 85 y_t^2 g^{\prime 2} \lambda/6 
- 144 y_t^2  \lambda^2 
+ (497/16 - 4 n_G) g^6
\nonumber \\ &&
-(97/48 + 4 n_G/3) g^4 g^{\prime 2} 
-(239/48 + 20 n_G/9) g^2 g^{\prime 4} 
-(59/48 + 20 n_G/9) g^{\prime 6} 
\nonumber \\ &&
+ (10 n_G -313/8) g^4 \lambda 
+ 39 g^2 g^{\prime 2} \lambda/4 
+ (229/24 + 50 n_G/9) g^{\prime 4} \lambda 
+ 108 g^2 \lambda^2 
\nonumber \\ &&
+ 36 g^{\prime 2} \lambda^2 
- 312 \lambda^3 
,
\label{eq:betak2SM} 
\\
\beta_{g'}^{(1)} &=&  (20 n_G/9 + 1/6) g^{\prime 3}
,
\label{eq:betagpSM}
\\
\beta_{g}^{(1)} &=&  (4 n_G/3 - 43/6) g^3
,
\label{eq:betagSM}
\\
\beta^{(1)}_\phi &=&
\bigl [-3 y_t^2 + 9 g^2/4 + 3 g^{\prime 2}/4
+ (\xiprime_W - \xi_W/2) g^2 + (\xiprime_Z/2 - \xi_Z/4) (g^2 + g^{\prime 2})
\bigr ] \phi,
\label{eq:betaphi1SM} 
\\
\beta^{(2)}_\phi &=& \bigl \{
y_t^2 [27 y_t^2/4 -20 g_3^2 - 45 g^2/8 - 85 g^{\prime 2}/24]
- 6 \lambda^2
\nonumber \\ &&
+ (511/32 - 5 n_G/2) g^4 
- 9 g^2 g^{\prime 2}/16 - (25 n_G/18 + 31/96) g^{\prime 4}
\nonumber \\ &&
- 3 y_t^2 [ \xiprime_W g^2 + \xiprime_Z (g^2 + g^{\prime 2}) /2]
+ \xi_W [(\xiprime_W/4 - 2 - \xi_W/8 - \xi_Z/4 ) g^4 
+ \xiprime_Z g^2 g^{\prime 2} /4]  
\nonumber \\ &&
+ \xi_\gamma (\xiprime_W - 3 ) e^2 g^2/4
+ \xi_Z [ 
-g^4 c_W^2 - e^2 g^2/4 
+\xiprime_W e^2 g^{\prime 2}/4 
+\xiprime_Z (g^2 + g^{\prime 2})^2/8]
\nonumber \\ &&
+ \xiprime_W (17 g^4 + g^2 g^{\prime 2})/4
+ \xiprime_Z (17 g^4 + 4 g^2 g^{\prime 2} + g^{\prime 4})/8
\bigr \} \phi
,
\label{eq:betaphi2SM}
\\
\beta_{\xi_W}^{(1)} &=& \xi_W g^2 \left [25/3 - 8 n_G/3  - \xi_W 
- c_W^2 \xi_Z - s_W^2 \xi_\gamma 
\right ]
,
\label{eq:xiWSM}
\\
\beta_{\xiprime_W}^{(1)} &=& \xiprime_W \bigl [
6 y_t^2 + (41/6 - 8 n_G/3) g^2 - 3 g^{\prime 2}/2
+ \xi_\gamma e^2/2 
+  (\xi_W/2 - \xiprime_W) g^2 
\nonumber \\ &&
+ \xi_Z g^{\prime 2} s_W^2/2
- \xiprime_Z (g^2 + g^{\prime 2})/2
\bigr ]
,
\label{eq:xiprimeWSM}
\\
\beta_{\xi_Z}^{(1)} &=& \xi_Z \left [
g^2 c_W^2 (25/3 - 8 n_G/3 - 2 \xi_W) - g^{\prime 2} s_W^2 (1/3 + 40 n_G/9)  
\right ]
,
\label{eq:xiZSM}
\\
\beta_{\xiprime_Z}^{(1)} &=& \xiprime_Z \bigl [
6 y_t^2  
+ (41/6 - 8 n_G/3) g^2 c_W^2 - (11/6 + 40 n_G/9) g^{\prime 2} s_W^2 - 6 e^2
+ \xi_W e^2 
\nonumber \\ &&
- \xiprime_W g^2
+ (\xi_Z - \xiprime_Z) (g^2 + g^{\prime 2})/2  
\bigr ]
\label{eq:xiprimeZSM}
\\
\beta_{\xi_\gamma}^{(1)} &=& \xi_\gamma e^2 \left [ 
8 - 64 n_G/9 - 2 \xi_W
\right ]
.
\label{eq:xiASM}
\eeq
Equations (\ref{eq:betaLambda1SM}) and (\ref{eq:betaLambda2SM})
were obtained in ref.~\cite{Martin:2017lqn}, and 
eqs.~(\ref{eq:betam21SM})-(\ref{eq:betagSM}) and the parts of
eqs.~(\ref{eq:betaphi1SM}) and (\ref{eq:betaphi2SM}) 
that do not depend on the gauge-fixing parameters
$\xi_W$, $\xi_Z$, $\xiprime_W$, $\xiprime_Z$ can be found in the literature, 
for example in 
refs.~\cite{MVI}-\cite{MVIII}. The results dependent on the gauge-fixing parameters in
eqs.~(\ref{eq:betaphi1SM})-(\ref{eq:xiprimeZSM}) were obtained 
here by requiring that $V_{\rm eff}$
satisfies renormalization group invariance.
Again we note that any equality among any subset of the parameters
$\xi_W$, $\xi_Z$, $\xiprime_W$, $\xiprime_Z$, and $\xi_\gamma$ 
will not be preserved under renormalization group evolution, except in the special case that the 
corresponding parameters vanish.
Also, if the \MSbar gauge fixing parameters 
obey $\xi_W=\xiprime_W$ and/or $\xi_Z=\xiprime_Z$ at some particular choice
of renormalization scale, then the corresponding bare parameters will not 
obey these conditions, and vice versa.

\section{Numerical results for the Standard Model\label{sec:numerical}}
\setcounter{equation}{0}
\setcounter{figure}{0}
\setcounter{table}{0}
\setcounter{footnote}{1}

Consider the Standard Model with the following input parameters as a benchmark 
(the same as in refs.~\cite{Martin:2015lxa,Martin:2015rea,Martin:2016xsp,Martin:2015eia,Martin:2017lqn}, but with various other approximations 
for the effective potential):
\beq
Q &=& M_t = 173.34\>\mbox{GeV},
\label{eq:QSMsample}
\\
y_t(Q) &=& 0.93690,
\\
g_3(Q) &=& 1.1666,
\\
g(Q) &=& 0.647550,
\\
g'(Q) &=& 0.358521,
\\
\lambda(Q) &=& 0.12597,
\\
m^2(Q) &=& -(92.890\>\mbox{GeV})^2
\\
\Lambda(Q) &=& 0.
\label{eq:LambdaSMsample}
\eeq
Then, in the Landau gauge, 
the minimum of the (real part of the) 
2-loop effective potential is at
\beq
v_0 \equiv \phi_{\rm min}^{(\xi = 0)} &=& 246.950\>\mbox{GeV}
,
\\
V_{\rm eff}^{(\xi = 0)} (v_0)
&=& -(105.560\>\mbox{GeV})^4 .
\eeq
With this choice of input parameters, the Landau gauge Goldstone boson
\MSbar squared mass is $G = -(\mbox{30.763 GeV})^2$, so that $V_{\rm eff}$ is actually
complex at its minimum. For simplicity we do not apply the Goldstone boson resummation
procedure \cite{Martin:2014bca,Elias-Miro:2014pca} 
to eliminate the spurious imaginary part here. Instead, we simply
minimize the real part of $V_{\rm eff}$,
and it should be understood below that the spurious imaginary part is always dropped.
As shown in ref.~\cite{Martin:2014bca},
the practical numerical difference between the VEV obtained by minimizing the
real part of the non-resummed effective potential and the VEV obtained by minimizing the
Goldstone boson-resummed effective potential, which is always real, is very small.

\begin{figure}[!pt]
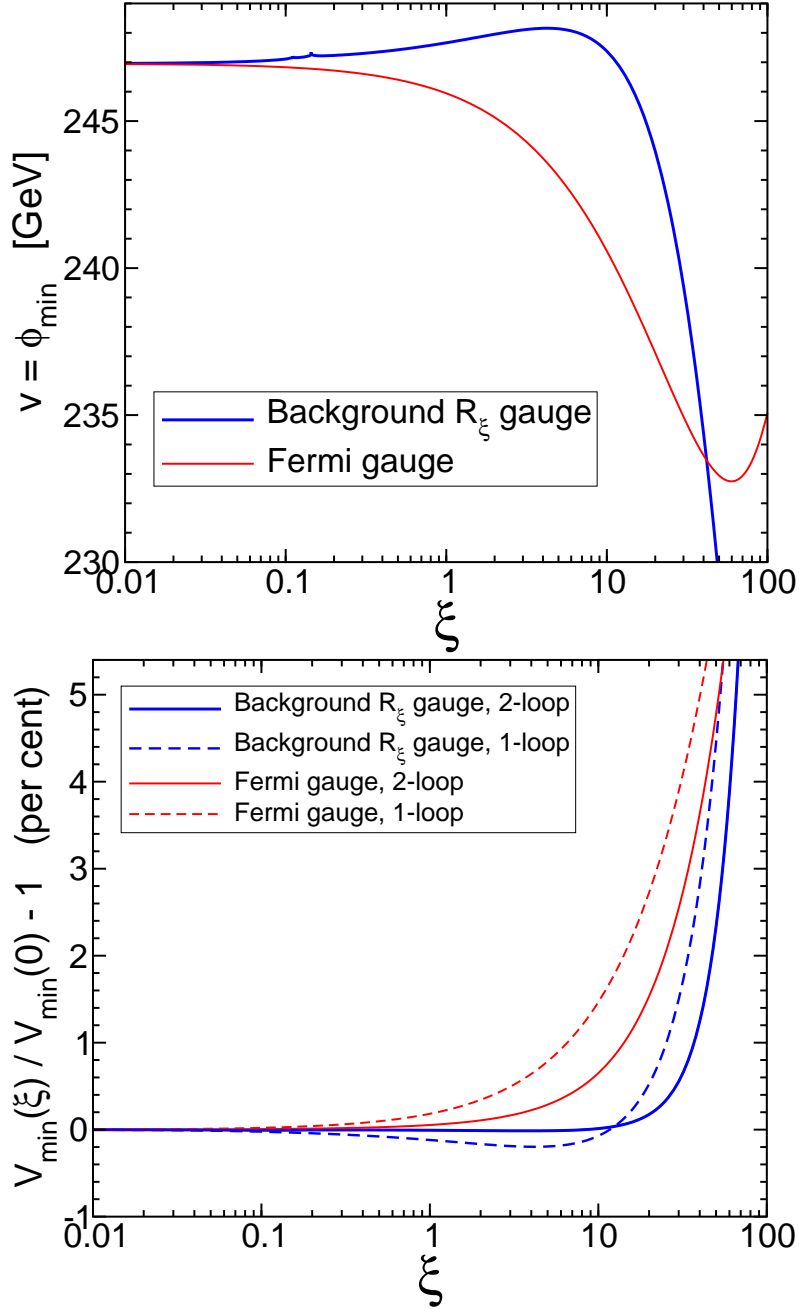

\begin{minipage}[]{0.635\linewidth}
\epsfxsize=\linewidth
\epsffile{vev_xi.eps}
\end{minipage}
\begin{minipage}[]{0.635\linewidth}
\epsfxsize=\linewidth
\epsffile{Vmin_xi.eps}
\end{minipage}
\caption{\label{fig:vevxi}
The Higgs VEV (top panel), and the resulting 
fractional change in the vacuum energy compared to the Landau gauge $\xi=0$ 
result, in per cent (bottom panel), at the minimum of the 
2-loop Standard Model effective potential, as a function of the
gauge-fixing parameter $\xi$. The solid blue (thicker) curves show the result
for the background field $R_\xi$ gauges (with 
$\xi \equiv \xi_W = \xiprime_W = \xi_Z = \xiprime_Z = \xi_\gamma$),
and the solid red (thinner) curves are the results for the Fermi gauges (with
$\xi \equiv \xi_W = \xi_Z = \xi_\gamma$ and $\xiprime_W = \xiprime_Z = 0$.)
The other input parameters are as given in 
eqs.~(\ref{eq:QSMsample})-(\ref{eq:LambdaSMsample}) of the text.
In the top panel, very small 
cusps are barely visible in the background field $R_\xi$ gauge $v$ 
curve at the points 
$\xi = 0.11112$ and $0.14388$ below which 
$M^2_{Z,+}$ and $M^2_{W,+}$, respectively, are negative. 
In the bottom panel, we also show for comparison the 
results from the 1-loop approximations,
as dashed lines.
The dependence of the VEV on $\xi$ is expected, but in principle
the minimum value of the vacuum energy is an observable and 
should be independent of $\xi$.
The significant deviation from this idealized behavior shown 
in the bottom panel is 
due to a breakdown in perturbation theory truncated at 2-loop order  for large $\xi$.}
\end{figure}

In Figure \ref{fig:vevxi}, we show the results for $v = \phi_{\rm min}$ 
and $V_{\rm eff}(v)$
as a function of $\xi$ for the cases:
\beq
\mbox{background field $R_\xi$ gauge}:&& \xi \equiv \xi_W = \xiprime_W = \xi_Z = \xiprime_Z = \xi_\gamma,
\\
\mbox{Fermi gauges}:&&\xi \equiv \xi_W = \xi_Z = \xi_\gamma,\quad\mbox{and}\quad
\xiprime_W = \xiprime_Z = 0.
\eeq
In the background field $R_\xi$ gauge, 
for small $\xi$ one finds that
$M^2_{Z,+}$ and $M^2_{W,+}$ are negative and so
$V_{\rm eff}(v)$
has a spurious imaginary part, but
$M^2_{Z,+}$ becomes positive for $\xi > 0.11112$, and $M^2_{W,+}$ is positive
for $\xi > 0.14388$, so that there is no spurious imaginary part at the minimum
of the two-loop effective potential for $\xi$ larger than this. 
(Very small cusps are visible on the background field $R_\xi$ gauge curve for $v$,
corresponding to the points where $M^2_{Z,+}$ and $M^2_{W,+}$ go through 0.)
In the Fermi gauge, $M^2_{Z,+}$ and $M^2_{W,+}$ are positive but 
$M^2_{Z,-}$ and $M^2_{W,-}$ are negative for all positive $\xi$, so that
the effective potential always has a spurious imaginary part, which again is ignored
in the minimization. 

Although $v$ is a non-trivial function of $\xi$, the minimum vacuum energy 
$V_{\rm eff}(v)$
is a physical observable (for example, by weakly coupling to gravity) 
and in principle should be completely independent of $\xi$ when 
computed to all orders in perturbation theory. 
In the second panel of 
Figure \ref{fig:vevxi}, it can be seen that the latter property indeed holds 
in the background field $R_\xi$ gauge to 
better than 1 part per mille for $\xi \lsim 16$
and to better than 1\% for $\xi \lsim 37$, but
the situation rapidly deteriorates for larger $\xi$.
In Fermi gauge, the deviation is larger, but $V_{\rm eff}(v)$ differs from 
its Landau gauge value by less than 1 part per mille for all
$\xi \lsim 1.88$ and by less than 1\% for $\xi \lsim 14$; 
the deviation again grows rapidly for larger $\xi$.
In the second panel of Figure~\ref{fig:vevxi}
the results from the 1-loop effective potential approximations are also shown,
as dashed lines; the deviations are significantly worse than at 2-loop order.

\begin{figure}[!pt]
\epsfxsize=0.635\linewidth
\epsffile{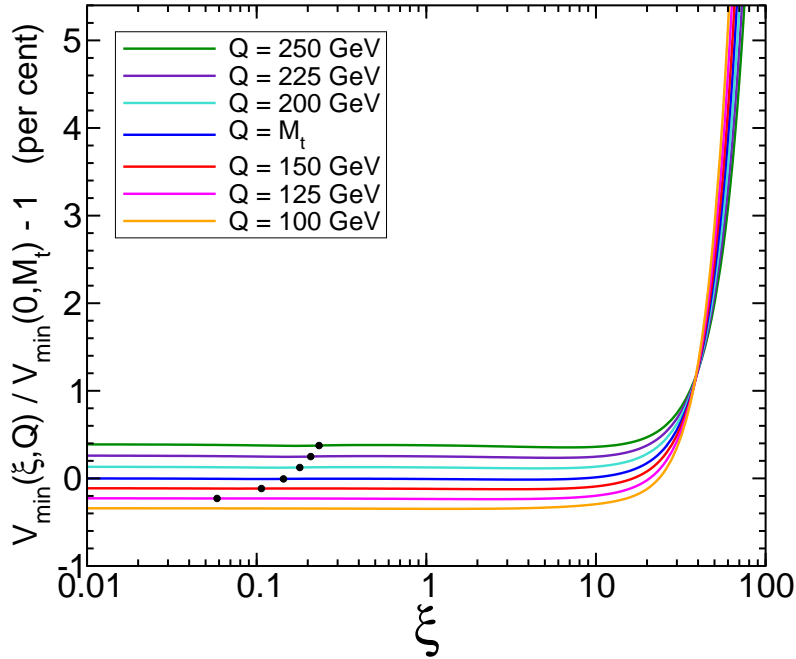}
\caption{\label{fig:vevxiQ}
The fractional change in the vacuum energy in per cent, at the minimum of the 
2-loop Standard Model effective potential, as a function of the
background $R_\xi$ 
gauge-fixing parameter 
$\xi \equiv \xi_W = \xiprime_W = \xi_Z = \xiprime_Z = \xi_\gamma$. 
The comparison value $V_{\rm min}(0,M_t)$ is
the Landau gauge result with the input parameters of 
eqs.~(\ref{eq:QSMsample})-(\ref{eq:LambdaSMsample}) at $Q = M_t$.
To find $V_{\rm min}(\xi,Q)$,
these parameters are then run using their 2-loop renormalization group equations 
to the scale $Q$, the gauge fixing is then imposed with parameter $\xi$, and the
2-loop effective potential is minimized. 
From top to bottom on the left, the
curves are  $Q = 250$ GeV, $225$ GeV, $200$ GeV, $M_t = 173.34$ GeV,
$150$ GeV, $125$ GeV, and $100$ GeV.
On each curve, the black dot is the point to the right of which the effective potential
is real at its minimum; to the left of the dot, it is actually the real part 
of the effective potential that is minimized.}
\end{figure}

In Figure \ref{fig:vevxiQ}, we show results for the background field $R_\xi$
gauge for seven different choices of the renormalization scale $Q$. In each case we show
the deviation of $V_{\rm eff,min}(\xi,Q)$ compared to the benchmark value 
$V_{\rm eff,min}(0,M_t)$ obtained in Landau gauge and with $Q$ set equal to
$M_t = 173.34$ GeV. To make this graph, the parameters in 
eqs.~(\ref{eq:QSMsample})-(\ref{eq:LambdaSMsample}) are first
run\footnote{Background field $R_\xi$ gauge is not respected by
renormalization group running, so we do not run $\xi$. 
Instead, the value of $\xi$ is the one imposed at $Q$. 
Also, note that the running
of $\Lambda$ is crucial for getting the correct $V_{\rm eff,min}(\xi,Q)$.}  
according to their 2-loop renormalization group equations to the scale $Q$,
and then the minimum value of the two-loop effective potential $V_{\rm eff,min}(\xi,Q)$
is obtained.
Since $V_{\rm eff,min}(\xi,Q)$ is a physical observable, it should
in principle be independent of both $\xi$ and $Q$ if calculated to all orders.
We see that for $\xi$ less than of order roughly 30, in the 2-loop approximation the 
dependence  on $\xi$ is much smaller than 
the dependence on the renormalization scale, 
but for larger $\xi$ this is no longer true as perturbation theory breaks down. 

The increasingly strong 
deviation of $V_{\rm eff,min}(\xi)/V_{\rm eff,min}(0)$ from 1 
is evidently due to the
failure of the 2-loop truncation of the perturbative 
expansion for large $\xi$. 
The fact that the $\xi \rightarrow \infty$ 
limit of the effective potential  
is problematic when calculated 
at finite loop order in Fermi gauges has been noted already 
in \cite{Andreassen:2014eha,Espinosa:2016uaw,Irges:2017ztc}.
In ref.~\cite{Andreassen:2014eha}, it was shown how a resummation of a class
of diagrams to all orders in perturbation theory restores the 
gauge-fixing independence within Fermi gauges. The Fermi gauge fixing also has 
IR divergence
problems \cite{Aitchison:1983ns,Loinaz:1997td,Andreassen:2014eha} 
in the limit that the minimum of the tree-level potential coincides
with the minimum of the full effective potential.
Ref.~\cite{Espinosa:2016uaw} showed that the same resummation 
that fixes the IR problems of Fermi gauges also cures the gauge dependence
issue. We expect that a suitable resummation of higher-order 
diagrams will also eliminate the problematic behavior for large $\xi$
in more general gauge-fixing schemes, including the background field $R_\xi$
gauge-fixing scheme illustrated here. However, this is beyond the scope of the
present paper. In any case, it is worth noting that for a range of reasonable values of
$\xi$ (say, $0.25 < \xi<10$) the background field 
$R_\xi$ gauge does not have infrared subtleties or spurious imaginary parts
(which can occur at smaller $\xi$, depending on $Q$) 
and the minimum value does not have a significant dependence on the 
gauge-fixing parameter (which occurs at larger $\xi$).

\section{Outlook\label{sec:Outlook}}
\setcounter{equation}{0}
\setcounter{figure}{0}
\setcounter{table}{0}
\setcounter{footnote}{1}

In this paper, we have obtained the two-loop effective potential for a general renormalizable
theory, using a generalized gauge fixing scheme that includes the Landau gauge, the Fermi gauges,
and the background-field $R_\xi$ gauges as special cases. The essential results are given as 37 
loop integral functions in eqs.~(\ref{eq:deffSSS})-(\ref{eq:defffFG}), with special cases 
arising for vanishing vector boson masses given in eqs.~(\ref{eq:fSSVxy0})-(\ref{eq:fFFvxy0}). 
For convenience, these results are also provided in an ancillary electronic file
called {\tt functions}.

In the most general case, these 37 functions contribute to the two-loop effective potential
as in eqs.~(\ref{eq:V2formSS})-(\ref{eq:V2formfFG}).  
The practical implementation of this result is sometimes complicated by the fact that 
the squared masses appearing as arguments of the loop integral functions can be complex. 
As far as we know, a complete treatment of the two-loop vacuum integral
basis functions $I(x,y,z)$ for complex arguments does not yet exist, and would be a worthwhile
subject of future investigations.
In favorable cases such as the Standard Model or the Abelian Higgs model, 
the absence of Goldstone mixing with other scalars allows a 
significant simplification, as given in eqs.~(\ref{eq:V2formSSsimp})-(\ref{eq:V2formfFGsimp}),
because the squared masses are then always
solutions of quadratic equations. However even in these simplified
cases the squared masses can still be complex, depending on the choice of gauge-fixing parameters.
In the numerical examples of the present paper, we simply avoided choices that could lead to 
complex squared masses.

For softly broken supersymmetric theories the results above will need to be extended. 
This is because the \MSbar scheme based on 
dimensional regularization introduces an explicit violation of supersymmetry. For applications
to the Minimal Supersymmetric Standard Model or its extensions, it will be necessary to
instead use the \DRbarprime scheme based on dimensional reduction, which respects supersymmetry. 
This will require a slightly different calculation than the one here, as has already been done 
\cite{Martin:2001vx} in the Landau gauge special case.
 
In our numerical study of the Standard Model case, we found that fixed-order 
perturbation theory breaks
down for sufficiently large $\xi$ (although moderately large choices $\xi\lsim 10$ 
seem to be fine, and introduce a smaller variation
than does the choice of renormalization scale, at least for the minimum vacuum energy as a test
observable). 
This is not unexpected, and given the results of 
e.g.~refs.~\cite{Andreassen:2014eha,Espinosa:2016uaw}
it seems likely that some appropriate
resummation to all orders in perturbation theory 
of selected higher-order corrections will cure that problem
in the most general cases.
This could also be a worthwhile subject of future work.

However, an alternate point of view, to which we are sympathetic, is that the complications
associated with generalized gauge-fixing schemes 
provide a strong motivation to simply stick to Landau gauge.
This avoids all possibilities of complex squared masses, 
kinetic mixing between Goldstone scalars and massive vector degrees of freedom,
as well as the non-trivial running of the gauge-fixing parameters. By sticking
only to Landau gauge, one does lose the
checks that come from requiring independence of physical observables with respect to varying
gauge-fixing parameters, but there are other powerful checks within Landau gauge 
coming from the cancellations of unphysical Goldstone contributions to physical quantities,
as shown for example in refs.~\cite{Martin:2014cxa}-\cite{Martin:2016xsp}.
From that point of view, the present paper might serve as a pointed
warning about the difficulties to be faced for those who would dare to 
venture outside of Landau gauge.

\section*{Appendix A: Basis integrals\label{appendix:basisintegrals}}
\renewcommand{\theequation}{A.\arabic{equation}}
\setcounter{equation}{0}
\setcounter{figure}{0}
\setcounter{table}{0}
\setcounter{footnote}{1}

In this Appendix, we review the conventions and notations for the 1-loop and 2-loop basis integrals,
which follow refs.~\cite{Martin:2001vx,Martin:2017lqn,Martin:2016bgz}.

Define the Euclidean integral notation in
\beq
d = 4 - 2 \eps
\eeq
dimensions:
\beq
\int_p &\equiv& (16 \pi^2) \frac{\mu^{2\eps}}{(2\pi)^d} \int d^d p .
\eeq
Here $\mu$ is the regularization scale, related to the 
\MSbar renormalization scale $Q$ by
\beq
4 \pi \mu^2 &=& e^{\gamma_E} Q^2
.
\eeq
Then the basis integrals appearing in the two-loop effective potential 
in terms of bare parameters are defined as:
\beq
\AAx  &=& \int_p \frac{1}{p^2 + x} \>=\> 
x (e^{\gamma_E} Q^2/x)^\eps \Gamma(-1+\eps) ,
\\
\IIxyz  &=& \int_p \int_q \frac{1}{[p^2 + x][q^2 + y][(p-q)^2 + z]}.
\eeq
Expanding in small $\eps$, we write:
\beq
\AAx  = -\frac{x}{\eps} + A(x) + \eps A_\eps(x) + \ldots ,
\label{eq:defineAA}
\eeq
with
\beq
A(x) &=& x \lnbar(x) - x ,
\eeq
where
\beq
\lnbar(x) = \ln(x/Q^2),
\eeq
and $A_\eps(x)$ is known, but we won't ever need its explicit form and 
it won't appear in the final expressions for the renormalized effective 
potential. Sometimes the following identities can be useful:
\beq
\frac{d}{dx} A(x) &=& A(x)/x + 1
,
\\
\frac{d}{dx} A_\eps (x) &=& [A_\eps (x) - A(x)]/x
.
\eeq
We also expand:
\beq
\IIxyz  &=& -(x+y+z)/2 \eps^2 + [A(x) + A(y) + A(z) -(x+y+z)/2]/\eps
\nonumber \\ &&
+ I(x,y,z) + A_\eps(x) + A_\eps(y) + A_\eps(z) + {\cal O}(\eps),
\label{eq:defineII}
\eeq
where $I(x,y,z)$ is known in terms of dilogarithms. 
The basis integrals needed for the 2-loop effective potential contribution 
written in terms of \MSbar parameters 
are just the non-bold-faced integrals $A(x)$ and $I(x,y,z)$.
In any 2-loop quantity written 
in terms of \MSbar parameters, all of the $A_\eps$ functions 
always cancel against 1-loop contributions; this is a useful check.

Below, define for convenience:
\beq
D = [p^2 + x][q^2 + y][(p-q)^2 + z].
\eeq
Then a useful integral table is: 
\beq
\int_p \int_q \>\frac{1}{D} &=& \IIxyz ,
\\
\int_p \int_q \>\frac{p^2}{D} &=& \AAy  \AAz  - x \IIxyz ,
\\
\int_p \int_q \> \frac{p \cdot q}{D} &=& 
\frac{1}{2} \left [ (z-x-y) \IIxyz  
- \AAx  \AAy  
+ \AAx  \AAz  
+ \AAy  \AAz  
\right ],
\\
\int_p \int_q \> \frac{(p^2)^2}{D} &=& 
x^2 \IIxyz   
- (x + y + z) \AAy  \AAz  
,
\\
\int_p \int_q \> \frac{p^2 q^2}{D} &=& 
x y \IIxyz   
- x \AAx  \AAz  
- y \AAy  \AAz  
,
\\
\int_p \int_q \> \frac{p^2 (p \cdot q)}{D} &=& 
\frac{1}{2} \bigl [ x (x+y-z) \IIxyz  
+ x \AAx  \AAy  
- x \AAx  \AAz  
\nonumber \\ &&
- (x+ 2 y) \AAy  \AAz  
\bigr ],
\\
\int_p \int_q \> \frac{(p \cdot q)^2}{D} &=& 
\frac{1}{4} \bigl [ (x+y-z)^2 \IIxyz  
+ (x + y - z) \AAx  \AAy  
\nonumber \\ &&
+ (z-3x-y) \AAx  \AAz  
+ (z-x-3y) \AAy  \AAz  
\bigr ],
\\
\int_p \int_q \> \frac{(p^2)^3}{D} &=&
-x^3 \IIxyz  + \left [
x^2 + y^2 + z^2 + x y + x z + (2 + 4/d) y z 
\right ]
\AAy  \AAz 
,
\\
\int_p \int_q \> \frac{(p^2)^2 q^2}{D} &=&
-x^2 y \IIxyz  + x^2 \AAx  \AAz 
+ y (x+y+z) \AAy  \AAz 
,
\\
\int_p \int_q \> \frac{(p^2)^2 (p \cdot q)}{D} &=&
\frac{1}{2} \Bigl [
x^2 (z-x-y) \IIxyz  
- x^2 \AAx  \AAy 
+ x^2 \AAx  \AAz 
\nonumber \\ &&
+ [x^2 + 2 x y + 2 y^2 + (2 + 4/d) y z] \AAy  \AAz 
\Bigr ]
,
\\
\int_p \int_q \> \frac{p^2 q^2 (p \cdot q)}{D} &=&
\frac{1}{2} \Bigl [
x y (z-x-y) \IIxyz  - x y \AAx  \AAy 
+ x (2x+y) \AAx  \AAz  
\nonumber \\ &&
+ y (x + 2 y) \AAy  \AAz 
\Bigr ]
,
\\
\int_p \int_q \> \frac{p^2 (p \cdot q)^2}{D} &=&
\frac{1}{4} \Bigl [
-x (x+y-z)^2 \IIxyz  
+ (x^2 + 3 x y + 4 y^2 - x z + 4 y z/d) \AAy  \AAz 
\nonumber \\ &&
+ x (z-x-y) \AAx  \AAy 
+ x (3x + y-z) \AAx  \AAz 
\Bigr ]
,
\\
\int_p \int_q \> \frac{(p \cdot q)^3}{D} &=&
\frac{1}{8} \Bigl [
(z-x-y)^3 \IIxyz 
\nonumber \\ &&
+ [-x^2-y^2-z^2-(2 + 4/d) x y + 2 x z + 2 y z] \AAx  \AAy 
\nonumber \\ &&
+ [7 x^2 + y^2 + z^2 + 4 x y + (4/d-4) x z - 2 y z] \AAx  \AAz 
\nonumber \\ &&
+ [7 y^2 + x^2 + z^2 + 4 x y + (4/d-4) y z - 2 x z] \AAy  \AAz 
\Bigr ]
,
\eeq
and others obtained by $p \leftrightarrow q$ and $x \leftrightarrow y$.
Other integrals can be obtained from the above by e.g. 
\beq
\frac{1}{p^2 (p^2 + x)} &=& 
\frac{1}{x} \left [ \frac{1}{p^2} - \frac{1}{p^2 + x} \right ]
.
\eeq
We also make use of the notation:
\beq
\lambda(x,y,z) &=& x^2 + y^2 + z^2 - 2 x y - 2 x z - 2 y z.
\eeq

\section*{Appendix B: Derivatives with respect to the 
renormalization scale\label{appendix:QdQ}}
\renewcommand{\theequation}{B.\arabic{equation}}
\setcounter{equation}{0}
\setcounter{figure}{0}
\setcounter{table}{0}
\setcounter{footnote}{1}

In this Appendix, we collect the derivatives of the loop integral functions with respect
to the \MSbar renormalization scale $Q$.
\beq
Q\frac{\partial}{\partial Q} f_{SS}(x,y) &=& -2 y A(x) - 2 x A(y)
,
\\
Q\frac{\partial}{\partial Q} f_{SSS}(x,y,z) &=& 2 [x + y + z - A(x) - A(y) - A(z)]
,
\\
Q\frac{\partial}{\partial Q} f_{VS}(x,y) &=& -4 x y - 6 y A(x) - 6 x A(y)
,
\\
Q\frac{\partial}{\partial Q} f_{\Vbar S}(x,y) &=& -2 y A(x) - 2 x A(y)
,
\\
Q\frac{\partial}{\partial Q} f_{SSV}(x,y,z) &=& 
  -2 x^2 -12 x y -2 y^2 -6 x z -6 y z +10 z^2/3 
  + 6 x A(x) + 6 y A(y) 
  \nonumber \\ &&
   + (6 x + 6 y - 2 z) A(z)
,
\\
Q\frac{\partial}{\partial Q} f_{SS\Vbar}(x,y,z) &=& 
  2[-(x - y)^2 + (y - x + z) A(x) + (x - y + z) A(y)
  \nonumber \\ &&
   + (x + y) A(z)]
,
\\
Q\frac{\partial}{\partial Q} f_{VVS}(x,y,z) &=& 
  9 x/2 + 9 y/2 - z - 9 A(x)/2 - 9 A(y)/2 - 6 A(z)
,
\\
Q\frac{\partial}{\partial Q} f_{\Vbar VS}(x,y,z) &=& [x + 3 y + 6 z - 3 A(x) - 3 A(y)]/2
,
\\
Q\frac{\partial}{\partial Q} f_{\Vbar\Vbar S}(x,y,z) &=& 
  3 x/2 + 3 y/2 - z - A(x)/2 - A(y)/2 - 2 A(z)
,
\\
Q\frac{\partial}{\partial Q} f_{SSG}(x,y,z) &=& 
  2 (y - x) (x + y + z) + 2 (x - y - z) A(x) + 2 (x - y + z) A(y)
,
\\
Q\frac{\partial}{\partial Q} f_{GGS}(x,y,z) &=& 
  (z - x - y) (x + y + z) + x A(x) + y A(y) + (2 x + 2 y - z) A(z)
,
\\
Q\frac{\partial}{\partial Q} f_{SGG}(x,y,z) &=& 
  2 [(x-y) (x-z) (x + y + z) - y z A(y) - y z A(z)  
  \nonumber \\ && 
  + (2 x y + 2 x z -x^2 - y z - y^2 - z^2) A(x) 
  ]
,
\\
Q\frac{\partial}{\partial Q} f_{GSV}(x,y,z) &=& 
  x^2 + 6 x y + y^2 + 3 x z + 3 y z - 5 z^2/3 
  - 3 x A(x) - 3 y A(y) 
  \nonumber \\ && 
  + (z - 3 x - 3 y) A(z)
,
\\
Q\frac{\partial}{\partial Q} f_{GS\Vbar}(x,y,z) &=& 
  (x - y) (2 x + z) + 2 (y - x - z) A(y) - (x + y) A(z)
,
\\
Q\frac{\partial}{\partial Q} f_{GGG}(x,y,z) &=& 
  (x - y) (x + y - z) (x + y + z) + x (y - x + 2 z) A(x) 
  \nonumber \\ && 
  - y (x - y + 2 z) A(y) + (x - y) z A(z)
,
\\
Q\frac{\partial}{\partial Q} f_{GGV}(x,y,z) &=& 
  - x^3 - y^3 - 5 z^3/3 + x^2 y + x^2 z  + x y^2 + y^2 z 
  + 3 x z^2 + 3 y z^2
  \nonumber \\ && 
  + 6 x y z 
  + x (x - 3 y - 3 z) A(x) 
  + y (y - 3 x - 3 z) A(y)
  \nonumber \\ && 
  + z (z - 3 x - 3 y) A(z)
,
\\
Q\frac{\partial}{\partial Q} f_{VGG}(x,y,z) &=& \bigl [
  y^2 + 6 y z + z^2 
  + 3 x y + 3 x z 
  - 5 x^2/3 
  - 3 y A(y) - 3 z A(z)  
  \nonumber \\ && 
+ (x -3 y - 3 z) A(x) \bigr ]/2
,
\\
Q\frac{\partial}{\partial Q} f_{\Vbar GG}(x,y,z) &=& 
  [y^2 - 2 y z + z^2 
  - x y - x z - x^2  
  + y A(y) + z A(z) 
    \nonumber \\ && 
  + (x - y - z) A(x)]/2
,
\\
Q\frac{\partial}{\partial Q} f_{VVG}(x,y,z) &=& 
  [(y - x) (22 x + 22 y + 27 z)/3 + (8 x - 9 y - 9 z) A(x)
  \nonumber \\ && 
   + (9 x - 8 y + 9 z) A(y)]/2
,
\\
Q\frac{\partial}{\partial Q} f_{\Vbar VG}(x,y,z) &=& [
  -x^2 + x y + 3 y^2 + x z + 6 y z + z^2 +
  (x - 3 y - 3 z) A(x)
    \nonumber \\ && 
   - 3 y A(y) - 3 z A(z)]/2
,
\\
Q\frac{\partial}{\partial Q} f_{\eta\eta S}(x,y,z) &=& 2[-x -y - z + A(x) + A(y) + A(z)]
,
\\
Q\frac{\partial}{\partial Q} f_{\eta\eta G}(x,y,z) &=& 
  (x - y + z) (x + y + z) - x A(x) + (y -2 x - 2 z) A(y) - z A(z)
,
\\
Q\frac{\partial}{\partial Q} f_{VV}(x,y) &=& 
  -45 x y/2 - 27 y A(x)/2 - 27 x A(y)/2
,
\\
Q\frac{\partial}{\partial Q} f_{\Vbar V}(x,y) &=& 
  -9 x y/2 - 9 y A(x)/2 - 9 x A(y)/2
,
\\
Q\frac{\partial}{\partial Q} f_{\Vbar\Vbar}(x,y) &=& 
  -x y/2 - 3 y A(x)/2 - 3 x A(y)/2
,
\\
Q\frac{\partial}{\partial Q} f_{VVV}(x,y,z) &=& 
  (-28 x^2 - 243 x y - 28 y^2 - 243 x z - 243 y z - 28 z^2)/6 
  \nonumber \\ && 
  + (25 x + 18 y + 18 z) A(x) 
  + (18 x + 25 y + 18 z) A(y) 
  \nonumber \\ && 
  + (18 x + 18 y + 25 z) A(z)
,
\\
Q\frac{\partial}{\partial Q} f_{\Vbar VV}(x,y,z) &=& 
  x^2 + 4 x y - 6 y^2 + 4 x z + 9 y z - 6 z^2 
  + (6 y + 6 z - x) A(x) 
  \nonumber \\ && 
  + 3 (3 x - 2 y + 3 z) A(y)/2 
  + 3 (3 x + 3 y - 2 z) A(z)/2
,
\\
Q\frac{\partial}{\partial Q} f_{\Vbar\Vbar V}(x,y,z) &=& 
  [x y - x z - y z - 6 z^2 + 3 (y + z) A(x) + 3 (x + z) A(y)]/2
,
\\
Q\frac{\partial}{\partial Q} f_{\eta\eta V}(x,y,z) &=& 
  x^2 + 6 x y + y^2 + 3 x z + 3 y z - 5 z^2/3 - 3 x A(x) - 3 y A(y) 
  \nonumber \\ && 
  + (z -3 x - 3 y) A(z)
,
\\
Q\frac{\partial}{\partial Q} f_{\eta\eta \Vbar}(x,y,z) &=& 
  x^2 - 2 x y + y^2 - x z - y z - z^2 + x A(x) + y A(y)
  \nonumber \\ && 
   + (z - x - y) A(z)
,
\\
Q\frac{\partial}{\partial Q} f_{FFS}(x,y,z) &=& 
  2 [(z - x - y) (x + y + z) + x A(x) + y A(y)
  \nonumber \\ && 
   + (2 x + 2 y - z) A(z)]
,
\\
Q\frac{\partial}{\partial Q} f_{\Fbar\Fbar S}(x,y,z) &=& 
  4 [-x - y - z + A(x) + A(y) + A(z)]
,
\\
Q\frac{\partial}{\partial Q} f_{FFV}(x,y,z) &=& 
  z (6 x + 6 y + 8 z/3) + (6 x + 6 y - 4 z) A(z)
,
\\
Q\frac{\partial}{\partial Q} f_{FF\Vbar}(x,y,z) &=& 
  -8 x y -2 x z -2 y z + 4 x A(x) + 4 y A(y) + 2 (x + y) A(z)
,
\\
Q\frac{\partial}{\partial Q} f_{\Fbar\Fbar V}(x,y,z) &=& 
  -4 (x + y + 3 z) + 12 A(x) + 12 A(y) + 12 A(z)
,
\\
Q\frac{\partial}{\partial Q} f_{\Fbar\Fbar\Vbar}(x,y,z) &=& 
  4[-x - y - z + A(x) + A(y) + A(z)]
,
\\
Q\frac{\partial}{\partial Q} f_{\Fbar FG}(x,y,z) &=& 
  2 [(x - y - z) (x + y + z) + (2 y + 2 z - x) A(x)  
 \nonumber \\ && 
 + y A(y) + z A(z)]
.
\eeq
\newpage

{\it Acknowledgments:} 
This work was supported in part by the National Science Foundation grant number PHY-1719273.
HHP was supported in part by US Department of Energy Contract de-sc0011095.


\end{document}